\documentclass{article}

\usepackage{arxiv}
\usepackage[utf8]{inputenc} % allow utf-8 input
\usepackage[T1]{fontenc}    % use 8-bit T1 fonts
\usepackage{hyperref}       % hyperlinks
\usepackage{url}            % simple URL typesetting
\usepackage{booktabs}       % professional-quality tables
\usepackage{amsfonts}       % blackboard math symbols
\usepackage{nicefrac}       % compact symbols for 1/2, etc.
\usepackage{microtype}      % microtypography
\usepackage{lipsum}		% Can be removed after putting your text content
\usepackage{longtable}
\usepackage{multirow}
\usepackage{graphicx}
\usepackage{natbib}
\usepackage{doi}
\usepackage{lineno}
\usepackage{rotating}
\usepackage[tbtags]{amsmath}
\usepackage[dvipsnames]{xcolor}
\usepackage[width=1\textwidth]{caption}
\renewcommand{\figurename}{\textbf{Figure}}
\renewcommand{\tablename}{\textbf{Table}}
\renewcommand{\thetable}{\textbf{\arabic{table}}}
\renewcommand{\thefigure}{\textbf{\arabic{figure}}}
\usepackage{lscape}
\setcitestyle{authoryear,open={(},close={)}} %Citation-related commands
\usepackage{setspace}
\usepackage{tabularx}
\usepackage{threeparttable}
\usepackage{afterpage}
\usepackage{appendix}
\usepackage{tablefootnote}
\usepackage{anyfontsize}
\usepackage{float}
\title{Daily fluctuations in weather and economic growth at the subnational level: evidence from Thailand\thanks{ All data in this study are used only for research purposes. None of these were from or generated by the SCB. In the future, with written permission from the data provider, datasets that are not publicly available may be shared with other academic or commercial organizations I collaborate with to carry out further related research (where this is in the public interest or necessary for the purposes of another organization’s legitimate research interests). I will always make sure that any such sharing for further research is done in compliance with data protection laws, and under terms that protect data privacy and the confidentiality of the data.}}

\author{%\href{https://orcid.org/0000-0000-0000-0000}
        {%\includegraphics[scale=0.06{orcid.pdf}\hspace{1mm}
        Sarun Kamolthip}\thanks{This paper was initially developed during my tenure at Khon Kaen University (KKU) before joining Siam Commercial Bank (SCB). I gratefully acknowledge all the support and excellent research resources while working at the Faculty of Economics, KKU. The views expressed are solely mine and should not be reported as representing the views of the SCB or the KKU.}
        \\
	\texttt{sarun.kamolthip@scb.co.th} \\
}

\renewcommand{\headeright}{Working paper}
\renewcommand{\shorttitle}{}
\newcommand{\undertitle}{}

\hypersetup{
pdftitle={Daily_fluctuation_and_economic_growth_Thailand},
pdfsubject={q-bio.NC, q-bio.QM},
pdfauthor={Sarun Kamolthip},
pdfkeywords={climate change, economic growth},
}

    \makeatletter
\def\@fnsymbol#1{\ensuremath{\ifcase#1\or \dagger\or \dagger\dagger\or
   \mathsection\or \mathparagraph\or \|\or **\or \dagger\dagger
   \or \ddaggers\ddaggers \else\@ctrerr\fi}}
    \makeatother

\begin{document}

\maketitle
\thispagestyle{empty}

\newpage
\begin{abstract}
This paper examines the effects of daily temperature fluctuations on subnational economic growth in Thailand. Using annual gross provincial product (GPP) per capita data from 1982 to 2022 and high-resolution reanalysis weather data, I estimate fixed-effects panel regressions that isolate plausibly exogenous within-province year-to-year variation in temperature. The results indicate a statistically significant inverted-U relationship between temperature and annual growth in GPP per capita, with adverse effects concentrated in the agricultural sector. Industrial and service outputs appear insensitive to short-term weather variation. Distributed lag models suggest that temperature shocks have persistent effects on growth trajectories, particularly in lower-income provinces with higher average temperatures.

I combine these estimates with climate projections under RCP4.5 and RCP8.5 emission scenarios to evaluate province-level economic impacts through 2090. Without adjustments for biases in climate projections or lagged temperature effects, climate change is projected to reduce per capita output for 63-86\% of Thai population, with median GDP per capita impacts ranging from -4\% to +56\% for RCP4.5 and from -52\% to -15\% for RCP8.5. When correcting for projected warming biases--but omitting lagged dynamics--median losses increase to 57-63\% (RCP4.5) and 80-86\% (RCP8.5). Accounting for delayed temperature effects further raises the upper-bound estimates to near-total loss. These results highlight the importance of accounting for model uncertainty and temperature dynamics in subnational climate impact assessments. All projections should be interpreted with appropriate caution. (JEL: O44, Q51, Q53, R11)
\end{abstract}

% keywords can be removed
%\keywords{First keyword \and Second keyword \and More}

\newpage
\setcounter{footnote}{0} 
\section{Introduction}

The calendar year 2024 was the first year in which the average global temperature exceeded 1.5$^{\circ}$C relative to pre-industrial levels \citep{copernicus2025}. Although such a single year above 1.5$^{\circ}$C does not mean that the long-term temperature has reached the temperature goals, as measured by 20-year averages relative to a pre-industrial baseline, of the Paris Agreement, as underlined by the Intergovernmental Panel on Climate Change (IPCC) \citep{lee2024climate}, \cite{bevacqua2025year} show that, without very stringent climate mitigation, this very first single year at 1.5$^{\circ}$C most probably occurs within the 20-year period.

%The economic consequences of rising temperatures are increasingly well documented at global and national levels, yet much less is known about how warming climate affects economic growth within countries--particularly in developing countries where vulnerability and climate exposure vary substantially across regions. Subnational assessments are especially important for identifying heterogeneity in climate impacts and informing targeted adaptation strategies.

An increasing number of empirical studies provide quantitative estimates of the economic and societal damage caused by climate change. Understanding the economic impacts attributable to global warming is important for a comprehensive projection of climate change risks. However, because studies on the consequences of climate change are highly dependent on data, most previous studies have been conducted either at the global level (e.g., \cite{dell2012temperature, burke2015global, kotz2021day, carleton2022valuing}, or regional level (\cite{ciscar2011physical}), or in large countries that have sufficient data variation (for example, \cite{deschenes2011climate, burgess2017weather, hsiang2017estimating, barwick2018morbidity, deryugina2019mortality}. Nonetheless, people’s vulnerability to climate change may differ substantially within national geographies. As underlined by the IPCC \citep{dubash2022national}, subnational institutions play a complementary role to national institutions and are important for mitigation efforts. Therefore, assessments at the subnational level are needed to identify vulnerability at different subnational administrative units for efficient planning and policy-making. 

This study aims to assess the impact of climate change on Thailand's aggregate economic output at the subnational level. The analysis has two main features. First, I examine the historical relationship between changes in a province's weather and its economic performance. The main identification strategy exploits the plausibly random year-to-year variation in weather within a province to estimate the effects of local temperatures on economic outcomes. Specifically, by conditioning on the province fixed effects, I isolate the within-province year-to-year variation in weather exposure. Because weather fluctuations are unpredictable, it seems reasonable to presume that this variation is as good as randomly assigned and therefore orthogonal to unobserved determinants of aggregate economic outputs. The year fixed effects account for any time-varying trends or shocks that are common to all provinces in Thailand. The use of annual variation to estimate the impact of climate change was pioneered by \cite{deschenes2007economic} and \cite{schlenker2009nonlinear}, who both used annual county-level US data to estimate the impacts of weather on US agricultural outputs.

The analysis is accomplished using highly spatiotemporal granular weather data and advances in statistical theory, particularly climate econometrics. I find a statistically significant relationship between the annual growth rate of provincial output per capita and 24-hour average temperatures, particularly on extremely cold and hot days. For example, in the case of the polynomial functional form, the results show that a day at 35$^{\circ}$C, relative to the reference day of 26$^{\circ}$C, leads to a decrease in the provincial growth rate of 0.07 percentage points on average, while a day at 20$^{\circ}$C decrease this rate by 0.04 percentage points. All results indicate an inverted-U shape relationship between the provincial growth rate and 24-hour average temperatures. The finding about the relationship between temperatures and provincial growth rate, as estimated using a polynomial functional form, is robust to alternative model specifications. I then examine whether there is heterogeneity in the growth-temperature across provinces. In doing so, the temperature is interacted with a dummy for a province having low income, defined as a province having below-median average inflation-adjusted GPP per capita across the sample period. The results suggest that we cannot reject the hypothesis that high- and low-income provinces have the same response functions to temperature changes. 

I then follow \cite{dell2012temperature} and investigate whether temperature affects the growth or level of GPP per capita to better understand the dynamics of these temperature effects. In doing so, I estimate the distributed lag models with up to five lags of temperature. When assuming that high- and low-income provinces respond identically to changes in temperature, the lagged temperature effects appear to persist statistically and significantly. The results suggest that changes in temperature may affect the rate of economic growth rather than the level of economic output. However, when assuming that high- and low-income provinces respond differently to changes in temperature, the growth effects are more pronounced only in low-income provinces.

To understand the potential channels through which warming temperatures may affect aggregate output, I investigate the impacts of temperature on three components of output growth: the agricultural, industrial, and service sectors. When assuming that high- and low-income provinces respond identically to changes in temperature, I find substantially negative and statistically significant temperature effects in the agricultural sector. The pattern of temperature effects is an inverted-U shape (that is, the growth rate of agricultural components increases as temperature rises from cool to moderate and then declines). The results provide no evidence of the impact of temperature on output growth in both the industrial and service sectors. Similarly, the results, as estimated assuming that high- and low-income provinces respond differently to temperature changes, suggest that temperature changes affect growth in the agricultural sector. These findings are robust to alternative functional forms of the model. As for the other two components of GPP, the results provide no evidence of the effects of temperature on the growth of both industrial and service outputs. Again, these results are consistent across all alternative functional forms.

The second feature is the projection of future damage for each province from 2023 to 2090. I combine the empirical results with the projected changes in the climate from various climate models and economic development scenarios to generate projections of economic output under climate change in Thailand. Without bias correction in the projected climate, province-level projections under both RCP4.5 and RCP8.5 emission scenarios show that climate change will negatively affect half of the Thai population in 2050. These negative impacts are statistically uncertain, with an average of 51-57\% likelihood that climate change will have positive impacts. An increasing share of the Thai population is projected to be affected by a warming climate, while negative impacts are more certain as we approach the end of this century. In 2090, projections under the "business as usual" RCP8.5 emission scenario show that climate change will make 86\% of Thai people poorer in per capita terms than they would be in the absence of climate change, while a more aggressive emission reduction RCP4.5, 63\% are. These projections come with an average probability of 0.30-0.47 across provinces that climate change will have positive impacts. These results provide suggestive evidence that climate change will most likely affect Thailand's economy to a certain extent. Additional projections also show that the differences in the projected impact of a warming climate are mainly due to geographic heterogeneity in the baseline temperatures. In particular, the provinces in the Upper-North region, which are typically colder than other regions of Thailand, benefit more from increased average temperatures. The projections that account for the lagged effects of temperature over time, however, display substantially more negative impacts in all 77 provinces.

Next, I examine the impacts, again, without bias-correction of climate change on Thailand's gross regional product (GRP) per capita. The point estimates are uniformly more negative in the higher average temperature regions (i.e., Central, East, South, and West). The damages are modest in colder regions, such as the Lower North and Northeast regions, while the Upper North region, which has the lowest average temperature, benefits the most from increased average temperatures. Then, I examine whether the region-level projections are sensitive to different combinations of emission scenarios, specifications, and output growth assumptions. The projections are broadly similar in structure for both the RCP4.5 and RCP8.5 emission scenarios. Projected impacts in models that allow the effects of temperature to persist on regional output growth are less uncertain whether high- and low-income provinces are assumed to respond identically or differently to temperature changes. In models that do not account for lagged effects, the projections become more uncertain whether high- and low-income provinces are assumed to respond identically or differently.

Then, I project the impacts without bias-correction of climate change on Thailand's gross domestic product (GDP) per capita under the RCP8.5 emission scenario between 2023-2090 for four different historical growth-temperature response functions and three output growth assumptions combination. Similar to both provincial and regional projections, output growth assumptions likely have little impact on the projected change in GDP per capita under the RCP8.5 emission scenario. As also observed in the regional projections, projections are less uncertain and fall off steeply in earlier future years in models that allow the effects of temperature to persist on output growth because colder provinces also suffer large damages. In models that do not account for lagged effects, the projections are more uncertain whether high- and low-income provinces are assumed to respond identically or differently because colder provinces benefit somewhat from increased average temperatures, while hotter provinces remain worse off under a warming climate. I then examine the sensitivity of the projected impacts on Thailand's GDP under the emission stabilization scenario RCP4.5. Projections under both emission scenarios are broadly similar in structure but vary in magnitude.

To examine the impacts of potentially upward biases in future climate projections, I then project the impacts on provincial output per capita using a bias-correction strategy. These corrected projections are substantially different from those without bias correction. Projections under both RCP4.5 and RCP8.5 emission scenarios show that climate change will substantially affect 89-94\% of the Thai population in 2050. These projections come with an average probability of 0.10-0.20 that climate change will have positive impacts on any province. The climate-affected share of the Thai population is substantially high throughout the projected future period. Projections show that almost all Thai people will be negatively affected by climate change in 2090 under either the RCP4.5 or RCP8.5 emission scenarios, with an average probability of 0.06-0.07 that climate change will have positive impacts. In models that do not consider the delayed effects of temperature, whether high- and low-income provinces are assumed to respond identically or differently, median projections in all three output growth scenarios show that climate change reduces Thailand's output per capita by 57-63\% under RCP4.5 and 80-86\% under RCP8.5, relative to its GDP per capita in the absence of climate change. All models with delayed impacts project that Thailand will lose 94-100\% of its output in the absence of climate change. All projections are certain with only 1-6\% likelihood of positive impacts. Taken together, these results show that the projected impacts of climate change are sensitive to potential biases in future climate projections, particularly in models that do not account for lagged effects, in which the colder provinces somewhat benefit from increased average temperatures, while hotter provinces remain worse off under a warming climate.

However, there are some important caveats to these projections. First, the projected impacts are likely overestimated because adaptability to permanent changes in climate is not considered. Second, the projections also reveal a certain degree of uncertainty caused by two distinct sources of uncertainty: climate projection uncertainty and economic pathway uncertainty. Future climate projections involve considerable uncertainty arising from an incomplete understanding of the Earth’s physical systems. Similarly, predicting accurate economic growth in any economy is challenging because of a variety of factors that introduce complexity and uncertainty, particularly as financial systems become increasingly integrated. Third, these projected future damages rely on a number of strong assumptions, including that the climate projections are correct, the future growth of province-level output will remain constant in the baseline scenario or follow the growth paths that are projected by SSP3 and SSP5 pathways, and the demographics of the Thai population and their geographic distribution will remain unchanged.

The remainder of this paper is organized as follows: Section 2 describes the
data used in the estimation of the economic growth-temperature relationship and in the impact projections of climate change. Section 3 describes the empirical framework and econometric specifications, presents the results of the econometric analysis, and examines the channels through which the warming temperature may affect aggregate output. Section 4 describes the projection results at different aggregate levels as well as the limitations. Finally, Section 5 concludes the paper.

\section{Data sources and descriptive statistics}
\label{sec:data}

\subsection{Data sources}
\textit{Economic data}. To study the effects on economic output, I used two series of Gross Provincial Product per capita ($GPP_{pc}$) at current market price between 1981-2022. The economic output data is compiled by the \href{https://www.nesdc.go.th/nesdb_en/main.php?filename=national_account}{Office of the National Economic and Social Development Council} (NESDC). The \textit{old} GPP series covers 11 economic sectors and is provided from 1981–1995. The \textit{new} GPP series covers 16 economic sectors and is available in both current market prices and chain linking real terms. The national-level consumer price index (CPI) at constant 2019 price was used to deflate the GPP per capita at current market prices. The CPI dataset was obtained from \href{https://tpso.go.th/economic-data}{The Trade Policy and Strategy Office, Ministry of Commerce}. 

\textit{Weather data}. As is commonly found when dealing with ground station data \citep{auffhammer2013using, dell2014we}, Thailand's weather station coverage is sparse, and a number of missing observations can also be observed due to the birth and death of weather stations. Reanalysis estimates datasets were used instead of weather observations from ground stations because of their temporal resolution, spatial resolution, and data continuity. 

The primary source of the temperature data is the ERA5-Land reanalysis dataset \citep{munoz2021era5}. The ERA5-Land is a replay of the land component of the ERA5, which uses a vast amount of historical satellite and in-situ observations as input to the advanced modelling and data assimilation systems \citep{hersbach2020era5}. ERA5-Land dataset comes with the high spatial resolution of 0.1$^{\circ}$ x 0.1$^{\circ}$ (approximately 9 km x 9 km at the equator) and is provided from 1950 onward. I obtained 2-meter temperature and precipitation estimates at hourly frequency between 1981-2022 from \href{https://climate.copernicus.eu/climate-datasets}{the Copernicus Climate Change Service (C3S) Climate Data Store} implemented by the European Centre for Medium-Range Weather Forecasts (ECMWF). 

Although precipitation data are also provided by the ERA5-Land dataset, I used the precipitation estimates from the Climate Hazards Group InfraRed Precipitation with Station data (CHIRPS) as the primary source because precipitation is far less smooth in space than temperature. The CHIRPS incorporates the in-house climatology, satellite imagery, and in-situ station data to create gridded rainfall time series \citep{funk2015climate} and is provided from 1981 onward at daily frequency at 0.05$^{\circ}$x0.05$^{\circ}$ resolution (approximately 5 km x 5 km at the equator).

The gridded climate data was aggregated to the province-year level, which corresponds to the first administrative level as indicated in the \href{https://gadm.org/}{GADM database of Global Administrative Areas}. All the main specifications used population-weighted temperature and precipitation. The weights were constructed from the \textit{polygonized population dataset}, which was developed specifically for this study using the annual land use map and annual population count datasets. \footnote{Conceptually, this is similar to using the gridded population data, for example, \href{https://landscan.ornl.gov/}{the LandScan dataset produced by the Oak Ridge National Laboratory} and \href{https://earthdata.nasa.gov/centers/sedac-daac}{the Gridded Population of the World dataset produced by Columbia University's Center for International Earth Science Information Network (CIESIN)}, which is derived with geostatistical methods using population counts and spatial datasets. While the gridded population dataset derives the distribution of human population counts on continuous raster grids, I instead derived the distribution of population counts onto the polygons. Appendix Figure \ref{fig:lscan_vs_polypop} illustrates differences between LandScan's gridded population and this paper's polygonized population datasets.} The land use map has a scale of 1:25000 (approximately 12.5m × 12.5m in raster resolution) and provides granular details of land use classification on three levels. The first level consists of five major types of land use: urban and built-up, agricultural, forest, water bodies, and miscellaneous. The second and third levels provide finer and finest details of land use classification, respectively. The annual land use map dataset in shapefile format between 2006-2019 was obtained from the Land Development Department (LDD). To calculate a spatially disaggregated population that is compatible with the human elements of the economy, I used only the polygons classified as urban and built-up land types. The land use map of 2006 was used as a proxy for the land use pattern before 2006, whereas that of 2019 was used as a proxy for the land use pattern of 2019 onward. The annual population counts dataset at the third administrative level (or, \textit{Tambon} in Thai) between 1997-2022 was obtained from the \href{https://stat.bora.dopa.go.th/new_stat/webPage/statByAge.php}{Official Statistics Registration Systems} (OSRS) operated by the Department of Provincial Administration (DOPA).\footnote{This administrative level corresponds to, for example, county in the United States.} The population count for each polygon was distributed based on the share of each polygon's area to the total area of all polygons located within the administrative boundary of a given Tambon. Since population counts are available from 1997 onward, the population weights for the periods between 1982-1996 were calculated using 1997 population data.

The resulting polygonized population dataset was used to calculate the weather experienced by the population in each polygon. The grid-level hourly temperature $T_{cj}$, for example, was first aggregated to the polygon level. The 24-hour average temperature at grid cell $c$ covered at least partially by polygon $j$, $\frac{1}{24}\sum_{h=1}^{24}T_{cj}$, was multiplied by the corresponding cell weight $w_{cj}$ and summed over all $\textbf{C}_j$ grid cells covered at least partially by polygon $j$. Note that the cell weight $w_{cj}$ is an approximate fraction of the grid cell $c$ that is covered at least partially by the polygon $j$ and normalized such that they add up to one for each polygon. The polygon-level average temperature was then aggregated to the provincial level using a population weighting scheme. The population weight $w_{jp}$ is equal to the population of polygon $j$ divided by the total population of province $p$. Given the population weightings $w_{jp}$ for all $\textbf{J}_{p}$ polygons located within the administrative boundary of province $p$, the \textit{24-hour} population-weighted average temperature on day $d$ is expressed as:
\begin{equation}
\label{24hr_temp}
    T_{pd} = \sum\limits_{j=1}^{\textbf{J}_p} \sum\limits_{c=1}^{\textbf{C}_{j}} w_{jp} w_{cj} \cdot \frac{1}{24} \sum\limits_{h=1}^{24} T_{cj}
\end{equation} 
A similar procedure was applied for precipitation, but the 24-hour average term was omitted because the CHIRPS dataset was already provided at a daily frequency. Weather and economic data were then matched at the province-year level for 1982–2022. All nominal economic output measures were inflation-adjusted to the 2019 Thai bath. The resulting combination of economic and climate data yielded 3,086 province-year observations. The boxplots in Figure \ref{fig:summary_weather} summarize the historical population-weighted temperature and precipitation exposure. Summary statistics for the key variables are presented in Table \ref{tab:summary_stat}.

\textit{Climate and socioeconomic projection data}. The main source of future climatic projections is the NASA Earth Exchange Global Daily Downscaled Projections (NEX-GDDP) dataset. The NEX-GDDP dataset was downloaded from \href{https://www.nccs.nasa.gov/services/data-collections/land-based-products/nex-gddp}{The NASA Center for Climate Simulation}. The downscaled projections for the Representative Concentration Pathways (RCP) 4.5 and 8.5 from the 21 models and scenarios are produced and distributed. The NEX-GDDP dataset provides maximum and minimum temperatures, and mean precipitation at daily frequency of 0.25$^{\circ}$ x 0.25$^{\circ}$ (~25 km x 25 km at the equator). Owing to the limitation in computational resources, only seven of the 21 models between 2023-2090 were used to project the future impacts on economic output. Both temperature and precipitation estimates were processed using the same procedure as the historical weather data, as described previously. To project future impacts on economic output, I used the 2022 population distribution together with the 2019 land use pattern to construct a population-weighted average of the climate projection.

Appendix Figure \ref{fig:summary_future_climate} summarizes the projections of the distribution of average annual exposure across 77 provinces in Thailand in three future periods: 2031-2050, 2051-2070, and 2071-2090. The shift in the temperature distribution was clearly observed under the RPC4.5 and RCP8.5 emission scenarios. In long-term (2071-2090), population-weighted average temperatures across all provinces in Thailand are projected to rise by 2.86$^{\circ}$ under RCP 4.5 emission scenario, and by 4.06$^{\circ}C$ under RCP8.5, relative to population-weighted average temperature observed between 2003-2022. Nonetheless, a comparison of the projected and observed averages and variability between 2018-2022 shows a tendency for biases in all seven climate models. Appendix Figure \ref{fig:bias_in_projected_climate} presents the full distributions of daily mean temperature bins across Thailand, again, between 2018-2022. All projections are consistently upward-biased relative to the observed temperature. Section \ref{sec:projection} discusses the strategy used to examine how the impact projections are sensitive to these potential biases.

To account for plausible alternative trends in economic development in the far future, I considered country-level projections of socioeconomic drivers for the Shared Socioeconomic Pathways (SSP) (Release 3.1, July 2024). The SSP data were downloaded from \href{https://data.ece.iiasa.ac.at/ssp/#/downloads}{International Institute for Applied Systems Analysis (IIASA)}. I focus on the projections from the OECD ENV-Growth 2023 model, as it is the only model that projects per capita growth rates at the country level. The available data provide projected GDP per capita in constant 2017 dollar purchasing power parity (PPP) at five-year intervals at the country level. To calculate the annual projected growth rates, I assumed that Thailand grows at the same growth rate over these five years.

\section{Relationship between temperature fluctuations and aggregate economic output}
\label{sec:econ_temp_relationship}

This section presents an empirical framework for the analysis of how temperature can affect aggregate economic output. I first discuss the empirical framework and specifications used to estimate the impact of temperature fluctuations on economic growth. The third subsection reports and discusses the results from estimating models that assume a common response across provinces to temperature changes. To uncover differences in the growth-temperature response functions across population groups, this subsection also reports and discusses the results of estimating the models in which the temperatures are interacted with a low-income province dummy, defined as a province having below median average inflation-adjusted GPP per capita across the sample period. The last subsection further investigates the impacts of temperature on the components of GPP to shed light on the potential channels through which temperature may affect the aggregate output.

\subsection{Empirical framework}
Following the empirical framework derived in \cite{hsiang2016climate} and \cite{deryugina2014does}, I focus on an economy in which agents reallocate capital and labor to respond to temperature changes in a way that maximizes profits. On a given day indexed by $d$, an economy in province $p$ is conceptualized as one where producers observe the weather, adjust capital and labor resources in response, and then produce output. If the same sequence occurs over the course of a year (i.e. 365 times), under the assumption of temporal and spatial separability, i.e. that the economy conditional on temperature is additively separable across locations and moments of time, the annual aggregate output $Y_{py}$ can then be written as the sum of daily outputs observed across all locations $l$ in the province $p$ from each day $d$ in the year $y$
\begin{equation} \label{conceptual}
    Y_{py} = \sum_{d \in y} \sum_{l \in p} Y_{sd} \\
\end{equation}
Although the output data are usually measured at a higher level of aggregation, it is possible to recover nonlinear relationships at the grid cell level at which climatic data are recorded \citep{hsiang2016climate}. Suppose $h(.)$ is the local and instantaneous nonlinear economic-temperature response function that best describes the daily output observed at the end of the day and can be approximated as a linear combination of $M$ simple nonlinear functions. For a given climate on day $d$ and location $l$,
\begin{equation} \label{dose_response}
    h(\textbf{c}_{ld}) \approx \sum_{m=1}^{M} \beta_m h_m (\textbf{c}_{ld})
\end{equation}

Substituting the approximation from Equation \ref{dose_response} into Equation \ref{conceptual} and interchanging the order of the summation yields:
\begin{eqnarray} \label{framework}
    Y_{py} &=& \sum_{d \in y} \sum_{l \in p} Y_{sd} \nonumber \\
           &\approx& \sum_{d \in y} \sum_{l \in p} \sum_{m=1}^{M} \beta_m h_m (\textbf{c}_{ld}) \nonumber \\
           &=& \sum_{m=1}^{M} \beta_m \underbrace{\Bigg(\sum_{d \in y} \sum_{l \in p} f_m  (\textbf{c}_{ld}) \Biggr)}_{\Bar{h}_{mpy}} \nonumber \\
           &=& \sum_{m=1}^{M} \beta_m \Bar{h}_{mpy}
\end{eqnarray}
The regressors $\Bar{h}_{mpy}$ are the nonlinear transformation of weather (e.g., temperature and precipitation) and, in practice, are computed at the grid cell level before aggregating the values to the same spatial level as that of the aggregate output data using population weights and summing over days within a year \citep{carleton2022valuing}.

As suggested by \cite{deryugina2014does}, the functional form analogous to Equation \ref{framework} was first introduced by \cite{deschenes2011climate}. In the context of aggregate economic outcomes, $\beta_m$ has been successfully modeled as an $M$th order polynomial or restricted cubic spline \citep{burke2015global}, or an $M$-stepwise constant or “binned” function \citep{deryugina2014does}. Under similar assumptions, an analogous approach has also been used in the agronomic literature \citep{schlenker2009nonlinear}, in the context of health \citep{carleton2022valuing, deryugina2019mortality}, and to study the impacts of higher temperatures on electricity consumption \citep{aroonruengsawat2011impacts}. 

\subsection{Econometric specifications}
Although the empirical framework Equation \ref{framework} focuses on the effect of temperature on the level of economic output, researchers practically use economic growth as a dependent variable, in part because aggregate economic output measures usually exhibit high serial correlation.\footnote{See \cite{burke2015global} for further discussion.} Moreover, as a given effect size on the growth path of output will ultimately exceed the same effect size on the level of economic output for the projected long-run effects of climate change, it is important for policymakers to understand whether the effects of temperature generate a permanent or temporary loss of output relative to the trend (i.e., growth versus level effects) \citep{dell2014we}. Using economic growth as the dependent variable, \cite{dell2012temperature} demonstrate that a higher temperature can reduce economic growth, not just the level of output.\footnote{\cite{dell2012temperature} investigate further using Monte Carlo analysis and show that running a growth regression is an effective approach to produce unbiased estimates of both growth and level effects given the properties of their simulated data.} Following previous climate-econometric literature \citep{burke2015global, dell2012temperature}, I took the first difference of the logarithm of annual inflation-adjusted GPP per capita. Specifically, I estimated the following fixed-effects panel regression of the form
\begin{equation} \label{eqn:general_spec}
    g_{py} = \sum_{m} \beta_m h_m(\textbf{T}_{py}) + \sum_{n} \rho_n h_n(\textbf{R}_{py}) + \alpha_p + \alpha_y + \epsilon_{py}
\end{equation}
where $g_{py}$ is the annual growth rate of output per capita of province $p$ for year $y$, $\textbf{T}_{py}$ is the daily average temperature vector, and $\textbf{R}_{py}$ is the daily total precipitation vector. Both $\textbf{T}_{py}$ and $\textbf{R}_{py}$ were summed over the year. The effects of changes in precipitation were accounted for because changes in local annual temperatures and precipitation were historically correlated \citep{auffhammer2013using}. Unless otherwise specified, all the main specifications include both province fixed effects and year fixed effects. A set of province fixed effects $\alpha_p$ accounts for unobserved time-invariant differences that influence provinces’ average growth rates, such as their history, culture, or geography. $\alpha_y$ is a set of year fixed effects that accounts for time-varying differences that are common across provinces (e.g., technological innovations or changes in national economic policies) as well as year-specific shocks (e.g., a sudden rise in energy prices or a global economic crisis). Following \cite{dell2012temperature}, the standard error terms $\epsilon_{py}$ were clustered at the province level to account for temporal autocorrelation within each province.

To uncover possible nonlinear relationships between temperatures and the annual growth rate of output per capita, I estimated three different functional forms for $h(\textbf{T}_{py})$: $M$th order polynomial, binned regression, and heating/cooling degree days functional forms. The binned regression is an important functional form in previous climate-impacts research (e.g. \cite{schlenker2009nonlinear, aroonruengsawat2011impacts, deschenes2011climate, deryugina2014does, deryugina2019mortality}. However, \cite{deschenes2011climate} suggested that binned regression likely overestimates the impacts of climate change because it failed to account for adaptations undertaken in response to permanent changes in climate. Similarly, \cite{dell2012temperature} also indicated that the temperature effects, as estimated using a fixed effects panel regression with a linear temperature, may be affected by not accounting for adaptability to permanent changes in climate. \cite{burke2015global} argued that an estimation using a fixed effects model with higher-order temperature terms to identify the effects of temperature could bridge this gap somewhat because economic agents with different average temperatures were allowed for historical adaptation to longer-run temperature changes. The heating/cooling degree days functional form was additionally used to present possible variations in the estimated temperature effects. This functional form is similar to the concept of growing degree days in the agronomic literature.

The first functional form assumed that $h(\textbf{T}_{py})$ was an $M$th order polynomial of the form $h(\textbf{T}_{py}) = \sum_{m=1}^M \beta_m T_{py}^m$. Increasingly flexible specifications up to seventh-order polynomials were estimated to search for the most parsimonious specification that provides sufficient flexibility to capture nonlinearity patterns in the data.\footnote{Specifically, the gridded hourly temperatures $T_{cj}$ in Eq.\ref{24hr_temp} were first transformed to the desired polynomial order, averaged out over day $d$, and then summed across the year $y$ to create the annual temperature vector of province $p$ as follows:
\begin{equation}
\label{eqn:poly_calc}
    T_{py} = \Bigg[ \sum\limits_{d \in y} \sum\limits_{j=1}^{\textbf{J}_p} \sum\limits_{c=1}^{\textbf{C}_{j}} w_{jp} w_{cj} \cdot \frac{1}{24} \sum\limits_{h=1}^{24} T_{cj}, \sum\limits_{d \in y} \sum\limits_{j=1}^{\textbf{J}_p} \sum\limits_{c=1}^{\textbf{C}_{j}} w_{jp} w_{cj} \cdot \frac{1}{24} \sum\limits_{h=1}^{24} T_{cj}^{2}, ..., \sum\limits_{d \in y} \sum\limits_{j=1}^{\textbf{J}_p} \sum\limits_{c=1}^{\textbf{C}_{j}} w_{jp} w_{cj} \cdot \frac{1}{24} \sum\limits_{h=1}^{24} T_{cj}^{7} \Bigg] \nonumber
\end{equation}
A similar procedure was applied for precipitation but omitted $\frac{1}{24} \sum\limits_{h=1}^{24}$ term as the CHIRPS dataset was already provided at a daily frequency. 
} 
Quadratic polynomials of precipitation were used to account for the effects of changes in precipitation in all examined specifications. Figure\ref{fig:model_specs} shows the resulting response functions for the alternative specifications. The preferred specification used a 24-hour average quadratic polynomial in daily temperature (i.e., $h(T) = \beta_1 T_{py} + \beta_2 T_{py}^2$). Substituting the function $h(\textbf{T}_{py})$ and $h(\textbf{R}_{py})$ in Equation \ref{eqn:general_spec}, the model becomes
\begin{equation} \label{eqn:polynomial_function}
    g_{py} = \beta_1 T_{py} + \beta_2 T_{py}^2 + \rho_1 R_{py} + \rho_2 R_{py}^2 + \alpha_p + \alpha_y + \epsilon_{py}
\end{equation}

The second functional form estimated $h(\textbf{T}_{py})$ using binned regression. Following the notation of \cite{deryugina2014does}, $h(\textbf{T}_{py})$ was a stepwise function of the form $h(\textbf{T}_{py}) = \sum_{m=1}^M \beta_m \Tilde{T}_{m,py}$, where $\Tilde{T}_{m,py}$ was calculated by summing the fraction of days in which the population in province $p$ was exposed to the $m$th temperature bin across year $y$.\footnote{To calculate the number of days the population in province $p$ in year $y$ were exposed to $m$th temperature bin, $\Tilde{T}_{m,py}$, with lower bound $\underline{m}^{\circ}C$ and upper bound $\overline{m}^{\circ}C$ in the binned regression, I summed the fraction of day in which the population in province $p$ was exposed to temperature bin $m$ in day $d$ (i.e. number of hours of the gridded hourly temperature $T_{cj}$ in temperature bin $m$ $\times$ $\frac{1}{24}$) across the year as follows:
\begin{equation}
\label{eqn:binned_calc}
    \Tilde{T}_{m,py} = \Bigg[ \sum\limits_{d \in y} \sum\limits_{j=1}^{\textbf{J}_p} \sum\limits_{c=1}^{\textbf{C}_{j}} w_{jp} w_{cj} \cdot \frac{1}{24} \sum\limits_{h=1}^{24} 1\cdot[T_{cj} \in [\underline{m},\overline{m})] \Bigg] \nonumber
\end{equation}
The similar procedure applied for precipitation but omitted $\frac{1}{24} \sum\limits_{h=1}^{24}$ term as the CHIRPS dataset was already provided at daily frequency. 
} This stepwise function can be considered an important benchmark, as it regresses the outcome on year-total time within each temperature bin; therefore, it is closest to being fully non-parametric \citep{carleton2022valuing}. I explored various combinations of temperature intervals and omitted bins. All combinations used six bins of daily total precipitation to account for the effects of different precipitation levels. The specification selection procedure is further detailed in Appendix \ref{si:methods}. The preferred specification used 5$^{\circ}C$ interval and seven bins of 24-hour average of hourly temperature. 

\begin{equation} \label{eqn:binned_function}
    g_{py} = \sum_{m=1}^{7} \beta_m \Tilde{T}_{m,py} + \sum_{n=1}^6 \beta_n \Tilde{R}_{m,py} + \alpha_p + \alpha_y + \epsilon_{py}
\end{equation}

The bottom bin ($\Tilde{T}_{m=1}$) lumps the fraction of days in which province $p$ is exposed to temperatures below 13$^{\circ}C$ into one bin ($\Tilde{T}_{m=1}$), summed across the year. $\Tilde{T}_{m=2}$ sums the fraction of days when $T \in [13,18)^{\circ}C$ across the year, $\Tilde{T}_{m=3}$ sums the fraction of days when $T \in [18,23)^{\circ}C$ across the year, and so on. The top bin ($\Tilde{T}_{m=7}$) lump the fraction of day in which the temperatures are above 38$^{\circ}C$, summed across the year. The omitted bin, which is captured by the model's intercept, is $\Tilde{T}_{m=4}$ when $T \in [23,28)^{\circ}C$, summed across the year. The $\beta_m$ estimates represent the marginal effects on the annual growth rate of output per capita resulting from one additional day in the $m$th temperature bin compared to a day in the temperatures in the omitted bin. The daily total precipitation bins $\Tilde{R}_{m,py}$ were similarly constructed across six bins. The bottom bin ($\Tilde{R}_{m=1}$) corresponds to the number of days with no precipitation, and the top bin ($\Tilde{R}_{m=6}$) corresponds to the number of days when the daily total precipitation is greater than 40 mm. The remaining bins cover a 10 mm span of the daily total precipitation.

The third functional form applied the concept of heating/cooling degree days, which is commonly found in the agronomic literature \citep{schlenker2009nonlinear}. I estimated $h(\textbf{T}_{py})$ using a piecewise linear function in which the impacts of beneficial temperatures and output-decreasing temperatures were separated around the critical temperature threshold. As discussed previously, binned regression is closest to being non-parametric and is therefore an important benchmark. I directly used the lower and upper edges of the omitted bin $\Tilde{T}_{m=4}$ of the preferred specification in Equation \ref{eqn:binned_function} as the thresholds for the heating and cooling degree days, respectively, to allow the results to be comparable with those of the binned regression.
\begin{equation} \label{eqn:hdd_cdd_function}
    g_{py} = \beta_{HDD} HDD_{py} + \beta_{CDD} CDD_{py} + \rho_1 R_{py} + \alpha_p + \alpha_y + \epsilon_{py}
\end{equation}
 where the heating degree hours below 23$^{\circ}C$ and the cooling degree hours above 28$^{\circ}C$ were each summed over day across the year.\footnote{Specifically, the heating degree days ($HDD_{py}$) below 23$^{\circ}C$ and the cooling degree days above 28$^{\circ}C$ ($CDD_{py}$) were calculated as follows:
\begin{equation}
\label{eqn:hdd_calc}
    HDD_{py} = \Bigg[ \sum\limits_{d \in y} \sum\limits_{j=1}^{\textbf{J}_p} \sum\limits_{c=1}^{\textbf{C}_{j}} w_{jp} w_{cj} \cdot \frac{1}{24} \sum\limits_{h=1}^{24} (23-T_{cj}) \Bigg], \forall T_{cj} < 23^{\circ}C \nonumber
\end{equation}

\begin{equation}
\label{eqn:cdd_calc}
    CDD_{py} = \Bigg[ \sum\limits_{d \in y} \sum\limits_{j=1}^{\textbf{J}_p} \sum\limits_{c=1}^{\textbf{C}_{j}} w_{jp} w_{cj} \cdot \frac{1}{24} \sum\limits_{h=1}^{24} (T_{cj} - 28) \Bigg], \forall T_{cj} > 28^{\circ}C \nonumber
\end{equation}
}
Linear precipitation was used to account for the effects of changes in precipitation. Figure\ref{fig:model_specs} shows the resulting response functions of the alternative specifications. All alternative specifications exhibit an inverted-U pattern in the data. While the threshold temperatures of HDD ($28^{\circ}$C) and CDD ($28^{\circ}$C) are simply chosen because they are analogous to the omitted bin of the preferred binned functional form, the preferred specification is also more conservative, as the slope of the cooling degree days above $28^{\circ}$C is less steep than other CDD thresholds, and the slope of the heating degree days below $23^{\circ}$C is indifferent to those of the other HDD thresholds. Appendix \ref{si:methods} provides full details of the specification selection of each alternative functional form, using the same set of fixed effects and controls as described in Equations \ref{eqn:polynomial_function}, \ref{eqn:binned_function}, and \ref{eqn:hdd_cdd_function}, respectively.

Note that the identification strategy of all three functional forms relies on the presumably random year-to-year local variation in temperature, so the omitted variable bias is likely less vulnerable. Specifically, by conditioning on the province fixed effects $\alpha_p$, this ensures that the systematic patterns of weather in each province are isolated from the within-location year-to-year variation in temperature exposure. Because weather fluctuations are unpredictable and potentially difficult for economic agents to anticipate, it seems reasonable to presume that the variation is as good as randomly assigned. The year fixed effects $\alpha_y$ account for any time-varying trends or shocks that are common across all provinces in Thailand. Following \cite{deschenes2011climate, schlenker2009nonlinear}, I assumed that the detrended year-to-year variations within each province were uncorrelated with year-to-year variations in potentially important factors that might affect economic growth. 

\subsection{Empirical results}
This subsection presents the impact of temperature on the growth of provincial aggregate output, as observed in the historical data. First, I discuss a common growth-temperature response function, as estimated using Equations \ref{eqn:polynomial_function}, \ref{eqn:binned_function}, and \ref{eqn:hdd_cdd_function}. The next results uncover differences in the growth-temperature response functions across population groups by interacting the temperature with a dummy for a province having \textit{low income}, defined as a province having below median average inflation-adjusted GPP per capita across the sample period. To better understand the dynamics of the temperature effects in both low-income and no dummy models, I further discuss whether temperature affects growth through level effects or growth effects \citep{dell2012temperature}, using more flexible models with up to five lags of temperature. The results of various robustness checks on alternative specifications are also reported. 

\subsubsection{A common growth-temperature response function} \label{sec:pooled_model}
I begin by estimating Equations \ref{eqn:polynomial_function}, \ref{eqn:binned_function}, and \ref{eqn:hdd_cdd_function}. These simple models examine the null hypothesis that temperature does not affect the growth. The results demonstrate an average growth-temperature response function across 77 provinces in Thailand. Specifically, the estimation uncovers the average growth-temperature response function $h(T)$ in the general panel specification in Equation \ref{eqn:general_spec}.

Each panel in Figure \ref{fig:baseline_model} presents the main estimation results of each alternative functional form assumption. \footnote{As described by \cite{burke2015global}, testing temperature effects at different temperatures is different than typical testing for significant parameter estimates because the models are nonlinear, so the temperature effects vary at different temperature and thus must be evaluated at specific temperatures in order to have meaning.} The graphs visualize a common temperature response function across 77 provinces in Thailand, evaluated at each 24-hour average temperature. Interestingly, all three functional forms exhibit analogous patterns in the data, suggesting the inverted-U shape of the response function: the growth rate of output per capita increases as temperature rises from cool to moderate (specifically, the reference temperature), and then declines. The point estimates can be interpreted as the effect of a single day at each 24-hour average temperature on the growth rate of output per capita relative to a day with an average reference temperature. For example, the results indicate that a day at 35$^{\circ}$C leads to a decrease in the GPP growth rate of approximately 0.07 percentage points, relative to the reference day of 26$^{\circ}$C for the polynomial functional form (Panel A). Similarly, the results for the degree days functional form (Panel C) indicate that a day at 35$^{\circ}$C, which is equivalent to 7 degree days relative to a temperature threshold of 28$^{\circ}$C, leads to a decrease in the GPP growth rate of approximately 0.07 percentage points. As for the binned regression (Panel B), a day at 35$^{\circ}$C, of which the effect is captured by the coefficient on the 33-38$^{\circ}$C temperature bin, leads to a decrease in the GPP growth rate of approximately 0.10 percentage points, relative to the omitted $[23,28)^\circ$C temperature bin. These effects are statistically significant at the 95\% confidence level.

Column 1 in Table \ref{tab:growth_level_effects_pooled} additionally details the estimated growth-temperature responses at specific temperatures, relative to a day with the respective reference temperature in each functional form, to facilitate comparison of the estimation results. The point estimates in the polynomial (Panel A) and degree days (Panel C) functional forms are statistically significant, while those in the binned regression (Panel B) are broadly comparable in magnitude but only statistically significant at some temperatures. Regardless of statistical significance, the estimated impacts of temperature in all functional forms decrease as the temperature rises from cool to moderate and then increase. These results make it difficult to reject the hypothesis that the growth-temperature response is nonlinear.

For a quick understanding of these temperature effects, here, I adapt the procedure described by \cite{deryugina2014does} to calculate a marginal change in the annual growth rate of the economic output of 1$^{\circ}$C warming. Suppose the GPP growth was uniform across 365 days in a year, then a decrease of 0.014 percentage points of an annual growth from a day at 30$^{\circ}$C\footnote{Since the population-weighted average temperatures that are projected by seven climate models under RCP4.5 and RCP8.5 between 2071-2090 are 29.37$^{\circ}$C and 30.57$^{\circ}$C (see Appendix Figure \ref{fig:summary_future_climate}), respectively, I therefore take the estimates at 30$^{\circ}$C to showcase this.}, relative to the reference day of 26$^{\circ}$C for the polynomial functional form, indicates the growth of that day is roughly $0.014\% \cdot 365 = 4.990\%$ less than the growth of an average day of 26$^{\circ}$C. Linearizing the effect of temperature relative to the approximate zero effect at 26$^{\circ}$C, a marginal change in an annual GPP growth rate of 1$^{\circ}$C rise in temperature is $\frac{-4.990\%}{4^{\circ}C}= -1.248\%$. Analogously, for the binned regression, a decrease of 0.041 percentage points of an annual GPP growth from a day at 30$^{\circ}$C indicates the growth of that day is roughly 14.801\% less than the growth of an average day in the omitted temperature bin. The effects of a 1$^{\circ}$C rise in temperature on an annual GPP growth rate is -3.700\%, relative to the approximate zero effect at 26$^{\circ}$C.\footnote{To be comparable with the calculation of the polynomial functional form, I use 26$^{\circ}$C, which is also the mid-point of the omitted bin, as the reference. If we use 23$^{\circ}$C or 28$^{\circ}$C as the references, the effects of 1$^{\circ}$C rise in temperature on an annual GPP growth rate are -2.114\% and -7.401\%, respectively.} As for the degree days functional form, the effects of a 1$^{\circ}$C rise in temperature on an annual GPP growth rate is -3.799\%, relative to the approximate zero effect at the CDD threshold of 28$^{\circ}$C. Taken together with the estimated growth-temperature responses at various temperatures, these results suggest that estimates using the second-order polynomial functional form are, if anything, more conservative in estimating the impacts of warming temperatures on economic output.

In addition to robustness to alternative functional forms, I considered robustness to alternative model specifications and samples. Appendix Table \ref{tab:robustness_results_common} reports the results of estimating a variety of robustness checks for the average growth-temperature response function $h(T)$ in which the average effect across 77 provinces was recovered. To be comparable with the main estimation results, the regression estimates of the models in Figure \ref{fig:baseline_model} are repeated in the first column of Appendix Table \ref{tab:robustness_results_common}. Still, three functional form assumptions were estimated for each robustness check. I first dropped precipitation (column 2) to investigate whether changes in precipitation affect temperature estimates. Columns 3-6 demonstrate models that used an alternative set of controls relative to the baseline model. In column 3, I added region-by-year FE; in column 4, I adapted the specification of \cite{burke2015global} by adding quadratic province-specific time trends; in column 5, I adapted the specification of \cite{dell2012temperature} by replacing year FE with region-by-year FE and poor-year FE; and in column 6, I replaced year FE with quadratic time trends. Column 7 estimated the baseline model using a balanced sample in which all provinces were present in the sample for the entire period. Column 8 estimated the baseline model plus one lag of the per capita growth rate to account for potential time-varying omitted variables \citep{burke2015global}.

\textit{Role of precipitation.} When not controlling for changes in precipitation, the point estimates in column 2 show that the temperature effects on growth are similar in magnitude to those in the baseline results. These results suggest that whether or not accounting for precipitation does not substantially affect temperature estimates. Notably, the effects of precipitation are typically statistically insignificant and broadly consistent across a range of alternative model specifications for both the polynomial (Panel A) and temperature bins (Panel B) functional forms. Nonetheless, the effects of precipitation in quadratic province-specific time trends (column 4), balanced sample (column 7), and lagged dependent variable (column 8) specifications are statistically significant and almost identical to the baseline results for the degree days functional form (Panel C). \footnote{The effects of annual precipitation on growth are not shown to conserve space. Estimates are available upon request.}

\textit{Alternative model specifications and samples.} The results in columns 3-6 of the polynomial functional form (Panel A) are broadly consistent in magnitude and statistical significance as I change the set of controls. In contrast, the temperature estimates and statistical significance in the binned functional form (columns 3-6 of Panel B) appear to be sensitive to alternative sets of controls, suggesting that they should be interpreted with caution. These results are not unexpected since the binned model is the method that is demanding of the data and columns 3-6 all have more control variables, compared to the baseline specification. The results in columns 3-6 of the degree days functional form (Panel C) are somewhat mixed. The HDD estimates in columns 3, 5, and 6 are comparable in magnitude to the baseline specification. However, for the CDD estimates, only the results in columns 3 and 5 are similar in magnitude to the baseline specification. The results in column 4, which is demanding of the data as the model incorporates province-specific time trends as well as province FE and year FE, appear to be statistically insignificant and smaller in magnitude than the baseline results for both HDD and CDD estimates. The results in column 8 of all three functional forms are close in both magnitude and standard errors to those in the baseline results, suggesting that the results are not affected by the lack of control for time-varying omitted variables. Column 7 shows how the temperature estimates change for alternative samples. Both the point estimates and standard errors of all three functional forms are close to those in the baseline specification.

I further investigated the alternative formulation of the growth-temperature response function $h(T)$ to understand the nonlinear response observed in Figure \ref{fig:baseline_model}. Following \cite{burke2015global}, I substituted temperature interacted with average temperature and temperature interacted with average GPP per capita for the quadratic temperature term in Equation \ref{eqn:polynomial_function}. Appendix Table \ref{tab:alternative_formulation} presents these estimation results. In the absence of interaction with the province’s average income, the results show strong evidence of a nonlinear and concave temperature response. However, when the temperature-income interaction was included, the results indicate that the growth-temperature responses are driven by differences in average temperature and are affected by differences in income. See Appendix \ref{si:alternative_formulation} for further details.

\subsubsection{Heterogeneity in the growth-temperature response function}
The above results, as estimated from the pooled model where response functions were assumed to be the same across provinces, reject the null hypothesis that temperature has no effect on growth. Earlier studies at the global scale \citep{nordhaus2006geography, dell2009temperature, dell2012temperature, burke2015global} suggested that hot countries tended to be poor and cold countries were rich. \cite{dell2012temperature} found that hot countries exhibited the larger (negative) temperature effect than that of the cold countries, but being poor was a critical factor that determined this relationship. However, \cite{burke2015global} found limited evidence of heterogeneity in temperature response between rich and poor countries. This section further investigates by allowing high- and low-income provinces to respond differently to temperature changes. To uncover the differential growth-temperature response functions across provinces, the temperature was interacted with a dummy for a province having \textit{low income}, defined as a province having below-median average inflation-adjusted GPP per capita across the sample period.

Each column in Figure \ref{fig:response_xPoor_GPP} presents the estimation results for each alternative functional form assumption. The graphs display temperature response functions for both low-income (Panels D, E, and F) and high-income (Panels A, B, and C) provinces at each 24-hour average temperature. The point estimates can be interpreted as the effect of a single day at each 24-hour average temperature on the growth rate of output per capita, relative to a day with an average reference temperature. Both high- and low-income provinces broadly exhibit an inverted-U pattern in the data, except for the flatter response function in the high-income subsample in the binned regression (Panel B). The point estimates between the high- and low-income subsamples in the polynomial and degree days functional forms are broadly similar in magnitude, especially at high temperatures. However, these estimates between the high- and low-income subsamples are substantially different in the binned regression. Because the data are broken into subsamples and binned regression is a method that is demanding of the data, these results are not unexpected. In no case are the response functions in both high- and low-income provinces statistically different from the point estimates of the benchmark model, as estimated assuming a common response across high- and low-income provinces to temperature changes. These results suggest that we cannot reject the hypothesis that high- and low-income provinces have the same response functions to temperature changes.

To facilitate the comparison of the estimation results, columns 1-2 in Table \ref{tab:growth_level_effects_xPoor} detail the estimates of the temperature response functions at specific temperatures, relative to a day with the respective reference temperature in each functional form for both high- and low-income provinces. The results in the polynomial (Panel A) and degree days (Panel C) functional forms show that the temperature responses of both high- and low-income provinces are negative and statistically significant at high temperatures. The estimates in the binned regression (Panel B) are consistent in sign but not always statistically significant. These results suggest that we cannot reject the hypothesis that the growth-temperature responses of either low- or high-income provinces are zero at all points in the temperature distribution.

\textit{Robustness checks.} Appendix Table \ref{tab:robustness_heterogenous_response} reports the results of estimating the baseline specification (column 1) and a variety of robustness checks (columns 2-8) for differential growth-temperature response functions. All alternative specification checks are analogous to those in Appendix Table \ref{tab:robustness_results_common} (see Section \ref{sec:pooled_model} for details). Regardless of statistical significance, the estimated parameters in the baseline specification (column 1) of all three alternative functional forms exhibit a nonlinear and concave structure of the growth-temperature response function $h(T)$ in both low- and high-income provinces. No interaction terms in the baseline specification of the polynomial (Panel A) and the degree days (Panel C) functional forms are statistically significant, but the interaction terms of the binned functional form (Panel B) are only statistically significant in the two highest temperature bins ($[33,38)^{\circ}$C and $>38^{\circ}$C bins). These results suggest, at least for the polynomial and the degree days functional forms, that we cannot reject the hypothesis that high- and low-income provinces respond identically to changes in temperature.

\subsubsection{Growth effects versus level effects}
The above sections discuss the results of estimating simple models with no lags. This section investigates further by considering more flexible models with up to five lags of temperature to test whether temperature affects the growth or level of GPP per capita to better understand the dynamics of these temperature effects \citep{dell2012temperature,dell2014we}. I followed \cite{dell2012temperature} and estimated the generalized forms of Equations \ref{eqn:polynomial_function}, \ref{eqn:binned_function}, and \ref{eqn:hdd_cdd_function} discussed previously by adding lags of temperature and precipitation. The growth versus level effects are identified by adding both the immediate and lagged effects of temperature fluctuations across years. If the cumulative effect of temperature and its lags shrinks to zero, it indicates that temperature affects the level of aggregate economic output. However, if the summed effects are similar to or even larger in absolute magnitude than the immediate effect, these results suggest growth effects. 

Table \ref{tab:growth_level_effects_pooled} reports the estimated temperature responses at various temperatures, assuming that high- and low-income provinces respond identically to changes in temperature. Column 1 shows the immediate effects of temperature, while columns 2-4 shows the cumulative effect of temperature and its lags (up to five lags). Again, these estimates can be interpreted as the effect of a single day at each 24-hour average temperature on the growth rate of output per capita relative to a day with an average reference temperature. Virtually every point estimate in the polynomial (Panel A) and degree days (Panel C) functional forms is statistically significant and appears to persist (that is, the cumulated effects of the contemporaneous temperature and all lags do not sum to zero) as more lags are included. The only exception is the cumulative effects with five lags (column 4) at the cold end of the temperature distribution, which are still negative but no longer statistically significant. Because binned regression is a method that is already demanding of the data and the statistical uncertainty also increases as more lags are included, the estimated effects in the binned model are rarely statistically significant. Notably, the cumulative effects of a day at 35$^{\circ}$C in the binned regression (Panel B) are all negative but no longer statistically significant after 5 lags are accounted for.

I further tested the growth effects versus the level effects when the temperature response was assumed to differ between high- and low-income provinces. Table \ref{tab:growth_level_effects_xPoor} presents the temperature responses at various temperatures. In low-income provinces, most of the point estimates at the hot end of the temperature distribution are statistically significant when estimated using the polynomial (Panel A) and degree days (Panel C) functional forms. These estimates remain substantially negative as more lags are included, indicating growth effects at the hot end of the temperature distribution. However, most of the estimates in the binned regression (Panel B) are not statistically significant. Both the immediate effect (column 1) and the cumulative effects (columns 3, 5, and 7) in the binned regression are, however, negative and statistically significant at 35$^{\circ}$C in low-income provinces. For the high-income provinces, most of the temperature responses, as estimated using the polynomial (Panel A) and degree days (Panel C) functional forms, are substantially negative, particularly at the hot end of the temperature distribution. Nonetheless, only a few of these estimates are statistically significant. %Somehow, these results provide suggestive evidence that, when the dynamic %effects of temperature over time are factored in, the high-income provinces may adapt to changes in %climate more successfully than the low-income provinces, and, as a result, the output growth of the %high-income provinces are potentially less affected with additional warming.\footnote{Applying what %described in \cite{burke2015global} to our case, in a fixed effects model with higher-order %temperature terms, both within-province and cross-province variation are used to identify the %temperature effects. This identification strategy of using within-province and cross-province %variation implicitly allows for more historical adaptation to a longer-run climate. Hence, the results %provide some suggestive evidence that the higher income provinces may have more climate adaptability.} 

In summary, the lagged temperature effects appear to statistically and significantly persist when assuming that high- and low-income provinces identically respond to changes in temperature. The results suggest that changes in temperature may affect the rate of economic growth, rather than the level of economic output. When assuming that high- and low-income provinces respond differently to changes in temperature, the estimates of low-income provinces using the polynomial and degree days functional forms show suggestive evidence of the growth effects on the output, particularly at the hot end of the temperature distribution. While most of the estimates using the binned regression show no evidence of the growth effects, the results at 35$^{\circ}$C provide somewhat substantial evidence of the growth effects on the output growth of low-income provinces. Adding increasing numbers of lags tends to make the cumulative effects for high-income provinces more negative, particularly at the hot end of the temperature distribution. However, these estimates are also more uncertain when more lags are included.

\subsection{Impacts of temperature at the sector level}
This subsection further investigates the impact of temperature on the components of output growth to understand the potential channels through which warming temperatures may affect aggregate output. Specifically, I tested the null hypotheses of no effects of temperature on the growth in real value added of the agricultural, industrial, and service sectors. First, I discuss the relationship between temperature changes and growth in each economic sector, as estimated by assuming that both high- and low-income provinces respond identically to temperature changes. Then I discuss the results as estimated, assuming that high- and low-income provinces respond differently to temperature changes. Lastly, I reports the results as estimated using the distributed lag form of Equation \ref{eqn:polynomial_function} and discuss whether temperature affects the growth or level of the components of the output.

\subsubsection{A common growth-temperature response function on the components of GPP}
Table \ref{tab:sectoral_output_common} reports the temperature response functions for the components of GPP at various temperatures, assuming that high- and low-income provinces respond identically to changes in temperature. The growth-temperature responses in the agricultural (Panel A), industrial (Panel B), and service (Panel C) real value added were calculated using regression estimates from Equation \ref{eqn:polynomial_function} with up to five lags. Column 1 shows the immediate effects of temperature, while columns 2-4 shows the cumulative effect of temperature and its lags. The contemporaneous temperature effects, as estimated from the regression with no lags in the agricultural sector (column 1, Panel A), are substantially negative and statistically significant. The pattern of the temperature effects is an inverted-U shape. The growth rate of the agricultural component increases (that is, the estimated impacts of temperature become less pronounced) as the temperature rises from cool to moderate and then declines. I also find the inverted-U shape in the pattern of the contemporaneous temperature effects on the growth in service real value added (column 1, Panel C), although these estimates are not statistically significant. Column 1 of Panel B shows the impact of mixed temperature on the growth of the industrial component. The temperature effects are negative at the cold end of the temperature distribution and become positive at higher temperatures, and none of these effects are statistically significant.

Overall, the results suggest that we can reject the null hypothesis that temperature changes do not affect output growth in the agricultural sector. However, the results provide no evidence of the impact of temperature on output growth in the industrial sector. This finding is different from \cite{hsiang2010temperatures} which reports substantial impacts of temperature shocks in industrial production compared to those in agricultural production. Note that, as suggested by \cite{dell2012temperature}, the results of this study only indicate the net effects of temperature on aggregate industrial output, without providing specific insights into how different industries might be affected. I also find no evidence that temperature changes affect output growth in service real value added.

\subsubsection{Heterogeneity in the growth-temperature response function on the components of GPP}
The results in Panel A show substantially negative effects of contemporaneous temperature in both low-income (column 1) and high-income (column 2) provinces on the growth in agricultural real value added. The point estimates show an inverted-U shape in the pattern of temperature effects on the growth in agricultural output in both high- and low-income provinces. These effects are all statistically significant, suggesting that we can reject the null hypothesis of no temperature effects on the growth of agricultural output. Notably, the immediate effects of temperature in both subsamples are close in magnitude and statistically significant, suggesting that the growth-temperature response functions in high- and low-income provinces are more likely to be the same. Panel B shows the impact of temperature on the growth of industrial output. The results for both low-income (column 1) and high-income (column 2) provinces are not statistically significant, suggesting that we cannot reject the null hypothesis that temperature changes do not affect growth in industrial output. While the contemporaneous temperature effects on the growth in service real value added in high-income provinces (column 2, Panel C) are all negative and exhibit an inverted-U shape pattern, none of these effects are statistically significant at the conventional confidence level. Column 1 of Panel C presents the point estimates of the impact of temperature on the growth of industrial output in low-income provinces. The results show that low-income provinces have a slightly flatter response function, with none of these effects being statistically significant. These results suggest that we cannot reject the null hypothesis that temperature changes do not affect growth in service output in both high- and low-income provinces.

Appendix Figure \ref{fig:response_xPoor_sector} displays the robustness of the growth-temperature response functions on the components of GPP, as estimated using the polynomial (left column), temperature bins (middle column), and degree days (right column) functional forms, allowing high- and low-income provinces to respond differently to changes in temperature. The results in Panels A, B, and C consistently show an inverted-U shape in the pattern of temperature effects on the growth of agricultural output in both high- and low-income provinces. The temperature effects in all three functional forms are statistically significant at the hot end temperature distribution in both high- and low-income provinces, suggesting that for both low- and high-income provinces, we can reject the hypothesis that the growth-temperature responses in agricultural real value added are zero at all points in the temperature distribution. Panels D, E, and F show the temperature effects on the growth of industrial output. The pattern of temperature effects on the growth in industrial real value added in both high- and low-income provinces is somewhat convex in all three functional forms, although the estimates are not statistically significant. The results suggest that we cannot reject the hypothesis of no temperature effects on the growth of industrial output in both high- and low-income provinces. The results for the service sector are shown in Panels G, H, and I. The temperature effects on the growth in service output are inverted-U shape in high-income provinces and flatter in low-income provinces. The results are consistent across all three functional forms. These effects are not statistically significant at the conventional confidence level, suggesting that we cannot reject the hypothesis of no temperature effects on the growth in service output in both high- and low-income provinces.

Taken together, the results, as estimated assuming that high- and low-income provinces respond differently to temperature changes, suggest that we can reject the null hypothesis that temperature changes affect the growth in agricultural real value added. The findings are consistent whether the regression estimates used second-order polynomial, temperature bins, or degree days functional forms. As for the other components of GPP, the results provide no evidence of the impact of temperature on the growth of both industrial and service outputs. Again, these results are consistent across all three functional forms.

I further tested the growth effects versus the level effects on the growth of the components of GPP. Since the above results suggest that we cannot reject the null hypothesis of no temperature effects on the growth in both industrial output and service output, I then focused only on the growth in agricultural real value added and rule out both level effects and growth effects on the growth in industrial output and service output.\footnote{Following \cite{dell2012temperature}, a failure to reject the null hypothesis that temperature does not affect growth would indicate an absence of both level and growth effects.} Again, the growth versus level effects are identified by adding both the immediate and lagged effects of temperature fluctuations across years. This section discusses the results of estimating the second-order polynomial Equation \ref{eqn:polynomial_function} with 1, 3, and 5 lags. Panel A in Table \ref{tab:sectoral_output_common} reports the contemporaneous and cumulated lag temperature effects on the growth in agricultural output, as estimated by assuming that high- and low-income provinces respond identically to temperature changes. Almost all point estimates are substantially negative and statistically significant. The estimates are fairly stable as more lags are included, except for the cumulative effects with five lags (column 4, Panel A) at the cold end of the temperature distribution. The results provide suggestive evidence of the growth effects at the hot end of the temperature distribution. Table \ref{tab:sectoral_output_xPoor} reports the cumulative temperature effects on the components of GPP at various temperatures, as estimated by assuming different growth-temperature response functions between high- and low-income provinces. Similar to the case of the pooled growth-temperature response function, the results for agricultural real value added (Panel A) provide suggestive evidence of the growth effects, especially at the hot end of the temperature distribution in both high- and low-income provinces.

\section{Projections of future damages on economic output}\label{sec:projection}

The above sections discuss the relationship between historical temperature changes and aggregate economic output. This section combines the empirical results from Section \ref{sec:econ_temp_relationship} with the projected changes in climate to generate projections of economic output under climate change. The first two subsections outline how I used the estimation in Equation \ref{eqn:polynomial_function} to project these responses into the future to calculate the economic risk of climate change and how these projections were aggregated. Subsequently, the projection results without bias correction are discussed. The next subsection presents the bias-corrected projections and discusses the impacts of potentially upward biases in future climate projections. The last subsection discusses the limitations of these projections.

\subsection{Methods: Impact projections of climate change}
I followed \cite{burke2015global} and used historical response functions to predict the annual GPP per capita growth rate, relative to the growth rate absent climate change, in which temperatures are assumed to be fixed at their 2003-2022 average. Following the notation of \cite{burke2015global}, the projections of the future growth rate of economic output per capita (denoted $g_{py}^+$) under different climate scenarios consist of two parts: the growth rate absent climate change and the differential effect of temperature changes. As a baseline scenario, I first assumed that each province would grow in the future at its observed average growth rate over 2003-2022. The projected output per capita growth in province $p$ in year $y$ after 2022 is then given by
\begin{equation} \label{eqn:future_growth}
    g_{py}^+ = \bar{g}_{p} + \delta_{py}
\end{equation}
where $\bar{g}_{p}$ is the observed average growth rate of output per capita during 2003-2022 and $\delta_{py}$ denotes the differential effect of temperature changes on output growth of province $p$ in year $y$ and was estimated by
\begin{equation} \label{eqn:additional_growth}
    \delta_{py} = h(T_{py}^+) - h(\bar{T}_{p})
\end{equation}
where $T_{py}^+$ is the projected change in temperature in any year after 2022, and $\bar{T}_{p}$ is the average temperature in province $p$ across the baseline period of 2003-2022.

In addition to the above baseline scenario, I also considered the economic development projected by the SSP3 and SSP5 scenarios. The available data provide projected GDP per capita in constant 2017 dollars purchasing power parity (PPP) at five years interval at the country level. I assumed that Thailand grows at the same growth rates between these five years interval to calculate the annual projected growth rates.\footnote{As such, the annual projected growth rates of Thailand's GDP in the future year $y$ is given by $(GDPcap_{y+5}/GDPcap_{y})^{1/5} - 1$.} To disaggregate national output per capita into provincial output per capita and leverage the projected annual growth rates data derived from the SSPs, I assumed that the output growth of each province $p$ in year $y$ was influenced by country's growth in the previous year, and estimated panel regression using the sample between 2003-2022. I then used these derived coefficients to project the annual growth rate of each province in the absence of climate change ($\eta$). For a given SSP scenario, the projected output per capita growth in province $p$ in each future year $y$ under climate change is then given by 
\begin{equation} \label{eqn:future_growth_SSP}
    g_{py}^+ = \eta_{py} + \delta_{py}
\end{equation}
where $\eta_{py}$ is the projected annual growth rate absent climate change of province $p$ in each future year $y$ in a given SSP scenario. I follow \cite{burke2015global} and focus on only two SSPs scenarios: SSP3 and SSP5.

There are two other important sources of uncertainty in the projected impacts of climate change: physical uncertainty in climate projections and uncertainty in the estimate of $h(T)$. To account for uncertainty in the climate projections, I estimated the differential effects using seven climate projections (as described in section \ref{sec:data}) under both the RCP4.5 and RCP8.5 emission scenarios. This distribution of climate projections is expected to capture some uncertainties in the climate system through 2090. As for the uncertainty in the projected impacts arising from the econometric estimation of response functions $h(T)$, I executed a block bootstrapping simulation by adapting the procedure in \cite{burke2015global}. First, I randomly drew a set of sampling provinces with replacement to account for autocorrelation. Second, for each bootstrap, I fitted the second-order polynomial model in Equation \ref{eqn:polynomial_function} and stored the resulting parameters. Third, using these parameters in combination with province-specific values of output growth provided by the baseline scenario and the projected climate provided by a given emission scenario from each of the seven climate projections, I predicted a response function for each of the 77 provinces. Finally, this process was repeated 1,000 times for each province, climate projection model, emission scenario, and output growth assumption combination. These 1,000 response functions were then used as the resulting distribution of estimates to characterize projection uncertainty for each year between 2023-2090.

To understand the sensitivity of these impact projections to alternative specifications, I explored how projections change when accounting for lagged effects of temperature over time and when allowing high- and low-income provinces to respond differently to temperature changes. In the former setting, the growth of output per capita in a given year was assumed to be affected by temperature in that year and the previous five years to account for the dynamic effects of temperature over time. In the latter setting, provinces in the future year $y$ could move to the high-income response function if, in the previous period $y-1$, their output per capita rose above the median output per capita across all provinces (and vice versa if it falls). Hence, four separate estimates were generated (that is, "Common, No lags", "Common, 5 lags", "High/Low, No lags", and "High/Low, 5 lags") for each province, climate projection model, emission scenario, and output growth assumption combination. The same block bootstrapping simulation procedure, as described previously, was also executed to project the impacts arising from econometric estimation. In all projections, parameters from estimating the second-order polynomial Equation \ref{eqn:polynomial_function} were used in combination with province- and year-specific values of output growth provided by each output growth scenario.

Finally, I examined the impacts of potentially upward biases in future climate projections, as described in Section \ref{sec:data}. Because it was very challenging to directly correct the projected temperature and it was also not a focus of this study, I instead assumed that these biases found between the observed and projected temperatures during 2018-2022 (see Appendix Figure \ref{fig:bias_in_projected_climate}) were systematic and remain unchanged throughout the future impact projections. To correct for these assumed systematic biases, the differential effect of temperature changes on output growth, $\delta_{py}$, was subtracted by the presumably fixed impact correction observed during 2018-2022 (denote $\bar\delta_{p}$). Equation \ref{eqn:additional_growth} therefore becomes
\begin{eqnarray} \label{eqn:bias_corrected_additional_growth}
    \delta^{'}_{py} &=& \delta_{py} - \bar\delta_{p} \nonumber \\
           &=& [h(T_{py}^+) - h(\bar{T}_{p})] - [h(\bar{T}_{p,2018-2022}^+) - h(\bar{T}_{p,2018-2022})]
\end{eqnarray}
where $\delta^{'}_{py}$ denotes the bias-corrected differential effect of temperature changes on output growth, and $\bar\delta_{p}$ denotes the impact correction term used to \textit{offset} the upward biases in the climate projections. For a given climate model, the impact correction of each province $p$ is the differential effect of the average projected temperature between 2018-2022 (denotes $\bar{T}_{p,2018-2022}^+$) and the average observed temperatures between 2018-2022 (denote $\bar{T}_{p,2018-2022}$) on output growth. Again, I assumed that these impact correction terms remain fixed throughout the future impact projections.

\subsection{Aggregation: Impact projections of climate change at higher aggregate level}

To enable the calculation of the higher level of aggregate economic output, like Gross Regional Product and Gross Domestic Product, I additionally assumed that the population share of each province, relative to either the entire country or its respective region, was fixed at its observed median population share over 2003-2022.\footnote{Notably, the population share of each province have barely changed since 2003. Average standard deviation of population share during 2003-2022 is 0.00089.} Gross Regional Product (GRP) of region $R$ in each future year $y$ were then constructed by
\begin{equation} \label{eqn:grp}
    GRPcap_{Ry} = \sum_{i\in R} \omega_{i} \cdot GPPcap_{iy}
\end{equation}
where $\omega_{i}$ is province $i$'s population share in region $R$. Note that there are six regions based on the NESDC designation: North, Northeast, Central, East, West, and South. To demonstrate different impacts of climate change, however, the North region is additionally divided into Upper North (containing Chiangmai, Chiangrai, Lampang, Lamphun, Maehongson, Man, Phayao, Phrae, and Uttaradit) and Lower North (containing Kamphaengphet, Nakhonsawan, Phetchabun, Phichit, Phitsanulok, Sukhothai, Tak, and Uthaithani) regions following the National Geographical Committee designations.

Similarly, the Gross Domestic Product in each future year $y$ is given by
\begin{equation} \label{eqn:gdp}
    GDPcap_{y} = \sum_{p} \omega_{p} \cdot GPPcap_{py}
\end{equation}
where $\omega_{p}$ is province $p$'s population share relative to the entire country.

\subsection{Results: projected impacts of climate change at various aggregate levels \textit{without bias-correction}}

Figure \ref{fig:impact_GPP_comparison_bothRCPs}A shows the projected impacts "\textit{without bias-correction}" of climate change on output per capita in the "baseline" output growth scenario of each province, relative to its output per capita in the absence of climate change, between 2023-2090. These projections were estimated using Equations \ref{eqn:future_growth} and \ref{eqn:additional_growth} together with regression estimates from second-order polynomial Equation \ref{eqn:polynomial_function} with no lags, and assuming that both high- and low-income provinces respond identically to changes in temperature. The red and blue lines represent the median impacts under RCP8.5 and RCP4.5 scenarios, respectively. In addition, Figure \ref{fig:impact_GPP2090_map} displays median projections of the impacts "\textit{without bias-correction}" on province-level output per capita, as also estimated using Equations \ref{eqn:future_growth} and \ref{eqn:additional_growth} together with regression estimates from second-order polynomial Equation \ref{eqn:polynomial_function}, under the RCP8.5 emission scenario in 2090 for each combination of four historical growth-temperature response functions and three output growth assumptions.

Province-level projections under both RCP4.5 and RCP8.5 emission scenarios show that climate change will negatively affect half of the Thai population (Figure \ref{fig:impact_GPP_comparison_bothRCPs}B) in 2050. These negative impacts are statistically uncertain, with an average 51-57\% likelihood (Figure \ref{fig:impact_GPP_comparison_bothRCPs}C) that climate change will have positive impacts. Figure \ref{fig:impact_GPP_comparison_bothRCPs}B and C show that an increasing share of the Thai population is projected to be affected by a warming climate, while negative impacts are more certain as we approach the end of this century. In 2090, projections under the "business as usual" RCP8.5 emission scenario show that climate change will make 86\% of the Thai population poorer in per capita terms than they would be in the absence of climate change, while with more aggressive emission reductions RCP4.5, 63\% are (Figure \ref{fig:impact_GPP_comparison_bothRCPs}B). However, the projections are statistically uncertain, with an average probability of 0.30-0.47 across provinces that climate change will have positive impacts. 

These projections provide suggestive evidence that climate change will most likely affect Thailand's economy to certain extent. Projections in column 1 of Figure \ref{fig:impact_GPP2090_map} further show that the differences in the projected impact of a warming climate are mainly due to geographic heterogeneity in the baseline temperatures. In particular, the provinces in the Upper-North region, which are typically colder than the other regions of Thailand, benefit more from increased average temperatures. These province-level projections are broadly consistent in magnitude, whether high- and low-income provinces are assumed to respond identically (column 1) or differently (column 3) to temperature changes. The projections in columns 2 and 4 of Figure \ref{fig:impact_GPP2090_map}), which account for the lagged effects of temperature over time, display substantially more negative impacts in all 77 provinces. As shown in Tables \ref{tab:growth_level_effects_pooled}, as lagged effects of temperature are factored in, the estimated growth-temperature response functions are substantially more negative in models that allow growth effects to persist over subsequent years.

Next, I examined the impacts, again, "\textit{without bias-correction}" of climate change on Thailand's gross regional product (GRP) per capita. To illustrate the distribution of projections in each future year, projections were visually weighted \citep{hsiang2013visually} to present the probability that an economic output trajectory is observed in the 1,000 bootstrapped response functions. The darker areas represent a higher likelihood that the projected impact will pass through a given value in a given year. Figure \ref{fig:projected_impact_by_region_rcp85_density} presents the projected impacts of climate change on regional output per capita under the RCP8.5 emission scenario between 2023-2090 for baseline (left column), SSP3 (middle column), and SSP5 (right column) output growth assumptions. To derive GRP per capita projections, the province-level output per capita projections were aggregated following Equation \ref{eqn:grp}. Again, these province-level outputs per capita were estimated using Equations \ref{eqn:future_growth} and \ref{eqn:additional_growth} together with regression estimates from the second-order polynomial Equation \ref{eqn:polynomial_function} with no lags and assuming that both high- and low-income provinces respond identically to changes in temperature. Similar to provincial projections, the output growth assumptions likely have little impact on the structure of the projected change in GRP per capita under the RCP8.5 emission scenario. Point estimates are uniformly more negative in the higher average temperature regions (that is, Central, East, South, and West regions). The damages are modest in colder regions like Lower-North and Northeast regions, while the Upper-North region, which has the lowest average temperature, benefits the most from increased average temperatures.

To examine whether the region-level projections are sensitive to different combinations of emission scenarios, specifications, and output growth assumptions, Appendix Figure \ref{fig:projected_impact_GRP_allRCPs_allSpecs_allGrowths} presents regional projected impacts under both RCP4.5 and RCP8.5 emission scenarios between 2023-2090 using regression estimates from second-order polynomial Equation \ref{eqn:polynomial_function} for each of the four historical growth-temperature response functions with a baseline output growth scenario. The projections are broadly similar in structure for both emission scenarios. Projected impacts in models that allow the effects of temperature to persist on regional output growth are less uncertain whether high- and low-income provinces respond identically (column 2) or differently (column 4) to temperature changes. The estimates in Tables \ref{tab:growth_level_effects_pooled} and \ref{tab:growth_level_effects_xPoor} provide the reason for this: as the effects of temperature are assumed to be cumulative and affect growth in a given year, cold provinces could initially benefit on net, and hotter provinces remain worse off with further warming. However, as a given province's average temperature becomes warmer, the impacts of future warming climate worsen in each future, which makes net economic impacts negative for provinces that are initially cold. In models that do not account for lagged effects, the projections become more uncertain whether high- and low-income provinces are assumed to respond identically (column 1) or differently (column 3). This is because the estimated growth-temperature response functions are substantially flatter than the response function with lagged effects. Cold provinces increasingly benefit from increased average temperatures, whereas hotter provinces remain worse off under a warming climate. This explanation can also be applied to the different median estimates observed in the lower north and northeast regions (columns 1 and 3 in Appendix Figure \ref{fig:projected_impact_GRP_allRCPs_allSpecs_allGrowths}). Notably, differences in the response function drastically benefit provinces in the upper northern region of Thailand.

Lastly, I projected the impacts "\textit{without bias-correction}" of climate change on Thailand's gross domestic product (GDP) per capita under RCP8.5 emission scenario between 2023-2090 for four different historical growth-temperature response functions and three output growth assumptions combination (Figure \ref{fig:projected_impact_onGDP_rcp85_density}). The province-level output per capita projections were aggregated following Equation \ref{eqn:gdp} to derive the GDP per capita projections. These province-level outputs per capita were first estimated using Equations \ref{eqn:future_growth} and \ref{eqn:additional_growth} together with regression estimates from the second-order polynomial Equation \ref{eqn:polynomial_function}, under the RCP8.5 emission scenario in 2090 for each combination of four historical growth-temperature response functions and three output growth assumptions. Projections were visually weighted \citep{hsiang2013visually} to illustrate the distribution of projections and to present the density of projected impacts observed in the bootstrapped response functions. Median estimates are more uniformly negative in models with delayed impacts, whether high- and low-income provinces are assumed to respond identically (row 2 in Figure \ref{fig:projected_impact_onGDP_rcp85_density}) or differently (row 4 in Figure \ref{fig:projected_impact_onGDP_rcp85_density}). Similar to both provincial and regional projections, the output growth assumptions likely have little impacts on projected change in GDP per capita under RCP8.5 scenario. As observed in the regional projections, projections are less uncertain and fall off steeply at earlier future years in models that allow the effects of temperature to persist on output growth (Figure \ref{fig:projected_impact_onGDP_rcp85_density}D-F and J-L) because colder provinces also suffer large damages. In models that do not account for lagged effects, the projections are more uncertain whether high- and low-income provinces are assumed to respond identically (row 1 in Figure \ref{fig:projected_impact_onGDP_rcp85_density}) or differently (row 3 in Figure \ref{fig:projected_impact_onGDP_rcp85_density}) because colder provinces somewhat benefit from increased average temperature, while hotter provinces remain worse off under warming climate. Appendix Figure \ref{fig:projected_impact_onGDP_bothRCPs} presents the sensitivity of the projected impacts on Thailand's GDP under the emission stabilization scenario RCP4.5. Projections under both emission scenarios are broadly similar in structure, but vary in magnitude. Notable results are in the "Common, 5 lags" models under RCP4.5, in which projections are more uncertain than those under RCP8.5 for all three output growth assumptions (row 2 in Figure \ref{fig:projected_impact_onGDP_bothRCPs}). This is because emission stabilization scenario RCP 4.5, which results in a lower temperature rise than that of RCP8.5, makes net economic impacts less negative for cold provinces, especially provinces in the higher latitude regions (i.e., Upper-North, Lower-North, and Northeast regions). The same explanation can also be applied to the more positive median impacts observed in models that do not account for lagged effects, whether high- and low-income provinces are assumed to respond identically (row 1 in Figure \ref{fig:projected_impact_onGDP_bothRCPs}) or differently (row 3 in Figure \ref{fig:projected_impact_onGDP_bothRCPs}) to temperature changes under the RCP4.5 emission scenario. Note that the plots in both rows 1 and 3 also display the impacts of using different output growth assumptions in projections. In a given RCP emission scenario, a given model with the baseline scenario, in which each province is assumed to grow in the future at its observed average growth rate over 2003-2022, provide more negative projections than those predicted by the same model with either the SSP3 or SSP5 pathways. When considered alongside the impacts of the projected climate under the RCP4.5 emission scenario described above, these results are not unexpected because the output growth of provinces derived from either SSP3 or SSP5 is mostly larger than their corresponding "fixed" ones in the baseline scenario, particularly those colder provinces in the Upper-North region.

To summarize the projections numerically, Table \ref{tab:summary_impacts_on_GDP2090} reports some important percentiles in the bootstrapped distribution of climate-impact projections on GDP per capita in 2090 for each combination of model specifications and output growth assumptions under the RCP4.5 (Panel A) and RCP8.5 (Panel B) emission scenarios. With the "business as usual" RCP8.5 emission scenario (Panel A), models that do not take delayed effects of temperature into account (that is, "No lags" models) with baseline output growth assumption project that climate change reduces Thailand's output per capita by a half, relative to its GDP per capita in the absence of climate change, whether high- and low-income provinces are assumed to respond identically ("Common, No lags" model in Panel B) or differently ("High/Low, No lags" model in Panel B). Nonetheless, the inner 90\% credible interval, determined as the 5th and 95th percentile values, are substantially wide, with a probability of 0.32-0.37 for positive impacts. Using the "No lags" specification with SSP3 and SSP5 output growth assumptions, the models project smaller damages of 15\% for identical response functions across provinces and 22\% for different response functions between high- and low-income provinces. Projections are more uncertain, with a 46\% likelihood of positive impacts. Taken together, in models with no delayed effects of temperature, the estimates vary in magnitude depending on the output growth assumptions and model specification being used, although the structure (that is, a substantially wide range of the inner 90\% interval and the likelihood of positive impacts) are similar.

With the emission stabilization scenario RCP 4.5, all models provide projections similar to those under RCP8.5. Models that do not consider the delayed effects of temperature project small losses (3-5\%) in the baseline scenario and substantial gain (44-55\%) in SSP3 and SSP5 output growth assumptions, with a probability of 0.49-0.63 for positive impacts. As described previously, this is because the climate under the RCP4.5 emission scenario projects a lower temperature rise than that of RCP8.5, and cold provinces benefit on net, whereas hotter provinces remain worse off. Models allowing for delayed effects of temperature under both RCP4.5 and RCP8.5 emission scenarios project substantially large damages to economic output. All models with delayed impacts project that Thailand will lose 95-99\% of its output in the absence of climate change, with very unlikely positive impacts.

\subsection{Results: projected impacts of climate change at various aggregate levels \textit{with bias-correction}}

To examine the impacts of potentially upward biases in future climate projections, this subsection presents the projected impacts on provincial output per capita using the bias-correction strategy discussed previously. Analogous to Figure \ref{fig:impact_GPP_comparison_bothRCPs}, Figure \ref{fig:impact_GPP_comparison_bothRCPs_corrected}A displays the median projected impacts of climate change on provincial output per capita with baseline output growth assumption under RCP4.5 (blue lines) and RCP8.5 (red lines) emission scenarios between 2023-2090. All projections were based on the "\textit{bias-corrected}" differential effect of temperature changes on output growth $\delta^{'}_{py}$, as estimated using Equation \ref{eqn:bias_corrected_additional_growth} and regression estimates from the second-order polynomial Equation \ref{eqn:polynomial_function} with no lags and assuming that both high- and low-income provinces respond identically to changes in temperature. 

Figure \ref{fig:impact_GPP_comparison_bothRCPs_corrected}A shows the projected impacts of climate change "\textit{with bias-correction}" on output per capita in the "baseline" output growth scenario of each province, relative to its output per capita in the absence of climate change, between 2023-2090, with red lines representing median impacts under RCP8.5 and blue line under RCP4.5. These corrected projections are substantially different from those displayed in Figure \ref{fig:impact_GPP_comparison_bothRCPs}A, particularly in cold provinces that benefit from increased average temperatures. With bias correction, projections under both the RCP4.5 and RCP8.5 emission scenarios show that climate change will substantially affect 89-94\% of the Thai population (Figure \ref{fig:impact_GPP_comparison_bothRCPs_corrected}B) in 2050. These projections come with average probability 0.10-0.20 that climate change will have positive impacts on any provinces. The climate-affected share of Thai population are substantially high through out the projected future period. Projections show that almost all Thai people will be negatively affected by climate change in 2090 under either the RCP4.5 or RCP8.5 emission scenarios. Figure \ref{fig:impact_GPP_comparison_bothRCPs_corrected}C also show more certainty in projections. In 2090, the likelihood that climate change will have positive impacts is substantially low, with an average probability of 0.06-0.07. 

These corrected projections provide suggestive evidence that climate change will affect Thailand's economy to a certain extent. Compared with Figure \ref{fig:impact_GPP_comparison_bothRCPs}A, the impacts of potentially upward biases in the climate projections are removed, and cold provinces (for instance, \textit{Mae Hong Son}, \textit{Chiangmai}, and \textit{Chiangrai} in the Upper-North region) benefit much less on net, while hotter provinces remain worse off. A larger share of the Thai population is also expected to be affected by the warming climate in 2050 under both RCP4.5 and RCP8.5 emission scenarios, with a probability 0.10-0.20 that climate change will have positive impacts on any province. The projections are even more certain for 2090. The likelihood that climate change will have positive impacts is substantially low, with an average probability of 0.06-0.07.

Finally, to show how the projected impacts of climate change are sensitive to potential biases in future climate projections, I reproduced Table \ref{tab:summary_impacts_on_GDP2090} by replacing $\delta_{py}$ (Equation \ref{eqn:additional_growth}) with $\delta^{'}_{py}$ (Equation \ref{eqn:bias_corrected_additional_growth}) when estimating the differential effect of temperature changes on output growth. Analogous to Table \ref{tab:summary_impacts_on_GDP2090}, Table \ref{tab:summary_impacts_on_GDP2090_correction} reports important percentiles in the bootstrapped distribution of climate-impact projections on GDP per capita in 2090 for each combination of model specifications and output growth assumptions under the RCP4.5 (Panel A) and RCP8.5 (Panel B) emission scenarios.

In models that do not consider the delayed effects of temperature, whether high- and low-income provinces are assumed to respond identically ("Common, No lags" models) or differently ("High/Low, No lags" models), median projections in all three output growth scenarios show that climate change reduces Thailand's output per capita by 57-63\% under RCP4.5 and 80-86\% under RCP8.5, relative to its GDP per capita in the absence of climate change. With bias correction in the projections, these models project substantially larger damages than those without bias correction. Moreover, the inner 90\% credible interval, determined as the 5th and 95th percentile values, are also substantially narrower. Again, the reason for this is that as the impacts of potentially upward biases in the climate projections are removed, cold provinces benefit much less on net, while hotter provinces are affected less. These differences in the projected impacts can be easily observed in Figure \ref{fig:impact_GPP_comparison_bothRCPs}A and Appendix Figure \ref{fig:impact_GPP_comparison_bothRCPs_corrected}A. The cold provinces that benefit from increased average temperature (that is, negative damages) in Figure \ref{fig:impact_GPP_comparison_bothRCPs}A become worse off when the projected impacts are bias-corrected in Appendix Figure \ref{fig:impact_GPP_comparison_bothRCPs_corrected}A. However, the median projections of climate impacts with bias correction show little change in models that allow for delayed effects of temperature under both the RCP4.5 and RCP8.5 emission scenarios. These models still project substantially large damages to economic output. Projections show that Thailand will lose 94-100\% of its output in the absence of climate change. Nonetheless, the inner 90\% credible intervals are substantially narrower for the models with delayed temperature effects. All projections are very certain with only 1-6\% likelihood of positive impacts.

Overall, the results in this subsection show that the projected impacts of climate change are sensitive to potential biases in future climate projections, particularly in models that do not account for lagged effects. Colder provinces benefit from increased average temperatures, whereas hotter provinces remain worse off under a warming climate. Hence, all projections should be interpreted with caution.

\subsection{Limitation} \label{Limitation}
As described throughout this section, these projected future damages rely on a number of strong assumptions, including that the climate projections are correct, the future growth of province-level output will remain constant in the baseline scenario or follow the growth paths that are projected by SSP3 and SSP5 pathways, and the demographics of the Thai population and their geographic distribution will remain unchanged. The above impact projections with bias correction present the sensitivity of the projected impacts of climate change to the climate projections. Nonetheless, as suggested by \cite{deschenes2011climate},  although these assumptions are strong, they allow for a transparent projection based on the available data.

Apart from statistical uncertainty, the above projections also reveal a certain degree of uncertainty caused by two distinct sources: climate projection uncertainty and economic pathway uncertainty. Future climate projections involve considerable uncertainty arising from an incomplete understanding of the Earth’s physical systems.\footnote{See \cite{tebaldi2007use} for further discussion on the challenges in interpreting multi-model ensembles for climate projections} This paper are however unable to account for this uncertainty. I follow previous research on the impacts of climate change (e.g., \cite{deschenes2011climate, burke2015global, carleton2022valuing}) and use seven climate projections under two RCPs to provide some variation in the possible future climate. Similarly, predicting accurate economic growth in any economy is already challenging due to a variety of factors that introduce complexity and uncertainty, particularly as financial systems become increasingly integrated. Different output growths adopted in the projections are expected to provide a sense of the potential consequences of different future scenarios.

Finally, as indicated by \cite{deschenes2011climate} that their estimates using binned regression failed to account for permanent changes in climate and likely overestimated the impacts of climate change. Using a fixed effects panel regression with a linear temperature, \cite{dell2012temperature} also suggested that their estimated temperature effects may be affected by not accounting for adaptability to permanent changes in climate. To tackle such possible overestimation, this study estimated a fixed effects model with higher-order temperature terms to identify the effects of temperature. This identification strategy uses both within-province and cross-province variation and allows for historical adaptation to longer-run temperature changes because provinces with different average temperatures are permitted to respond differently to within-province temperature changes \citep{burke2015global}.

\section{Conclusion}

This study investigates the historical relationship between temperature fluctuations and aggregate economic output in the context of a tropical country such as Thailand. An identification strategy exploits the random year-to-year variation in temperature within a province, after controlling for shocks common to all provinces in Thailand, to estimate the effects of temperature changes on the annual growth rate of output per capita. Specifically, I used province fixed effects and year fixed effects to isolate within-location year-to-year variation in temperature exposure from the systematic patterns of weather in each province.

The results show inverted-U shape temperature effects on the annual growth rate of provincial output per capita in all estimated functional forms. The estimated marginal change in the annual growth rate of provincial output per capita of 1$^{\circ}$C temperature increase are -1.248, -3.700, and -3.799 percentage points for a second-order polynomial, temperature bins, and degree days functional forms, respectively. In all estimated functional forms, the estimates assuming different growth-temperature response functions between high- and low-income provinces are found to be statistically insignificant different from those estimated assuming a common response function across high- and low-income provinces. The estimates using the polynomial and degree days functional forms also suggest that warming temperatures may affect the growth of economic output rather than its level, while little evidence is found when estimated using binned regression. However, these growth effects are more pronounced in low-income provinces with higher average temperatures. I also investigate the impact of warming temperature on the three components of aggregate output. The results provide suggestive evidence of the impact of temperature on output growth only in the agricultural sector, but not in the industrial and service sectors. 

Finally, I combined the empirical results with the projected changes in the climate to generate projections of economic output under climate change. Province-level projections under both RCP4.5 and RCP8.5 emission scenarios show that climate change will affect half of the Thai population by 2050. Projections under the "business as usual" RCP8.5 emission scenario show that climate change will make 86\% of Thai people poorer in per capita terms than they would be in the absence of climate change in 2090, while a more aggressive emission reduction RCP4.5, 62\% are. In both cases, the likelihood of non-negative climate impacts is substantially low with a probability of 0.01-0.03. These projections provide suggestive evidence that climate change will most likely affect Thailand's economy to certain extent. Nonetheless, I find that the projected impacts of climate change are sensitive to potential biases in future climate projections, particularly in models that do not account for lagged effects. Hence, all projections should be interpreted cautiously.

\bibliographystyle{plainnat}
\bibliography{references.bib}  

\clearpage
\newpage
\appendix
\renewcommand{\appendixpagename}{\centering Figures}
\appendixpage
\setcounter{figure}{0}

%Figure1
\begin{figure}[htb]
\centering
\captionsetup{width=1\textwidth}
\includegraphics[scale=.115]{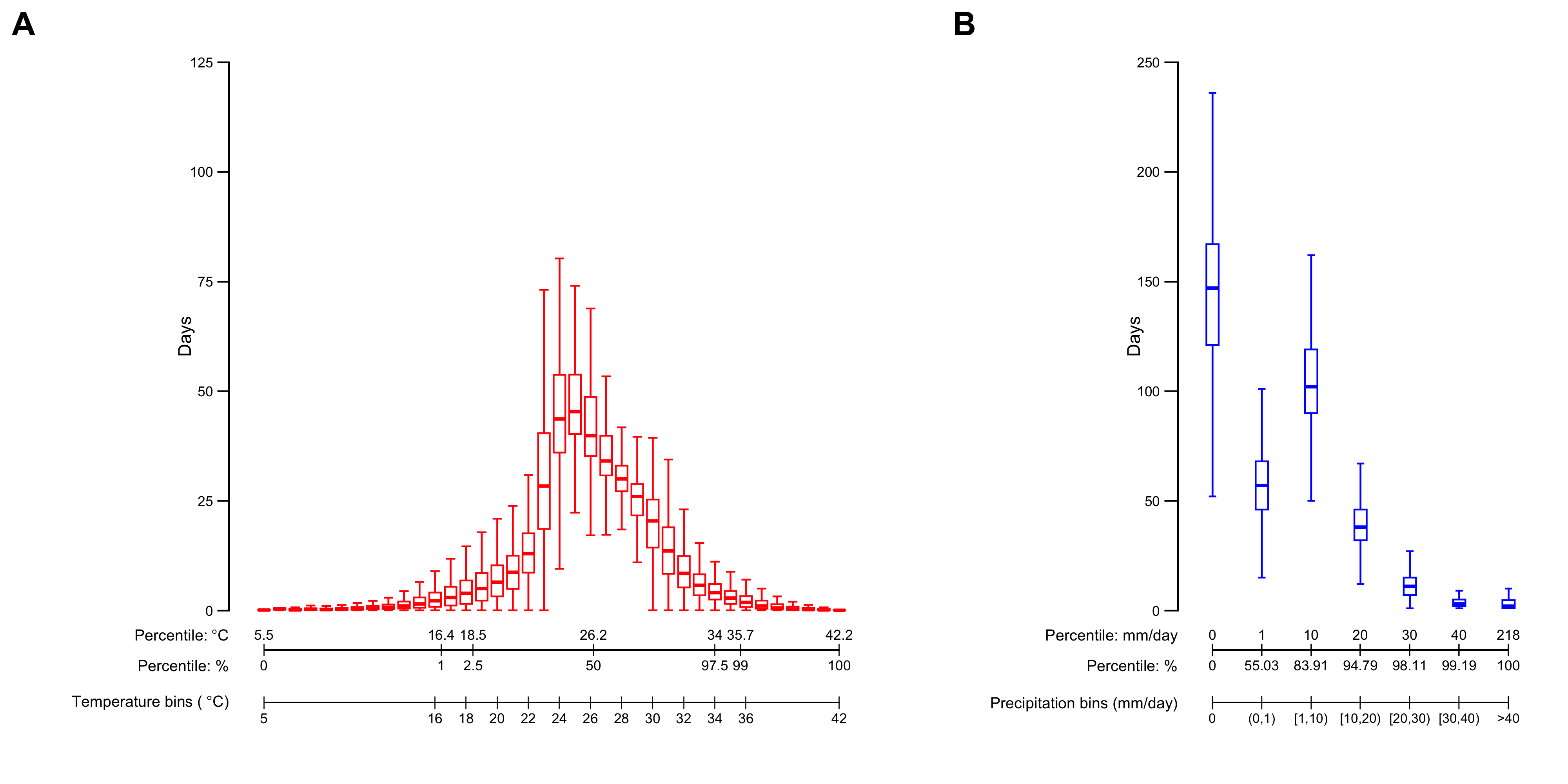}
\caption{\textbf{Descriptive Weather Statistics.} The boxplots present the distributions of time the population was exposed to the weather across provinces during 1982-2022. The box represents the 25\%-75\% range, and the middle line within each box indicates the median across all provinces and years. Whiskers indicate the minimum and maximum exposures for each interval. Important levels and their corresponding percentiles are also reported on the upper $x$-axis of each panel. \textit{Panel A} shows distribution of number of days the population were exposed to each 1$^\circ$C temperature bin. \textit{Panel B} shows distribution of number of days the population were exposed to seven precipitation bins. The first two bins correspond to days when the daily total precipitation was zero and 1 mm, respectively. The highest bin includes days when the total precipitation is above 40 mm. Other bins cover 10mm-span of daily total precipitation.}
\label{fig:summary_weather}
\end{figure}

%Figure2
\begin{figure}[t!]
\centering
\captionsetup{width=1\textwidth}
\includegraphics[scale=.078]{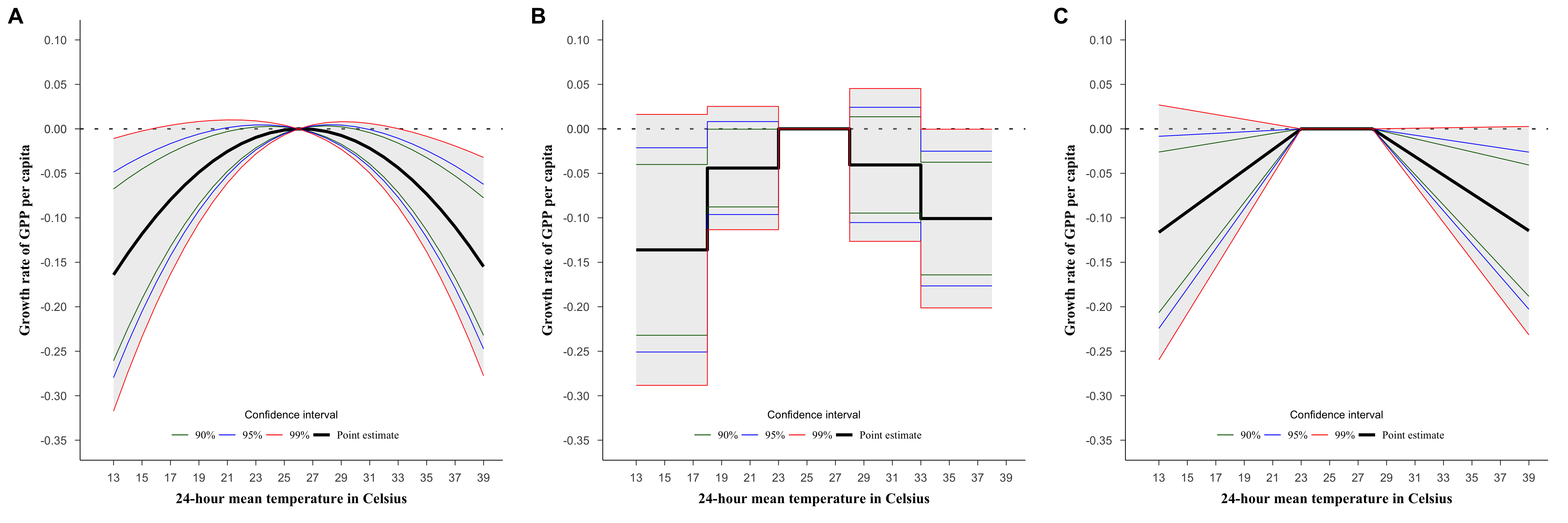}
\caption{\textbf{A common growth-temperature response function across provinces}. Graphs show the growth temperature response functions estimated using three different functional forms. The black line shows the temperature effect as estimated by assuming an identical response function across all provinces. The green, blue, and red lines show the 90\%, 95\%, and 99\% confidence intervals, respectively. The gray area highlights the 99\% confidence interval. \textit{Panel A} shows the response, using 24-hour average second-order polynomial in hourly average temperature, relative to a day with an average temperature of 26$^\circ$C. \textit{Panel B} shows the response, using a vector of 24-hour average temperatures with 5$^\circ C$ temperature bins, relative to the omitted $[23,28)^\circ C$ temperature bin. \textit{Panel C} shows the response, using 24-hour sums of heating degree hours below 23$^{\circ}C$ and 24-hour sums of cooling degree hours above 28$^{\circ}C$. The 90\%, 95\%, and 99\% confidence bands are added. All response functions are shown only for 24-hour average temperatures that are usually experienced across Thailand over the sample period, as illustrated in Figure \ref{fig:summary_weather}. All response functions were estimated with province- and year-specific fixed effects. See Section \ref{sec:econ_temp_relationship} for details of each functional form.}
\label{fig:baseline_model}
\end{figure}

%Figure3
\begin{figure}[t]
\centering
\captionsetup{width=1\textwidth}
\includegraphics[scale=.078]{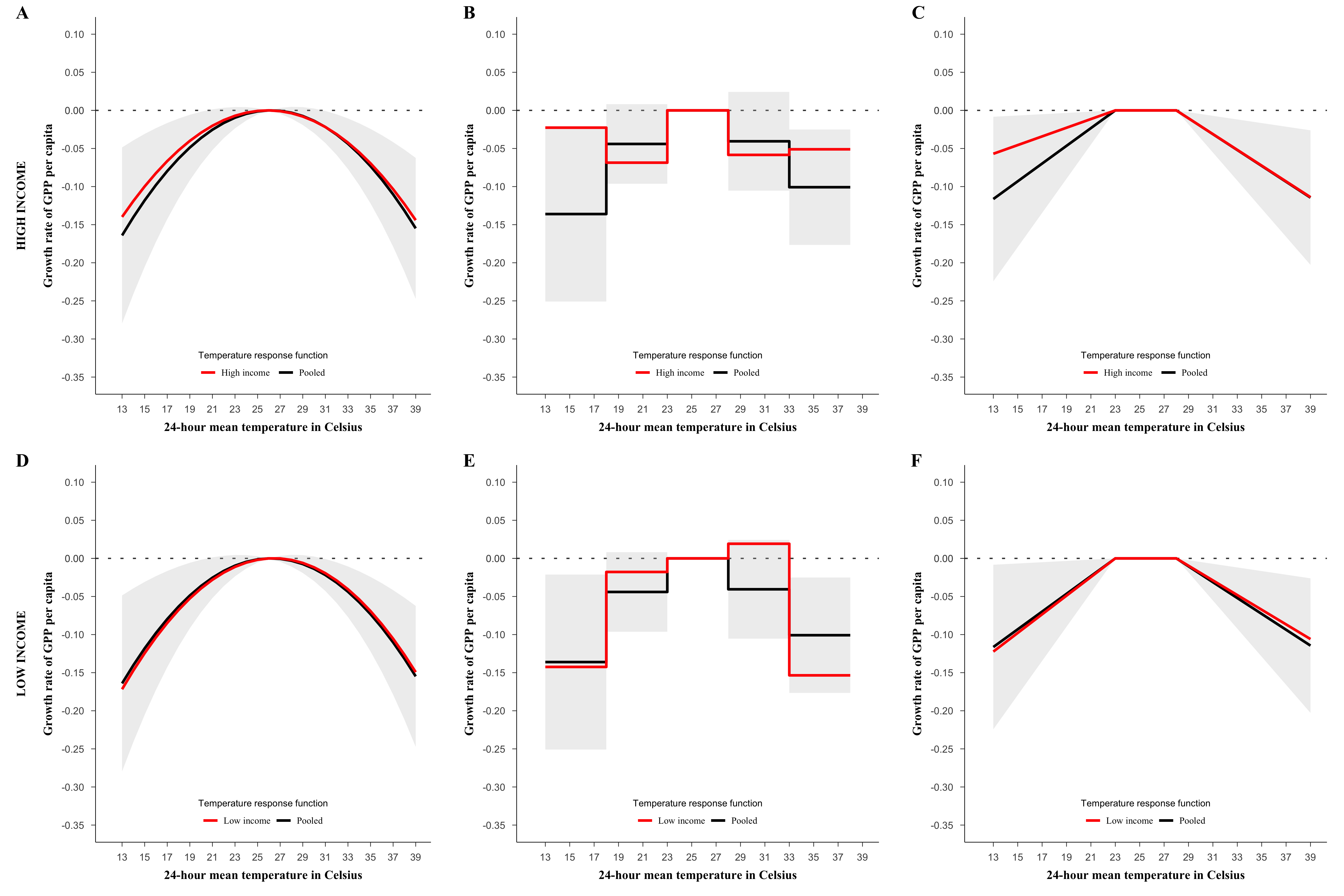}
\caption{\textbf{Heterogeneous growth-temperature response function in high- and low-income provinces}. The graphs show heterogeneity in the growth-temperature response functions as estimated using three different functional forms. The upper row (Panels A, B, and C) presents the estimates for higher-income provinces, and the lower row (Panels D, E, and F) presents the estimates for lower-income provinces. The black line shows the temperature effect as estimated by assuming a common response across high- and low-income provinces, and the gray areas represent the 95\% confidence interval. The red line shows the temperature effects as estimated by assuming different response functions across high- and low-income provinces. \textbf{Panel A} shows the response, using 24-hour average second-order polynomial in hourly average temperature, relative to a day with an average temperature of 26$^\circ$C. \textbf{Panel B} shows the response, using a vector of 24-hour average temperatures with 5$^\circ C$ temperature bins, relative to the omitted $[23,28)^\circ C$ temperature bin. \textbf{Panel C} shows the response, using 24-hour sums of heating degree hours below 23$^{\circ}C$ and 24-hour sums of cooling degree hours above 28$^{\circ}C$, summed across the year. All response functions are only shown for 24-hour average temperatures that are usually experienced across Thailand over the sample period as illustrated in Figure \ref{fig:summary_weather}. All response functions were estimated with province-specific and year-specific fixed effects. See Section \ref{sec:econ_temp_relationship} for details of each functional form.}
\label{fig:response_xPoor_GPP}
\end{figure}

%Figure4
\begin{figure}[t!]
\centering
\captionsetup{width=1\textwidth}
\includegraphics[scale=.115]{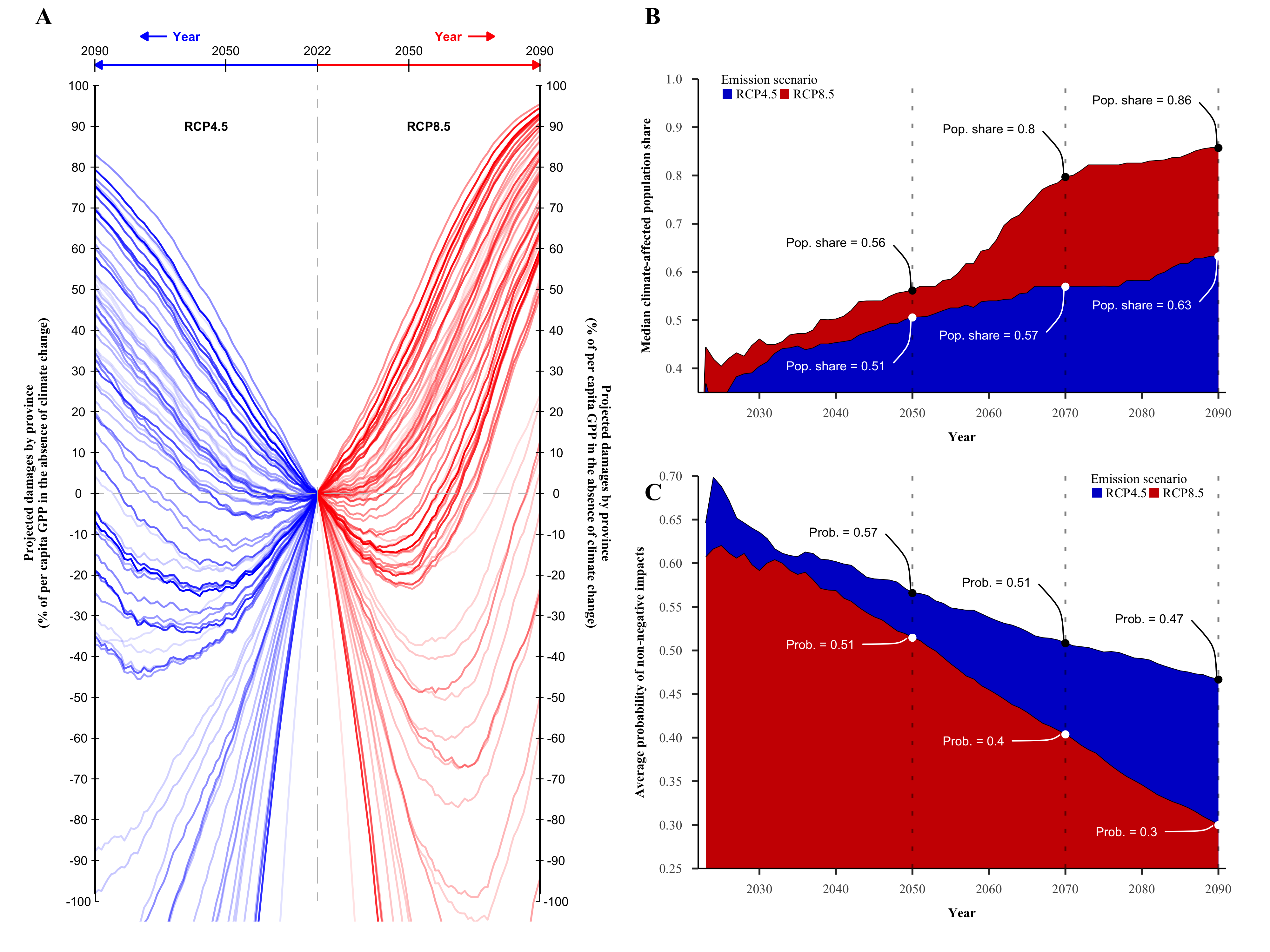}
\caption{\textbf{Province-level projections of temperature effects of climate change "\textit{without bias-correction}" on output per capita, relative to their GPP per capita absent climate change, based on uncorrected climate projections}. \textbf{Panel A} displays median projected impacts of climate change on provincial output per capita with baseline output growth assumption under RCP4.5 (blue lines) and RCP8.5 (red lines) emission scenarios between 2023-2090. The differential effect of temperature changes on output growth $\delta_{py}$ was estimated using Equation \ref{eqn:additional_growth} together with regression estimates from Equation \ref{eqn:polynomial_function} with no lags, assuming an identical response function across 77 provinces. The line thickness represents the population share of each province. \textbf{Panel B} presents median share of population which are negatively affected by warming climate under RCP4.5 and RCP8.5 emission scenarios. For a given future year, the median value was calculated based on 7,000 runs (7 climate models × 1,000 sets of bootstrapped coefficients). \textbf{Panel C} display average proportion of runs with non-negative climate-impacts to total runs across 77 provinces under RCP4.5 and RCP8.5. The probability of each province in a given future year was first calculated based on 7,000 runs. The resulting probabilities in a given future year were then averaged out across 77 provinces.}
\label{fig:impact_GPP_comparison_bothRCPs}
\end{figure}

%Figure5
\begin{figure}[t!]
\centering
\captionsetup{width=1\textwidth}
\includegraphics[scale=.115]{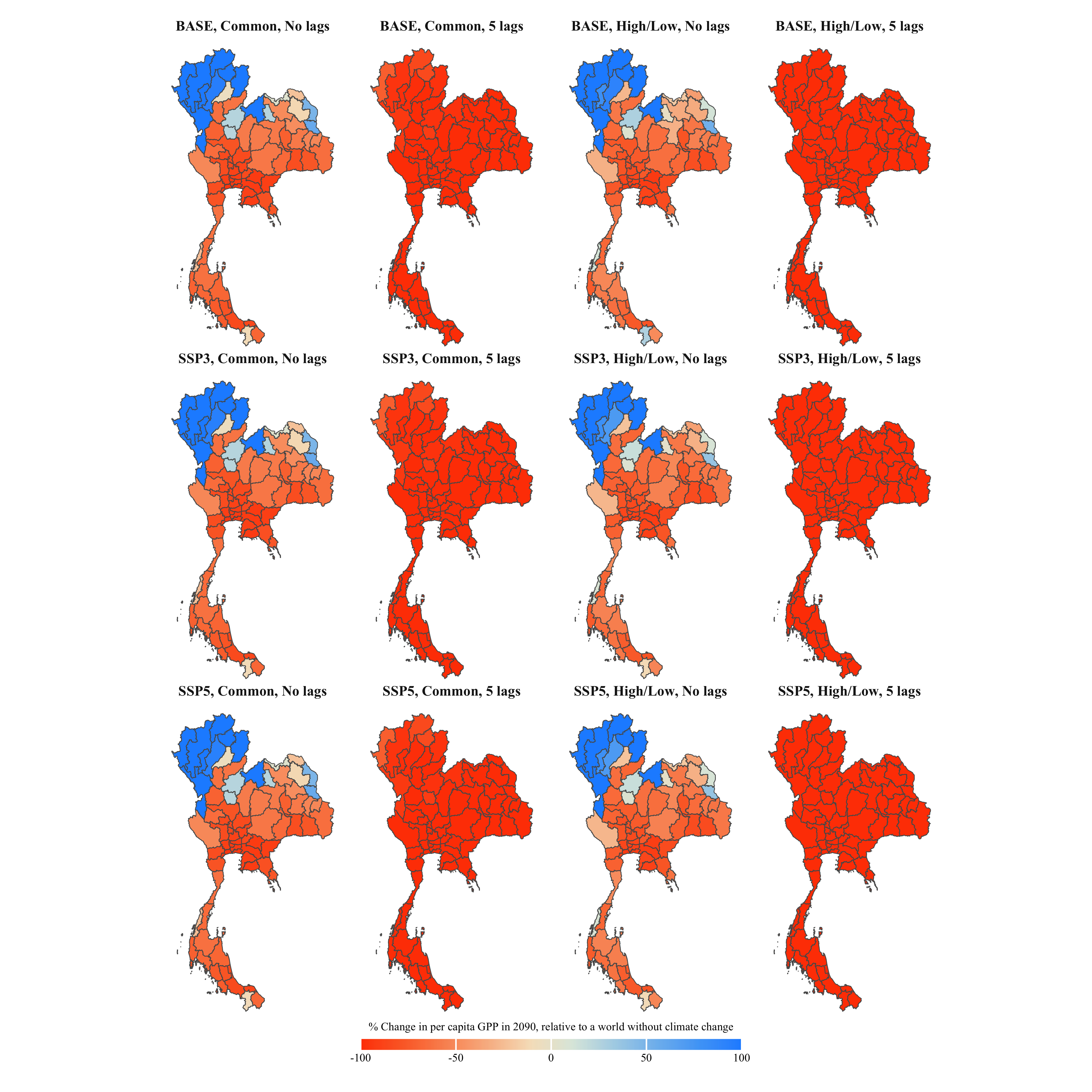}
\caption{\textbf{Projected impacts of climate change "\textit{without bias-correction}" on provincial output per capita of 77 provinces in 2090, relative to their GPP per capita absent climate change}. Graphs show the median projected impacts of climate change under the RCP8.5 emission scenario in 2090 for each combination of different historical growth-temperature response functions and different output growth scenarios.  Columns: (1) a common response function across high- and low-income provinces with no lags, (2) a common response function across high- and low-income with five lags, (3) the differentiated response functions between high- and low-income provinces with no lags, (4) the differentiated response functions between high- and low-income provinces with five lags. Rows: (1) baseline scenario, (2) SSP3, (3) SSP5.}
\label{fig:impact_GPP2090_map}
\end{figure}

%Figure6
\begin{figure}[t!]
\centering
\captionsetup{width=1\textwidth}
\includegraphics[scale=.115]{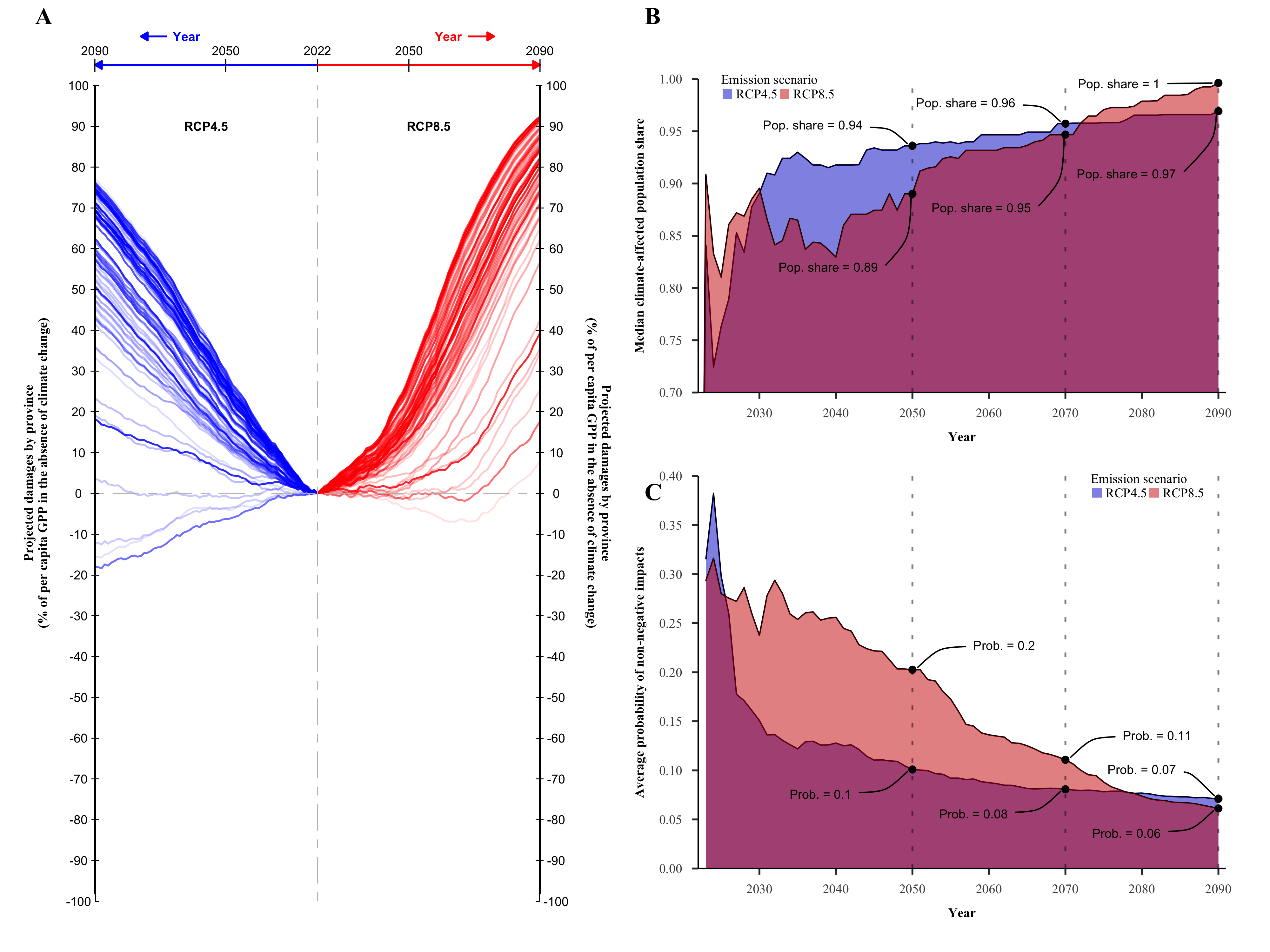}
\caption{\textbf{Province-level projections of temperature effects of climate change "\textit{with bias-correction}" on output per capita, relative to their GPP per capita absent climate change}. These graphs duplicate Figure \ref{fig:impact_GPP_comparison_bothRCPs}, but instead use the correction strategy discussed in Section \ref{sec:projection} in the main text. \textbf{Panel A} displays median projected impacts of climate change on provincial output per capita with baseline output growth assumption under RCP4.5 (blue lines) and RCP8.5 (red lines) emission scenarios between 2023-2090. The "\textit{bias-corrected}" differential effect of temperature changes on output growth $\delta^{'}_{py}$ was estimated using Equation \ref{eqn:bias_corrected_additional_growth} together with regression estimates from Equation \ref{eqn:polynomial_function} with no lags, assuming an identical response function across 77 provinces. Line thickness represents population share of each province. \textbf{Panel B} presents median share of population which are negatively affected by warming climate under RCP4.5 and RCP8.5 emission scenarios. For a given future year, the median value was calculated based on 7,000 runs (seven climate models × 1,000 sets of bootstrapped coefficients). \textbf{Panel C} display average proportion of runs with positive climate-impacts to total runs across 77 provinces under RCP4.5 and RCP8.5. The probability of each province in a given future year was first calculated based on 7,000 runs. The resulting probabilities in a given future year were then averaged out across 77 provinces.}
\label{fig:impact_GPP_comparison_bothRCPs_corrected}
\end{figure}

\newpage
\clearpage
%Figure7
\begin{figure}[t!]
\centering
\captionsetup{width=1\textwidth}
\includegraphics[scale=.1425]{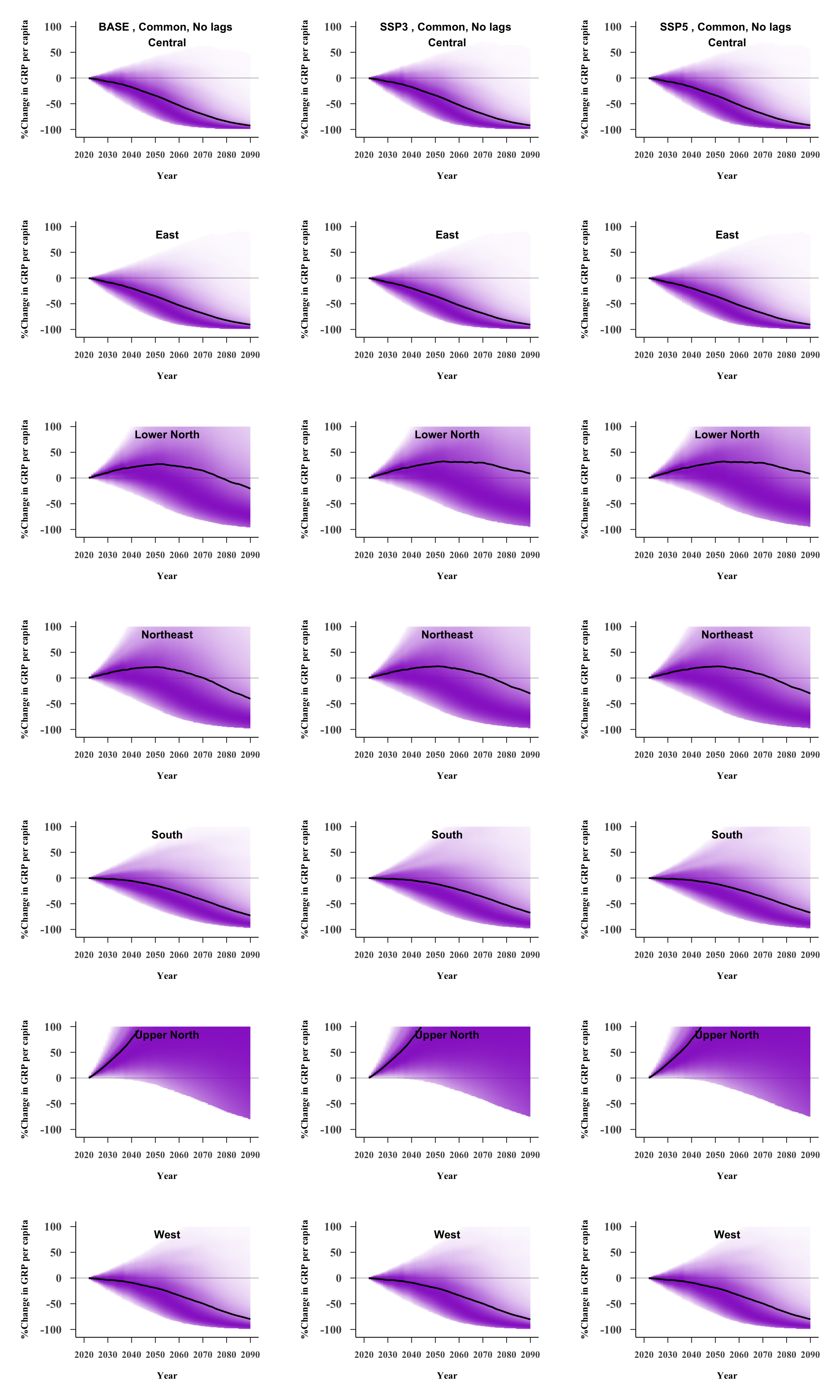}
\caption{\textbf{Projected impacts of climate change \textit{without} bias-correction on regional output per capita, relative to their GRP per capita absent climate change}. Graphs show the projected impacts of climate change, relative to projection using constant 2003-2022 average temperature, under RCP8.5 emissions scenario between 2023-2090 for three output growth assumptions. The projections were estimated using regression estimates from the second-order polynomial Equation \ref{eqn:polynomial_function} with no lags and assuming that both high- and low-income provinces respond identically to changes in temperature. Black lines are projected point estimates using the main specification. The projections were visually weighted \citep{hsiang2013visually}. Purple shaded area is 95\% confidence interval, colour opacity indicates estimated likelihood an impact trajectory passes through a value.}
\label{fig:projected_impact_by_region_rcp85_density}
\end{figure}

\newpage
\clearpage
%Figure8
\begin{figure}[H]
\centering
\captionsetup{width=1\textwidth}
\includegraphics[scale=.14]{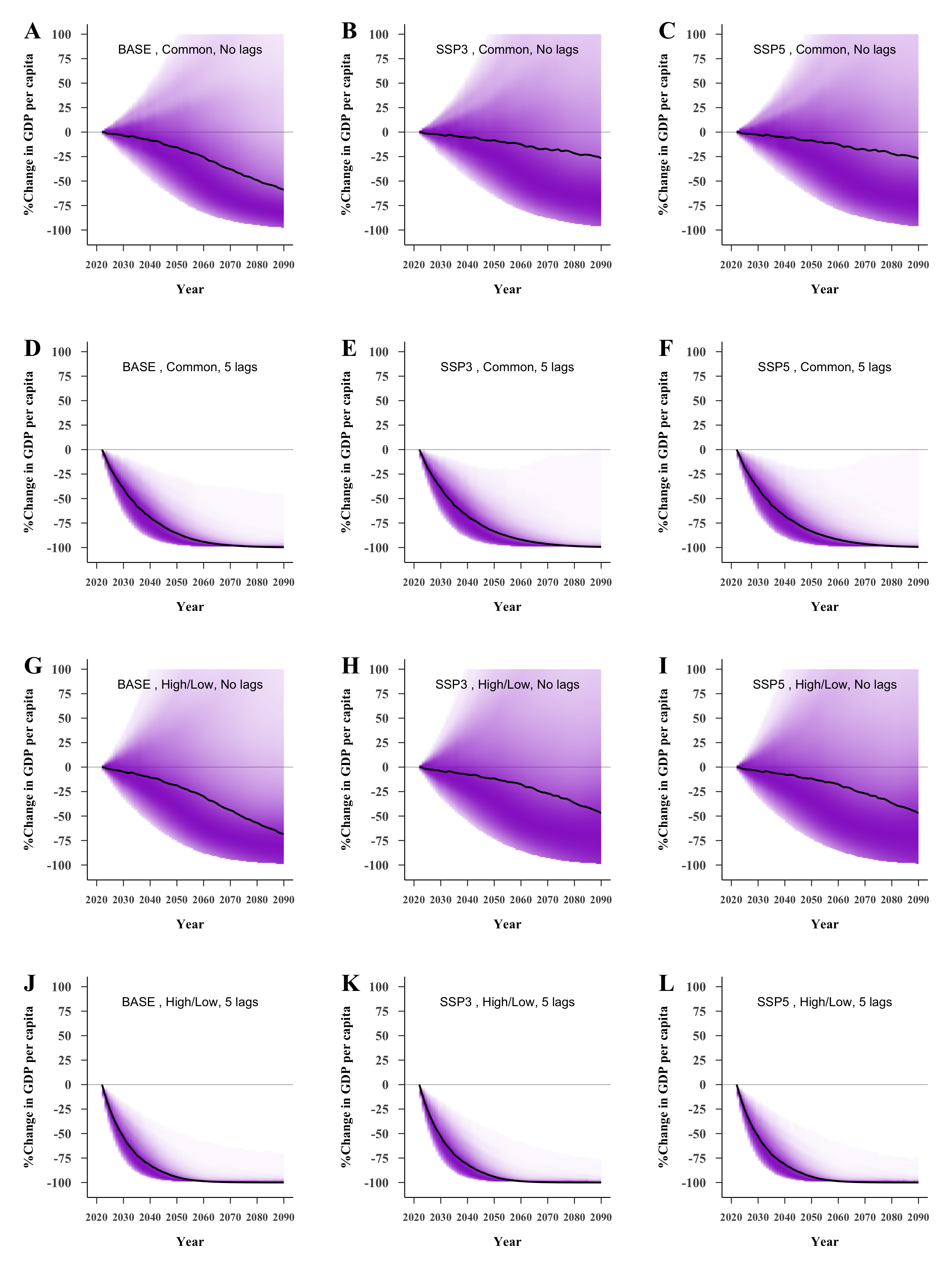}
\caption{\textbf{Projected impacts of climate change \textit{without} bias-correction on Thailand's gross domestic product (GDP) per capita, relative to GDP per capita absent climate change}. Graphs show the projected impacts of climate change, relative to projection using constant 2003-2022 average temperature, under the RCP8.5 emissions scenario between 2023-2090 for each combination of historical growth-temperature response functions and output growth assumptions. Black lines are projected point estimates using main specification. Projections are visually weighted following \cite{hsiang2013visually}. Purple shaded area is 95\% confidence interval, colour opacity indicates estimated likelihood an impact trajectory passes through a value.}
\label{fig:projected_impact_onGDP_rcp85_density}
\end{figure}

\clearpage
\newpage
\appendix
\renewcommand{\appendixpagename}{\centering Tables}
\appendixpage
\setcounter{table}{0}

%table1
\begin{table}[htb]
\small
\centering
\begin{threeparttable}
\captionsetup{width=1\textwidth}
\caption{\textbf{Summary statistics of key variables used in this study.}}
\label{tab:summary_stat}
\begin{tabularx}{1\textwidth}{lXXXXXX}
\toprule\toprule
 & N & Mean & Median & Min & Max & Std.Dev. \\ \midrule
Growth in GPP per capita$^a$ (\%) & 3,086 & 6.38 & 6.08 & -59.18 & 51.41 & 8.45 \\
Growth in GPP component$^a$ (\%) &   &   &   & \\
\hspace*{3mm}{Agriculture} & 3,086 & 5.70 & 6.13 & -63.53 & 81.98 & 16.14 \\
\hspace*{3mm}{Industry} & 3,086 & 8.05 & 7.63 & -115.83 & 114.38 & 16.73 \\
\hspace*{3mm}{Service} & 3,086 & 7.38 & 7.17 & -63.51 & 59.10 & 7.46 \\
Hourly temperature ($^\circ$C)$^b$ & 27,419,592 & 26.40 & 26.19 & 5.48 & 42.22 & 3.68 \\
Days in omitted bin$^c$ & 3,128 & 204.19 & 192.17 & 108.62 & 329.17 & 38.72 \\
Days in HDD (\textless 23$^\circ$C)$^d$ & 3,091 & 49.39 & 39.33 & 0.04 & 230.17 & 39.95 \\
Days in CDD (\textgreater 28$^\circ$C)$^d$ & 3,128 & 112.24 & 112.46 & 19.33 & 202.71 & 35.93 \\
Daily total precipitation (mm.) & 1,142,483 & 4.53 & 0.42 & 0.00 & 218.32 & 8.28 \\
Days with zero precipitation$^b$ & 3,128 & 143.19 & 147.00 & 30.00 & 242.00 & 34.97 \\ \bottomrule\bottomrule
\end{tabularx}
\begin{tablenotes}[flushleft]
\item \textit{Notes:}All weather variables used in this study are population-weighted, using the polygonized population dataset developed for the respective year as mentioned in the main text.\\
$^a$ Growth in GPP per capita and growth in the components of GPP values shown are inflation-adjusted to the 2019 Thai baht.\\
$^b$ Statistics of hourly temperature and daily total precipitation were calculated across 77 provinces without further population weighting.\\
$^c$ An omitted bin refers to $T \in [23,28)^{\circ}C$ (see Section \ref{sec:econ_temp_relationship} for details).\\
$^d$ The heating degree is defined as below 23$^{\circ}C$, while the cooling degree is defined as above 28$^{\circ}C$ (see Section \ref{sec:econ_temp_relationship} for details).
\end{tablenotes}
\end{threeparttable}
\end{table}

%table2
\begin{table}
\small
\centering
\captionsetup{width=1\textwidth}
\caption{\textbf{Growth-Temperature response function at different temperatures, assuming identical response across provinces.} The growth-temperature response functions were estimated using Equations \ref{eqn:polynomial_function}, \ref{eqn:binned_function}, and \ref{eqn:hdd_cdd_function} with up to five \textit{Panel A} shows the responses, using 24-hour average second-order polynomial in hourly average temperature, relative to a day with an average temperature of 26$^\circ$C. \textit{Panel B} shows the responses, using a vector of 24-hour average temperatures with 5$^\circ$C temperature bins, relative to the omitted $[23,28)^\circ$C bins. \textit{Panel C} shows the responses, using 24-hour sum of heating degree hours below 23$^{\circ}$C and 24-hour sum of cooling degree days above 28$^{\circ}$C. Point estimates indicate the effect of a single day at each 24-hour average temperature on the growth rate of output per capita, relative to a day with the respective reference temperature.}
\label{tab:growth_level_effects_pooled}
\begin{threeparttable}
\begin{tabularx}{1\textwidth}{lXXXXXXX}
\toprule\toprule
Temperature & \multicolumn{1}{l}{No lags} & \multicolumn{1}{c}{} & \multicolumn{1}{l}{1 lag} & \multicolumn{1}{c}{} & \multicolumn{1}{l}{3 lags} & \multicolumn{1}{c}{} & \multicolumn{1}{l}{5 lags} \\ 
 & \multicolumn{1}{l}{(1)} & \multicolumn{1}{c}{} & \multicolumn{1}{l}{(2)} & \multicolumn{1}{c}{} & \multicolumn{1}{l}{(3)} & \multicolumn{1}{c}{} & \multicolumn{1}{l}{(4)} \\ \midrule
%\multicolumn{8}{l}{} \\
\multicolumn{8}{l}{\textbf{Panel A: Second-order polynomial}} \\
%\multicolumn{8}{l}{} \\
15$^{\circ}$C & -0.1181*** &  & -0.1373** &  & -0.2017*** &  & -0.0766 \\
 & (0.0438) &  & (0.0522) &  & (0.0667) &  & (0.0692) \\
20$^{\circ}$C & -0.0361** &  & -0.0348* &  & -0.0479* &  & -0.0007 \\
 & (0.0173) &  & (0.0202) &  & (0.0250) &  & (0.0271) \\
30$^{\circ}$C & -0.0137* &  & -0.0303*** &  & -0.0509*** &  & -0.0543*** \\
 & (0.0081) &  & (0.0109) &  & (0.0162) &  & (0.0178) \\
35$^{\circ}$C & -0.0732*** &  & -0.1284*** &  & -0.2077*** &  & -0.1839*** \\
 & (0.0246) &  & (0.0346) &  & (0.0530) &  & (0.0557) \\
40$^{\circ}$C & -0.1800*** &  & -0.2933*** &  & -0.4681*** &  & -0.3820*** \\
 & (0.0532) &  & (0.0744) &  & (0.1134) &  & (0.1167) \\
\multicolumn{8}{l}{} \\
\multicolumn{1}{l}{Observations} & \multicolumn{1}{l}{3,086} & \multicolumn{1}{c}{} & \multicolumn{1}{l}{3,086} & \multicolumn{1}{c}{} & \multicolumn{1}{l}{2,940} & \multicolumn{1}{c}{} & \multicolumn{1}{l}{2,792} \\
\multicolumn{1}{l}{$R^2$} & \multicolumn{1}{l}{0.330} & \multicolumn{1}{c}{} & \multicolumn{1}{l}{0.332} & \multicolumn{1}{c}{} & \multicolumn{1}{l}{0.350} & \multicolumn{1}{c}{} & \multicolumn{1}{l}{0.370} \\
\multicolumn{1}{l}{Within $R^2$} & \multicolumn{1}{l}{0.008} & \multicolumn{1}{c}{} & \multicolumn{1}{l}{0.011} & \multicolumn{1}{c}{} & \multicolumn{1}{l}{0.021} & \multicolumn{1}{c}{} & \multicolumn{1}{l}{0.032} \\ \midrule
%\multicolumn{8}{l}{} \\
\multicolumn{8}{l}{\textbf{Panel B: Temperature bins}} \\
%\multicolumn{8}{c}{} \\
15$^{\circ}$C & -0.1361** &  & -0.0881 &  & -0.0845 &  & -0.0344 \\
 & (0.0576) &  & (0.0783) &  & (0.0946) &  & (0.0993) \\
20$^{\circ}$C & -0.0441* &  & -0.0510 &  & -0.0370 &  & -0.0719 \\
 & (0.0262) &  & (0.0364) &  & (0.0465) &  & (0.0452) \\
30$^{\circ}$C & -0.0406 &  & -0.0460 &  & -0.0605 &  & -0.1759** \\
 & (0.0325) &  & (0.0481) &  & (0.0552) &  & (0.0697) \\
35$^{\circ}$C & -0.1008*** &  & -0.2307*** &  & -0.2696*** &  & -0.1530 \\
 & (0.0380) &  & (0.0581) &  & (0.0857) &  & (0.1015) \\
40$^{\circ}$C & -0.5567 &  & -0.1015 &  & -0.3179 &  & 1.9948 \\
 & (0.3344) &  & (0.4062) &  & (0.8992) &  & (1.4556) \\
\multicolumn{8}{l}{} \\
\multicolumn{1}{l}{Observations} & \multicolumn{1}{l}{3,086} & \multicolumn{1}{c}{} & \multicolumn{1}{l}{3,086} & \multicolumn{1}{c}{} & \multicolumn{1}{l}{2,940} & \multicolumn{1}{c}{} & \multicolumn{1}{l}{2,792} \\
\multicolumn{1}{l}{$R^2$} & \multicolumn{1}{l}{0.332} & \multicolumn{1}{c}{} & \multicolumn{1}{l}{0.336} & \multicolumn{1}{c}{} & \multicolumn{1}{l}{0.359} & \multicolumn{1}{c}{} & \multicolumn{1}{l}{0.386} \\
\multicolumn{1}{l}{Within $R^2$} & \multicolumn{1}{l}{0.011} & \multicolumn{1}{c}{} & \multicolumn{1}{l}{0.016} & \multicolumn{1}{c}{} & \multicolumn{1}{l}{0.034} & \multicolumn{1}{c}{} & \multicolumn{1}{l}{0.057} \\ \midrule
%\multicolumn{8}{l}{} \\
\multicolumn{8}{l}{\textbf{Panel C: Heating/Cooling degree days}} \\
%\multicolumn{8}{c}{} \\
15$^{\circ}$C & -0.0931** &  & -0.1004* &  & -0.2052*** &  & -0.1425* \\
 & (0.0434) &  & (0.0557) &  & (0.0606) &  & (0.0762) \\
20$^{\circ}$C & -0.0349** &  & -0.0377* &  & -0.0769*** &  & -0.0534* \\
 & (0.0163) &  & (0.0209) &  & (0.0227) &  & (0.0286) \\
30$^{\circ}$C & -0.0208** &  & -0.0413*** &  & -0.0560*** &  & -0.0485** \\
 & (0.0081) &  & (0.0113) &  & (0.0165) &  & (0.0184) \\
35$^{\circ}$C & -0.0729** &  & -0.1447*** &  & -0.1959*** &  & -0.1698** \\
 & (0.0282) &  & (0.0397) &  & (0.0578) &  & (0.0646) \\
40$^{\circ}$C & -0.1249** &  & -0.2480*** &  & -0.3358*** &  & -0.2911** \\
 & (0.0484) &  & (0.0681) &  & (0.0990) &  & (0.1107) \\
\multicolumn{8}{l}{} \\
\multicolumn{1}{l}{Observations} & \multicolumn{1}{l}{3,086} & \multicolumn{1}{c}{} & \multicolumn{1}{l}{3,086} & \multicolumn{1}{c}{} & \multicolumn{1}{l}{2,940} & \multicolumn{1}{c}{} & \multicolumn{1}{l}{2,792} \\
\multicolumn{1}{l}{$R^2$} & \multicolumn{1}{l}{0.328} & \multicolumn{1}{c}{} & \multicolumn{1}{l}{0.330} & \multicolumn{1}{c}{} & \multicolumn{1}{l}{0.347} & \multicolumn{1}{c}{} & \multicolumn{1}{l}{0.363} \\
\multicolumn{1}{l}{Within $R^2$} & \multicolumn{1}{l}{0.005} & \multicolumn{1}{c}{} & \multicolumn{1}{l}{0.009} & \multicolumn{1}{c}{} & \multicolumn{1}{l}{0.016} & \multicolumn{1}{c}{} & \multicolumn{1}{l}{0.021} \\ \bottomrule\bottomrule
\end{tabularx}%
\begin{tablenotes}[flushleft]
\item \textit{Notes:} All models included province FE and year FE and account for the effects of changes in precipitation. When estimating the model with lags, the growth-temperature response functions were estimated using the distributed-lag form of Equations \ref{eqn:polynomial_function}, \ref{eqn:binned_function}, and \ref{eqn:hdd_cdd_function}, with precipitation and its lags also included. The standard errors in parentheses are clustered by province.\\
***, **, * indicate significance level of 1\%, 5\%, and 10\%, respectively.
\end{tablenotes}
\end{threeparttable}
%}
\end{table}

%table3
\begin{table}
\tiny
\centering
\captionsetup{width=1\textwidth}
\caption{\textbf{Growth-Temperature response function at different points in the temperature, allowing high- and low-income provinces to respond differently to temperature changes.} The growth-temperature response functions were estimated using Equations \ref{eqn:polynomial_function}, \ref{eqn:binned_function}, and \ref{eqn:hdd_cdd_function} with up to 5 lags. \textit{Panel A} shows the responses, using 24-hour average second-order polynomial in hourly average temperature, relative to a day with an average temperature of 26$^\circ$C. \textit{Panel B} shows the responses, using a vector of 24-hour average temperatures with 5$^\circ$C temperature bins, relative to the omitted $[23,28)^\circ$C bins. \textit{Panel C} shows the responses, using 24-hour sum of heating degree hours below 23$^{\circ}$C and 24-hour sum of cooling degree days above 28$^{\circ}$C. Point estimates indicate the effect of a single day at each 24-hour average temperature on the growth rate of output per capita, relative to a day with the respective reference temperature.}
\label{tab:growth_level_effects_xPoor}
\resizebox{\textwidth}{!}{%
\begin{threeparttable}
\begin{tabular}{@{}lllllllll@{}}
\toprule\toprule
 & \multicolumn{2}{c}{No lags} & \multicolumn{2}{c}{1 lag} & \multicolumn{2}{c}{3 lags} & \multicolumn{2}{c}{5 lags} \\ \cmidrule(l){2-9} 
\hspace*{2mm}Temperature & \multicolumn{1}{c}{Low inc.} & \multicolumn{1}{c}{High inc.} & \multicolumn{1}{c}{Low inc.} & \multicolumn{1}{c}{High inc.} & \multicolumn{1}{c}{Low inc.} & \multicolumn{1}{c}{High inc.} & \multicolumn{1}{c}{Low inc.} & \multicolumn{1}{c}{High inc.} \\
            & \multicolumn{1}{c}{(1)} & \multicolumn{1}{c}{(2)} & \multicolumn{1}{c}{(3)} & \multicolumn{1}{c}{(4)} & \multicolumn{1}{c}{(5)} & \multicolumn{1}{c}{(6)} & \multicolumn{1}{c}{(7)} & \multicolumn{1}{c}{(8)} \\ \midrule
%\multicolumn{9}{c}{} \\
\multicolumn{9}{l}{\textbf{Panel A: Second-order polynomial}} \\
%\multicolumn{9}{c}{} \\
\hspace*{2mm}15$^{\circ}$C & -0.1244** & -0.0998 & -0.1657*** & 0.0015 & -0.2456*** & -0.0528 & -0.1190 & 0.2352 \\
 & (0.0478) & (0.0992) & (0.0602) & (0.0994) & (0.0756) & (0.1254) & (0.0855) & (0.1413) \\
\hspace*{2mm}20$^{\circ}$C & -0.0394** & -0.0293 & -0.0443* & 0.0156 & -0.0580** & 0.0000 & -0.0174 & 0.1153 \\
 & (0.0180) & (0.0387) & (0.0224) & (0.0375) & (0.0276) & (0.0472) & (0.0307) & (0.0531) \\
\hspace*{2mm}30$^{\circ}$C & -0.0117 & -0.0141 & -0.0318*** & -0.0301** & -0.0626*** & -0.0385* & -0.0518** & -0.0595** \\
 & (0.0086) & (0.0115) & (0.0119) & (0.0143) & (0.0169) & (0.0210) & (0.0203) & (0.0229) \\
\hspace*{2mm}35$^{\circ}$C & -0.0692** & -0.0695** & -0.1407*** & -0.0898* & -0.2548*** & -0.1299* & -0.1878*** & -0.1143 \\
 & (0.0286) & (0.0340) & (0.0401) & (0.0484) & (0.0574) & (0.0706) & (0.0697) & (0.0776) \\
\hspace*{2mm}40$^{\circ}$C & -0.1741*** & -0.1669** & -0.3262*** & -0.1742 & -0.5736*** & -0.2693* & -0.4030*** & -0.1474 \\
 & (0.0638) & (0.0836) & (0.0885) & (0.1126) & (0.1249) & (0.1593) & (0.1513) & (0.1763) \\
\multicolumn{1}{l}{} &  &  &  &  &  &  &  &  \\
\multicolumn{1}{l}{Observations} & \multicolumn{2}{c}{3,086} & \multicolumn{2}{c}{3,086} & \multicolumn{2}{c}{2,940} & \multicolumn{2}{c}{2,792} \\
\multicolumn{1}{l}{$R^2$} & \multicolumn{2}{c}{0.331} & \multicolumn{2}{c}{0.335} & \multicolumn{2}{c}{0.358} & \multicolumn{2}{c}{0.384} \\
\multicolumn{1}{l}{Within $R^2$} & \multicolumn{2}{c}{0.009} & \multicolumn{2}{c}{0.015} & \multicolumn{2}{c}{0.032} & \multicolumn{2}{c}{0.054} \\ \midrule
%\multicolumn{9}{l}{} \\
\multicolumn{9}{l}{\textbf{Panel B: Temperature bins}} \\
%\multicolumn{9}{c}{} \\
\hspace*{2mm}15$^{\circ}$C & -0.1425*** & -0.0227 & 0.0014 & -0.0050 & -0.0397 & 0.0256 & 0.1748 & 0.0194 \\
 & (0.0529) & (0.2097) & (0.5003) & (0.1750) & (0.4992) & (0.2906) & (0.5460) & (0.2972) \\
\hspace*{2mm}20$^{\circ}$C & -0.0178 & -0.0686 & 0.0007 & -0.0792 & 0.0515 & -0.1010** & 0.0485 & -0.0761 \\
 & (0.0238) & (0.0469) & (0.1094) & (0.0543) & (0.1176) & (0.0492) & (0.1413) & (0.0669) \\
\hspace*{2mm}30$^{\circ}$C & 0.0192 & -0.0584* & 0.0803 & -0.0840* & 0.0779 & -0.0958** & 0.1043 & -0.2452*** \\
 & (0.0360) & (0.0341) & (0.1002) & (0.0426) & (0.1166) & (0.0453) & (0.1626) & (0.0487) \\
\hspace*{2mm}35$^{\circ}$C & -0.1535*** & -0.0511 & -0.3490*** & -0.1642** & -0.4355*** & -0.1717 & -0.4128** & -0.0390 \\
 & (0.0404) & (0.0486) & (0.1140) & (0.0706) & (0.1404) & (0.1208) & (0.1657) & (0.1494) \\
\hspace*{2mm}40$^{\circ}$C & -0.0882 & -1.2485*** & 0.4057 & -0.9372 & -0.6510 & -0.1927 & 0.6988 & 2.7781 \\
 & (0.3451) & (0.3786) & (1.1302) & (0.6263) & (1.5754) & (1.4418) & (1.5263) & (2.3777) \\
\multicolumn{1}{l}{} &  &  &  &  &  &  &  &  \\
\multicolumn{1}{l}{Observations} & \multicolumn{2}{c}{3,086} & \multicolumn{2}{c}{3,086} & \multicolumn{2}{c}{2,940} & \multicolumn{2}{c}{2,792} \\
\multicolumn{1}{l}{$R^2$} & \multicolumn{2}{c}{0.336} & \multicolumn{2}{c}{0.343} & \multicolumn{2}{c}{0.374} & \multicolumn{2}{c}{0.409} \\
\multicolumn{1}{l}{Within $R^2$} & \multicolumn{2}{c}{0.016} & \multicolumn{2}{c}{0.028} & \multicolumn{2}{c}{0.056} & \multicolumn{2}{c}{0.091} \\ \midrule
%\multicolumn{9}{c}{} \\
\multicolumn{9}{l}{\textbf{Panel C: Heating/Cooling degree days}} \\
%\multicolumn{9}{l}{} \\
\hspace*{2mm}15$^{\circ}$C & -0.0979** & -0.0456 & -0.1190 & 0.0459 & -0.2465 & -0.0631 & -0.1782 & 0.0664 \\
 & (0.0431) & (0.1448) & (0.3120) & (0.1146) & (0.3615) & (0.1337) & (0.3912) & (0.1532) \\
\hspace*{2mm}20$^{\circ}$C & -0.0367** & -0.0171 & -0.0446 & 0.0172 & -0.0924 & -0.0237 & -0.0668 & 0.0249 \\
 & (0.0162) & (0.0543) & (0.1170) & (0.0430) & (0.1356) & (0.0501) & (0.1467) & (0.0575) \\
\hspace*{2mm}30$^{\circ}$C & -0.0193** & -0.0207** & -0.0448** & -0.0362*** & -0.0754*** & -0.0354 & -0.0471* & -0.0444* \\
 & (0.0087) & (0.0102) & (0.0205) & (0.0132) & (0.0252) & (0.0214) & (0.0282) & (0.0237) \\
\hspace*{2mm}35$^{\circ}$C & -0.0674** & -0.0725** & -0.1566** & -0.1266*** & -0.2639*** & -0.1238 & -0.1648* & -0.1555* \\
 & (0.0305) & (0.0356) & (0.0718) & (0.0461) & (0.0881) & (0.0750) & (0.0987) & (0.0828) \\
\hspace*{2mm}40$^{\circ}$C & -0.1155** & -0.1243** & -0.2685** & -0.2171*** & -0.4523*** & -0.2122 & -0.2825* & -0.2666* \\
 & (0.0524) & (0.0611) & (0.1230) & (0.0790) & (0.1510) & (0.1285) & (0.1692) & (0.1420) \\
\multicolumn{1}{l}{} &  &  &  &  &  &  &  &  \\
\multicolumn{1}{l}{Observations} & \multicolumn{2}{c}{3,086} & \multicolumn{2}{c}{3,086} & \multicolumn{2}{c}{2,940} & \multicolumn{2}{c}{2,792} \\
\multicolumn{1}{l}{$R^2$} & \multicolumn{2}{c}{0.329} & \multicolumn{2}{c}{0.332} & \multicolumn{2}{c}{0.354} & \multicolumn{2}{c}{0.372} \\
\multicolumn{1}{l}{Within $R^2$} & \multicolumn{2}{c}{0.007} & \multicolumn{2}{c}{0.011} & \multicolumn{2}{c}{0.026} & \multicolumn{2}{c}{0.035} \\ \bottomrule\bottomrule
\end{tabular}%
\begin{tablenotes}[flushleft]
\item \textit{Notes:} All models included province FE and year FE and account for the effects of changes in precipitation. When estimating model with lags, the growth-temperature response functions were estimated using distributed-lag form of Equations \ref{eqn:polynomial_function}, \ref{eqn:binned_function}, and \ref{eqn:hdd_cdd_function}, with precipitation and its lags are also included. All temperature and precipitation variables interacted with a low-income province dummy defined as a province with below-median average inflation-adjusted GPP per capita across the sample period. The standard errors in parentheses are clustered by province.\\
***, **, * indicate significance level of 1\%, 5\%, and 10\%, respectively.
\end{tablenotes}
\end{threeparttable}
}
\end{table}

%table4
\begin{table}[t!]
\small
\centering
\captionsetup{width=1\textwidth}
\caption{\textbf{Temperature response function on the components of output growth, assuming identical response across provinces.} The growth-temperature response functions in the agricultural (Panel A), industrial (Panel B), and service (Panel C) sectors were estimated using Equation \ref{eqn:polynomial_function} with up to five lags. Point estimates indicate the effect of a single day at each 24-hour average temperature on the growth rate of the respective sectoral output, relative to a day with an average temperature of 26$^\circ$C.}
\label{tab:sectoral_output_common}
%\resizebox{\textwidth}{!}{%
\begin{threeparttable}
%\begin{tabular}{@{}cllll@{}}
\begin{tabularx}{1\textwidth}{lXXXX}
\toprule
Temperature & \multicolumn{1}{l}{No lags} & \multicolumn{1}{l}{1 lag} & \multicolumn{1}{l}{3 lags} & \multicolumn{1}{l}{5 lags} \\
 & \multicolumn{1}{l}{(1)} & \multicolumn{1}{l}{(2)} & \multicolumn{1}{l}{(3)} & \multicolumn{1}{l}{(4)} \\ \midrule
\multicolumn{5}{l}{\textbf{Panel A: Change in agricultural sector real value added}} \\
15$^{\circ}$C & -0.4241*** & -0.4533*** & -0.4434*** & -0.0211 \\
 & (0.0772) & (0.1159) & (0.1451) & (0.1647) \\
20$^{\circ}$C & -0.1237*** & -0.1068** & -0.1084* & 0.0520 \\
 & (0.0309) & (0.0441) & (0.0584) & (0.0664) \\
30$^{\circ}$C & -0.0611*** & -0.1161*** & -0.1056*** & -0.1192*** \\
 & (0.0166) & (0.0202) & (0.0277) & (0.0349) \\
35$^{\circ}$C & -0.2990*** & -0.4718*** & -0.4378*** & -0.3634*** \\
 & (0.0499) & (0.0660) & (0.0807) & (0.1032) \\
40$^{\circ}$C & -0.7164*** & -1.0617*** & -0.9924*** & -0.7134*** \\
 & (0.1050) & (0.1472) & (0.1715) & (0.2165) \\
\multicolumn{1}{l}{} &  &  &  &  \\
Observations & 3,086 & 3,086 & 2,940 & 2,792 \\
$R^2$ & 0.298 & 0.303 & 0.317 & 0.323 \\
Within $R^2$ & 0.022 & 0.028 & 0.046 & 0.058 \\ \midrule
\multicolumn{5}{l}{\textbf{Panel B: Change in industrial sector real value added}} \\
15$^{\circ}$C & -0.0277 & -0.1643 & 0.0169 & -0.0745 \\
 & (0.0967) & (0.1205) & (0.1653) & (0.1766) \\
20$^{\circ}$C & -0.0180 & -0.0838* & 0.0082 & -0.0193 \\
 & (0.0383) & (0.0475) & (0.0605) & (0.0658) \\
30$^{\circ}$C & 0.0160 & 0.0482 & -0.0041 & -0.0155 \\
 & (0.0159) & (0.0206) & (0.0302) & (0.0328) \\
35$^{\circ}$C & 0.0404 & 0.0997 & -0.0078 & -0.0668 \\
 & (0.0472) & (0.0621) & (0.1051) & (0.1106) \\
40$^{\circ}$C & 0.0697 & 0.1417 & -0.0098 & -0.1535 \\
 & (0.1041) & (0.1365) & (0.2359) & (0.2459) \\
\multicolumn{1}{l}{} &  &  &  &  \\
Observations & 3,086 & 3,086 & 2,940 & 2,792 \\
$R^2$ & 0.205 & 0.207 & 0.221 & 0.238 \\
Within $R^2$ & 0.001 & 0.002 & 0.012 & 0.019 \\ \midrule
\multicolumn{5}{l}{\textbf{Panel C: Change in service sector real value added}} \\
15$^{\circ}$C & -0.0330 & -0.0465 & -0.1378** & -0.1151* \\
 & (0.0257) & (0.0311) & (0.0593) & (0.0590) \\
20$^{\circ}$C & -0.0120 & -0.0171 & -0.0447* & -0.0378 \\
 & (0.0097) & (0.0129) & (0.0232) & (0.0233) \\
30$^{\circ}$C & 0.0001 & 0.0004 & -0.0109 & -0.0082 \\
 & (0.0052) & (0.0093) & (0.0096) & (0.0103) \\
35$^{\circ}$C & -0.0086 & -0.0115 & -0.0702** & -0.0560* \\
 & (0.0173) & (0.0277) & (0.0295) & (0.0310) \\
40$^{\circ}$C & -0.0273 & -0.0372 & -0.1803*** & -0.1456** \\
 & (0.0378) & (0.0562) & (0.0659) & (0.0677) \\
\multicolumn{1}{l}{} &  &  &  &  \\
Observations & 3,086 & 3,086 & 2,940 & 2,792 \\
$R^2$ & 0.492 & 0.493 & 0.464 & 0.476 \\
Within $R^2$ & 0.009 & 0.011 & 0.019 & 0.033 \\ \bottomrule
\end{tabularx}%
\begin{tablenotes}[flushleft]
\item \textit{Notes:} All models included province FE and year FE and account for the effects of changes in precipitation. When estimating the model with lags, the growth-temperature response functions were estimated using the distributed-lag form of Equations \ref{eqn:polynomial_function}, with precipitation and its lags included. The standard errors in parentheses are clustered by province.\\
***, **, * indicate significance level of 1\%, 5\%, and 10\%, respectively.
\end{tablenotes}
\end{threeparttable}
%}
\end{table}

%table5
\begin{table}[t!]
\centering
\captionsetup{width=1\textwidth}
\caption{\textbf{Temperature response function in the components of output growth, allowing high- and low-income provinces to respond differently to temperature changes.} The growth-temperature response functions in agricultural (Panel A), industrial (Panel B), and service (Panel C) sectors were estimated using Equation \ref{eqn:polynomial_function} with up to 5 lags. Point estimates indicate the effect of a single day at each 24-hour average temperature on the growth rate of the respective sectoral output, relative to a day with an average temperature of 26$^\circ$C.}
\label{tab:sectoral_output_xPoor}
\resizebox{\textwidth}{!}{%
\begin{threeparttable}
\begin{tabular}{@{}lllllllll@{}}
\toprule\toprule
 & \multicolumn{2}{c}{No lags} & \multicolumn{2}{c}{1 lag} & \multicolumn{2}{c}{3 lags} & \multicolumn{2}{c}{5 lags} \\ \cmidrule(l){2-9} 
\hspace*{2mm}Temperature & \multicolumn{1}{c}{Low income} & \multicolumn{1}{c}{High income} & \multicolumn{1}{c}{Low income} & \multicolumn{1}{c}{High income} & \multicolumn{1}{c}{Low income} & \multicolumn{1}{c}{High income} & \multicolumn{1}{c}{Low income} & \multicolumn{1}{c}{High income} \\
 & \multicolumn{1}{c}{(1)} & \multicolumn{1}{c}{(2)} & \multicolumn{1}{c}{(3)} & \multicolumn{1}{c}{(4)} & \multicolumn{1}{c}{(5)} & \multicolumn{1}{c}{(6)} & \multicolumn{1}{c}{(7)} & \multicolumn{1}{c}{(8)} \\ \midrule
\multicolumn{9}{l}{\textbf{Panel A: Change in agricultural sector real value added}} \\
\hspace*{2mm}15$^{\circ}$C & -0.4299*** & -0.4760** & -0.4403*** & -0.4432** & -0.5173*** & -0.3694 & -0.1076 & 0.3773 \\
 & (0.0836) & (0.1820) & (0.1227) & (0.2190) & (0.1704) & (0.2597) & (0.1855) & (0.2039) \\
\hspace*{2mm}20$^{\circ}$C & -0.1259*** & -0.1448** & -0.1062** & -0.1035 & -0.1231* & -0.0990 & 0.0271 & 0.1923 \\
 & (0.0318) & (0.0702) & (0.0444) & (0.0818) & (0.0643) & (0.1036) & (0.0669) & (0.0809) \\
\hspace*{2mm}30$^{\circ}$C & -0.0608*** & -0.0566*** & -0.1078*** & -0.1152*** & -0.1300*** & -0.0707** & -0.1325*** & -0.1101*** \\
 & (0.0185) & (0.0206) & (0.0265) & (0.0223) & (0.0331) & (0.0332) & (0.0446) & (0.0376) \\
\hspace*{2mm}35$^{\circ}$C & -0.2996*** & -0.2995*** & -0.4435*** & -0.4665*** & -0.5310*** & -0.3129*** & -0.4268*** & -0.2275** \\
 & (0.0596) & (0.0655) & (0.0912) & (0.0839) & (0.1093) & (0.0895) & (0.1520) & (0.1127) \\
\hspace*{2mm}40$^{\circ}$C & -0.7193*** & -0.7338*** & -1.0026*** & -1.0482*** & -1.1972*** & -0.7259*** & -0.8642** & -0.3223 \\
 & (0.1283) & (0.1620) & (0.2001) & (0.2130) & (0.2403) & (0.2101) & (0.3292) & (0.2432) \\
\multicolumn{1}{l}{} &  &  &  &  &  &  &  &  \\
\multicolumn{1}{l}{Observations} & \multicolumn{2}{c}{3,086} & \multicolumn{2}{c}{3,086} & \multicolumn{2}{c}{2,940} & \multicolumn{2}{c}{2,792} \\
\multicolumn{1}{l}{$R^2$} & \multicolumn{2}{c}{0.298} & \multicolumn{2}{c}{0.304} & \multicolumn{2}{c}{0.323} & \multicolumn{2}{c}{0.333} \\
\multicolumn{1}{l}{Within $R^2$} & \multicolumn{2}{c}{0.022} & \multicolumn{2}{c}{0.030} & \multicolumn{2}{c}{0.054} & \multicolumn{2}{c}{0.073} \\ \midrule
\multicolumn{9}{l}{\textbf{Panel B: Change in industrial sector real value added}} \\
\hspace*{2mm}15$^{\circ}$C & -0.0110 & 0.2734 & -0.1753 & 0.3188 & 0.0051 & 0.3224 & -0.1984 & 0.3463 \\
 & (0.1015) & (0.1610) & (0.1294) & (0.1684) & (0.1552) & (0.2527) & (0.1906) & (0.2963) \\
\hspace*{2mm}20$^{\circ}$C & -0.0189 & 0.1064 & -0.0942* & 0.1123 & 0.0083 & 0.1218 & -0.0757 & 0.1497 \\
 & (0.0383) & (0.0634) & (0.0498) & (0.0658) & (0.0578) & (0.0897) & (0.0704) & (0.1075) \\
\hspace*{2mm}30$^{\circ}$C & 0.0299 & -0.0140 & 0.0609 & 0.0072 & -0.0130 & -0.0091 & 0.0072 & -0.0476 \\
 & (0.0176) & (0.0198) & (0.0244) & (0.0265) & (0.0295) & (0.0402) & (0.0312) & (0.0461) \\
\hspace*{2mm}35$^{\circ}$C & 0.0866 & 0.0326 & 0.1348 & 0.1086 & -0.0376 & 0.0605 & -0.0326 & -0.0483 \\
 & (0.0586) & (0.0565) & (0.0779) & (0.0814) & (0.0996) & (0.1511) & (0.1091) & (0.1683) \\
\hspace*{2mm}40$^{\circ}$C & 0.1649 & 0.1503 & 0.2063 & 0.3126 & -0.0714 & 0.2203 & -0.1265 & 0.0163 \\
 & (0.1314) & (0.1354) & (0.1705) & (0.1827) & (0.2206) & (0.3499) & (0.2488) & (0.3895) \\
\multicolumn{1}{l}{} &  &  &  &  &  &  &  &  \\
\multicolumn{1}{l}{Observations} & \multicolumn{2}{c}{3,086} & \multicolumn{2}{c}{3,086} & \multicolumn{2}{c}{2,940} & \multicolumn{2}{c}{2,792} \\
\multicolumn{1}{l}{$R^2$} & \multicolumn{2}{c}{0.208} & \multicolumn{2}{c}{0.211} & \multicolumn{2}{c}{0.229} & \multicolumn{2}{c}{0.251} \\
\multicolumn{1}{l}{Within $R^2$} & \multicolumn{2}{c}{0.004} & \multicolumn{2}{c}{0.008} & \multicolumn{2}{c}{0.022} & \multicolumn{2}{c}{0.036} \\ \midrule
\multicolumn{9}{l}{\textbf{Panel C: Change in service sector real value added}} \\
\hspace*{2mm}15$^{\circ}$C & -0.0206 & -0.0939* & -0.0423 & -0.0161 & -0.1264* & -0.1809** & -0.0813 & 0.0587 \\
 & (0.0289) & (0.0524) & (0.0357) & (0.0679) & (0.0657) & (0.0698) & (0.0601) & (0.1095) \\
\hspace*{2mm}20$^{\circ}$C & -0.0097 & -0.0336 & -0.0167 & -0.0031 & -0.0412 & -0.0627** & -0.0307 & 0.0359 \\
 & (0.0104) & (0.0202) & (0.0142) & (0.0253) & (0.0251) & (0.0284) & (0.0240) & (0.0425) \\
\hspace*{2mm}30$^{\circ}$C & 0.0043 & -0.0011 & 0.0026 & -0.0055 & -0.0096 & -0.0061 & 0.0022 & -0.0291** \\
 & (0.0061) & (0.0067) & (0.0097) & (0.0095) & (0.0084) & (0.0142) & (0.0115) & (0.0145) \\
\hspace*{2mm}35$^{\circ}$C & 0.0073 & -0.0291 & -0.0038 & -0.0209 & -0.0633** & -0.0677* & -0.0156 & -0.0713 \\
 & (0.0213) & (0.0211) & (0.0300) & (0.0338) & (0.0277) & (0.0406) & (0.0343) & (0.0449) \\
\hspace*{2mm}40$^{\circ}$C & 0.0076 & -0.0864* & -0.0209 & -0.0457 & -0.1633** & -0.1891** & -0.0561 & -0.1199 \\
 & (0.0471) & (0.0503) & (0.0621) & (0.0793) & (0.0661) & (0.0849) & (0.0733) & (0.1057) \\
\multicolumn{1}{l}{} &  &  &  &  &  &  &  &  \\
\multicolumn{1}{l}{Observations} & \multicolumn{2}{c}{3,086} & \multicolumn{2}{c}{3,086} & \multicolumn{2}{c}{2,940} & \multicolumn{2}{c}{2,792}\\
\multicolumn{1}{l}{$R^2$} & \multicolumn{2}{c}{0.493} & \multicolumn{2}{c}{0.496} & \multicolumn{2}{c}{0.470} & \multicolumn{2}{c}{0.491}\\
\multicolumn{1}{l}{Wihtin $R^2$} & \multicolumn{2}{c}{0.011} & \multicolumn{2}{c}{0.017} & \multicolumn{2}{c}{0.031} & \multicolumn{2}{c}{0.060} \\ \bottomrule\bottomrule
\end{tabular}%
\begin{tablenotes}[flushleft]
\item \textit{Notes:} All models included province FE and year FE and account for the effects of changes in precipitation. When estimating model with lags, the growth-temperature response functions were estimated using distributed-lag form of Equations \ref{eqn:polynomial_function}, with precipitation and its lags are also included. All temperature and precipitation variables were interacted with a low-income province dummy defined as a province having below median average inflation-adjusted GPP per capita across the sample period. The standard errors in parentheses are clustered by province.\\
***, **, * indicate significance level of 1\%, 5\%, and 10\%, respectively.
\end{tablenotes}
\end{threeparttable}
}
\end{table}

%table6
\begin{table}[t!]
\small
\centering
\begin{threeparttable}
\captionsetup{width=1\textwidth}
\caption{\textbf{Summary of projected impacts "\textit{without bias-correction}" of climate change on Thailand's GDP per capita in 2090.} This table presents the projected impacts of climate change, relative to projections using the constant 2003-2022 average temperature (i.e., Thailand in the absence of climate change), under RCP4.5 (Panel A) and RCP8.5 (Panel B) emission scenarios in 2090 for each combination of historical growth-temperature response functions ("Model specification" column) and output growth assumptions (baseline scenario, SSP3, and SSP5). The percentiles were calculated from the bootstrapped distribution of projections. The probability of positive impacts is the proportion of projected positive impacts to total runs.}
\label{tab:summary_impacts_on_GDP2090}
\begin{tabularx}{1\textwidth}{llXXX}
\toprule\toprule
Model specification & Statistic & \multicolumn{3}{c}{Output growth assumptions} \\ \cmidrule(l){3-5} 
 & \multicolumn{1}{c}{} & BASE & SSP3 & SSP5 \\ \midrule
 &  & \multicolumn{3}{c}{\textit{Panel A: RCP4.5 emission scenario}} \\ \cmidrule(l){3-5} 
Common, No lags & 50th percentile & -3.51\% & 55.82\% & 54.45\% \\
 & {[}5th, 95th percentiles{]} & {[}-86.4\%, 727.2\%{]} & {[}-78.7\%, 1,428.4\%{]} & {[}-78.6\%, 1,386.6\%{]} \\
 & Probability of positive impacts & 0.49 & 0.63 & 0.63 \\ %\midrule
Common, 5 lags & 50th percentile & -97.79\% & -95.64\% & -95.61\% \\
 & {[}5th, 95th percentiles{]} & {[}-99.8\%, -43.7\%{]} & {[}-99.8\%, 11.5\%{]} & {[}-99.8\%, 10.0\%{]} \\
 & Probability of positive impacts & 0.03 & 0.06 & 0.06 \\ %\midrule
High/Low, No lags & 50th percentile & -5.17\% & 45.78\% & 44.42\% \\
 & {[}5th, 95th percentiles{]} & {[}-93.6\%, 4,923.0\%{]} & {[}-91.6\%, 8,496.9\%{]} & {[}-91.5, 8,175.5{]} \\
 & Probability of positive impacts & 0.49 & 0.58 & 0.58 \\ %\midrule
High/Low, 5 lags & 50th percentile & -99.78\% & -99.88\% & -99.87\% \\
 & {[}5th, 95th percentiles{]} & {[}-100.0\%, -91.9\%{]} & {[}-100.0\%, -91.6\%{]} & {[}-100.0\%, -94.4\%{]} \\
 & Probability of positive impacts & 0.01 & 0.01 & 0.01 \\ \midrule
 &  & \multicolumn{3}{c}{\textit{Panel B: RCP8.5 emission scenario}} \\ \cmidrule(l){3-5} 
Common, No lags & 50th percentile & -50.93\% & -15.01\% & -15.76\% \\
 & {[}5th, 95th percentiles{]} & {[}-96.0\%, 529.9\%{]} & {[}-93.1\%, 1,014.7\%{]} & {[}-93.0\%, 984.1\%{]} \\
 & Probability of positive impacts & 0.32 & 0.46 & 0.46 \\ %\midrule
Common, 5 lags & 50th percentile & -99.69\% & -99.32\% & -99.31\% \\
 & {[}5th, 95th percentiles{]} & {[}-100.0\%, -82.1\%{]} & {[}-100.0\%, -65.5\%{]} & {[}-100.0\%, -65.9\%{]} \\
 & Probability of positive impacts & 0.01 & 0.02 & 0.02 \\ %\midrule
High/Low, No lags & 50th percentile & -51.6\% & -21.7\% & -22.17\% \\
 & {[}5th, 95th percentiles{]} & {[}-98.3\%, 4,827.8\%{]} & {[}-97.4\%, 8,364.6\%{]} & {[}-97.4\%, 8,009.9\%{]} \\
 & Probability of positive impacts & 0.37 & 0.46 & 0.46 \\ %\midrule
High/Low, 5 lags & 50th percentile & -99.97\% & -99.99\% & -99.99\% \\
 & {[}5th, 95th percentiles{]} & {[}-100.0\%, -94.4\%{]} & {[}-100.0\%, -95.2\%{]} & {[}-100.0\%, -95.2\%{]} \\
 & Probability of positive impacts & 0.01 & 0.01 & 0.01 \\ \bottomrule\bottomrule
\end{tabularx}
\end{threeparttable}
\begin{tablenotes}[flushleft]
\item \textit{Notes:} All estimated models included province FE and year FE and account for the effects of changes in precipitation. When estimating the model with "5 lags,” the growth-temperature response functions were estimated using the distributed-lag form of Equations \ref{eqn:polynomial_function}, with precipitation and its lags included. All temperature and precipitation variables were interacted with a low-income province dummy defined as a province having below median average inflation-adjusted GPP per capita across the sample period. "Common" indicates models that assume a common response function across all provinces, while "High/Low" models assume that high- and low-income provinces respond differently to temperature changes. The differential effect of temperature changes on output growth $\delta_{py}$ was estimated using Equation \ref{eqn:additional_growth}.
\end{tablenotes}
\end{table}

%table7
\begin{table}[t!]
\small
\centering
\begin{threeparttable}
\captionsetup{width=1\textwidth}
\caption{\textbf{Summary of projected impacts of climate change "\textit{with bias-correction}" on Thailand's GDP per capita in 2090.} This table presents the projected impacts of climate change, relative to projection using constant 2003-2022 average temperature (that is, Thailand in the absence of climate change), under RCP4.5 (Panel A) and RCP8.5 (Panel B) emission scenarios in 2090 for each combination of historical growth-temperature response functions ("Model specification" column) and output growth assumptions (baseline scenario, SSP3, and SSP5). The percentiles were calculated from the bootstrapped distribution of projections. The probability of positive impacts is the proportion of projected positive impacts to total runs.}
\label{tab:summary_impacts_on_GDP2090_correction}
\begin{tabularx}{1\textwidth}{llXXX}
\toprule\toprule
Model specification & Statistics & \multicolumn{3}{c}{Output growth assumptions} \\ \cmidrule(l){3-5} 
 &  & \multicolumn{1}{c}{BASE} & \multicolumn{1}{c}{SSP3} & \multicolumn{1}{c}{SSP5} \\ \midrule
 &  & \multicolumn{3}{c}{\textit{Panel A: RCP4.5 emission scenario}} \\ \cmidrule(l){3-5} 
Common, No lags & 50th percentile & -62.91\% & -57.82\% & -57.68\% \\
 & {[}5th, 95th percentiles{]} & {[}-83.9\%,-28.3\%{]} & {[}-83.1\%,-22.5\%{]} & {[}-82.9\%,-22.5\%{]} \\
 & Probability of positive impacts & 0.01 & 0.01 & 0.01 \\
Common, 5 lags & 50th percentile & -96.00\% & -95.89\% & -95.82\% \\
 & {[}5th, 95th percentiles{]} & {[}-99.1\%,-79.1\%{]} & {[}-99.1\%,-77.3\%{]} & {[}-99.1\%,-77.2\%{]} \\
 & Probability of positive impacts & 0.00 & 0.00 & 0.00 \\
High/Low, No lags & 50th percentile & -60.55\% & -56.74\% & -56.60\% \\
 & {[}5th, 95th percentiles{]} & {[}-83.0\%,-8.5\%{]} & {[}-82.7\%,-5.8\%{]} & {[}-82.6\%,-5.8\%{]} \\
 & Probability of positive impacts & 0.04 & 0.03 & 0.03 \\
High/Low, 5 lags & 50th percentile & -92.89\% & -94.54\% & -94.42\% \\
 & {[}5th, 95th percentiles{]} & {[}-98.8\%,-17.3\%{]} & {[}-99.0\%,-36.8\%{]} & {[}-99.0\%,-36.2\%{]} \\
 & Probability of positive impacts & 0.04 & 0.03 & 0.03 \\ \midrule
 &  & \multicolumn{3}{c}{\textit{Panel B: RCP8.5 emission scenario}} \\ \cmidrule(l){3-5} 
Common, No lags & 50th percentile & -85.81\% & -82.7\% & -82.57\% \\
 & {[}5th, 95th percentiles{]} & {[}-95.9\%,-54.8\%{]} & {[}-94.8\%,-47.4\%{]} & {[}-94.8\%,-47.3\%{]} \\
 & Probability of positive impacts & 0.00 & 0.01 & 0.01 \\
Common, 5 lags & 50th percentile & -99.50\% & -99.42\% & -99.41\% \\
 & {[}5th, 95th percentiles{]} & {[}-99.9\%,-93.3\%{]} & {[}-100.0\%,-92.5\%{]} & {[}-99.9\%,-92.4\%{]} \\
 & Probability of positive impacts & 0.00 & 0.00 & 0.00 \\
High/Low, No lags & 50th percentile & -83.43\% & -80.51\% & -80.39\% \\
 & {[}5th, 95th percentiles{]} & {[}-95.6\%,-17.0\%{]} & {[}-94.7\%,-10.1\%{]} & {[}-94.6\%,-10.0\%{]} \\
 & Probability of positive impacts & 0.04 & 0.04 & 0.04 \\
High/Low, 5 lags & 50th percentile & -98.58\% & -98.99\% & -98.96\% \\
 & {[}5th, 95th percentiles{]} & {[}-99.9\%,16.5\%{]} & {[}-99.9\%,-17.2\%{]} & {[}-99.9\%,-16.6\%{]} \\
 & Probability of positive impacts & 0.06 & 0.05 & 0.05 \\ \bottomrule\bottomrule
\end{tabularx}
\end{threeparttable}
\begin{tablenotes}[flushleft]
\item \textit{Notes:} All estimated models included province FE and year FE and account for the effects of changes in precipitation. When estimating model with "5 lags", the growth-temperature response functions were estimated using distributed-lag form of Equations \ref{eqn:polynomial_function}, with precipitation and its lags are also included. All temperature and precipitation variables were interacted with a low-income province dummy defined as a province having below median average inflation-adjusted GPP per capita across the sample period. "Common" indicates models that assume a common response function across all provinces, while "High/Low" models assume that high- and low-income provinces respond differently to temperature changes. The "\textit{bias-corrected}" differential effect of temperature changes on output growth $\delta^{'}_{py}$ was estimated using Equation \ref{eqn:bias_corrected_additional_growth}.
\end{tablenotes}
\end{table}

\clearpage
\newpage
\appendix
\renewcommand{\appendixpagename}{\centering APPENDIX}
\appendixpage

\renewcommand{\thesection}{\Alph{section}}
\setcounter{table}{0}
\setcounter{figure}{0}
\setcounter{equation}{0}
\renewcommand{\thetable}{\textbf{\Alph{section}\arabic{table}}}
\renewcommand{\thefigure}{\textbf{\Alph{section}\arabic{figure}}}
\renewcommand{\figurename}{\textbf{Figure}}
\renewcommand{\tablename}{\textbf{Table}}
\renewcommand{\theequation}{\thesection.\arabic{equation}}

\section{Data and methods}
\subsection{Polygonized population dataset} \label{si:polygonized}
I developed a polygonized population dataset by combining the annual land use map and the annual population count datasets. The land use map comes with scale 1:25000 (approximately 12.5m x 12.5m in raster resolution) and provides granular detail of land use classification on three levels. The first level consists of five major types of land use: urban and built-up, agricultural, forest, water bodies, and miscellaneous. The second and third levels simply provide the finer and the finest detail of land use classification, respectively. The annual land use map dataset in the shapefile format between 2006-2019 was obtained from the Land Development Department (LDD). To calculate a spatially disaggregated population count that is compatible with the human elements of the economy, I used only polygons classified as urban and built-up land types. The land use map of 2006 was used as a proxy for the land use pattern before 2006, whereas that of 2019 was used as a proxy for the land use pattern of 2019 onward. The annual population counts dataset at the third administrative level (or, \textit{Tambon} in Thai) between 1997-2022 was obtained from the \href{https://stat.bora.dopa.go.th/new_stat/webPage/statByAge.php}{Official Statistics Registration Systems} (OSRS) operated by the Department of Provincial Administration (DOPA).\footnote{This administrative level corresponds to, for example, county in the United States.} 

\begin{figure}[t!]
\centering
\captionsetup{width=1\textwidth}
\includegraphics[scale=.145]{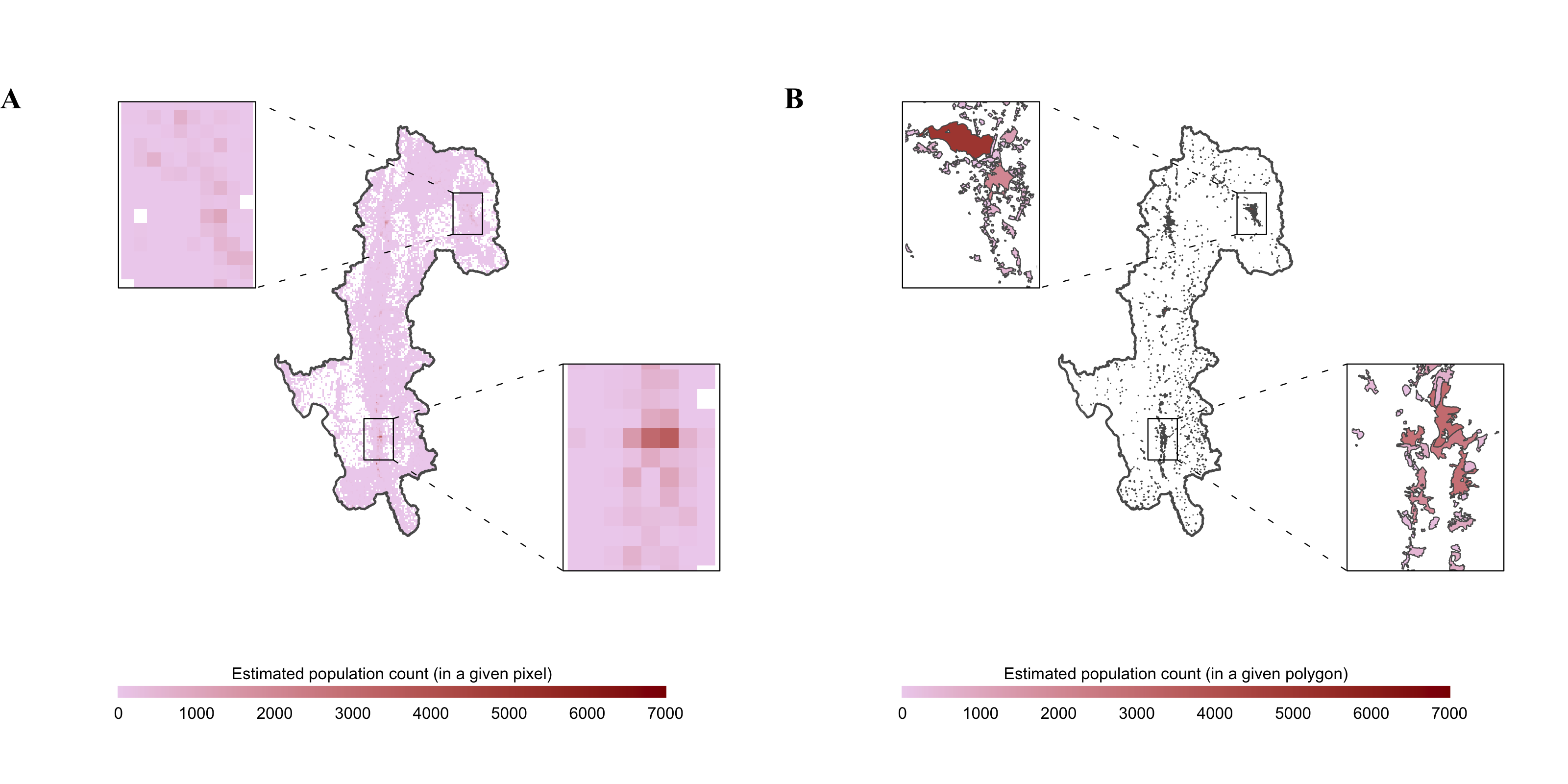}
\caption{\textbf{Comparison between LandScan's gridded population and polygonized popoluation datasets.} \textbf{Panel A} displays the LandScan's gridded population dataset. Each pixel contains a value that represents the number of people. \textbf{Panel B} displays the polygonized population which is developed for this paper. Similarly, each polygon contains the number of people, which was calculated based on the share of each polygon's area to the total area of all polygons located within the administrative boundary of a given Tambon. Both panels display the population distribution in the \textit{Mae Hong Son} province, which is located in the Upper North region of Thailand. To be comparable, both panels use the same colorbar scale. The inset plots in each panel zoom in on a specific regions of \textit{Mae Hong Son} province to provide more detailed views of the subsets of both data.}
\label{fig:lscan_vs_polypop}
\end{figure}

Population count for each polygon was distributed based on the share of each polygon's area to the total area of all polygons located within the administrative boundary of a given Tambon. Because population counts are available from 1997 onward, the polygonized data for the periods between 1982-1996 were calculated using 1997 population data. Conceptually, this polygonized population dataset is similar to using the gridded population data, for example, \href{https://landscan.ornl.gov/}{the LandScan dataset produced by the Oak Ridge National Laboratory} and \href{https://earthdata.nasa.gov/centers/sedac-daac}{the Gridded Population of the World dataset produced by Columbia University's Center for International Earth Science Information Network (CIESIN)}, which is derived with geostatistical methods using population counts and spatial datasets. While the gridded population dataset deriving the distribution of human population counts on a continuous raster grids, I instead derived the distribution of population counts onto the polygons.

Figure \ref{fig:lscan_vs_polypop} presents the differences between the LandScan gridded population (Panel A) and polygonized population (Panel B) datasets for \textit{Mae Hong Son} province, which is located in the Upper North region of Thailand. Each pixel of the LandScan data contains the number of people in that pixel, whereas the value contained in a given polygon represents the number of people allocated to that polygon, not the population density. The population in the LandScan data was distributed more widely over the given space. The lower-right inset plots in both panels show a broadly similar pattern of population distribution, whereas the upper-left inset plots in both panels show a somewhat different distribution of the population. Note that the upper-left inset plots in both panels cover \textit{Amphur Pai}, which is a highly populated area as it is one of the most popular destinations for both Thai and foreign tourists. Such differences in the upper-left inset plots present suggestive evidence that the LandScan data are somewhat inferior to the polygonized data.

\subsection{Future climate projections} \label{si:future_climate_projections}
\begin{figure}[t!]
\centering
\captionsetup{width=1\textwidth}
\includegraphics[scale=.1225]{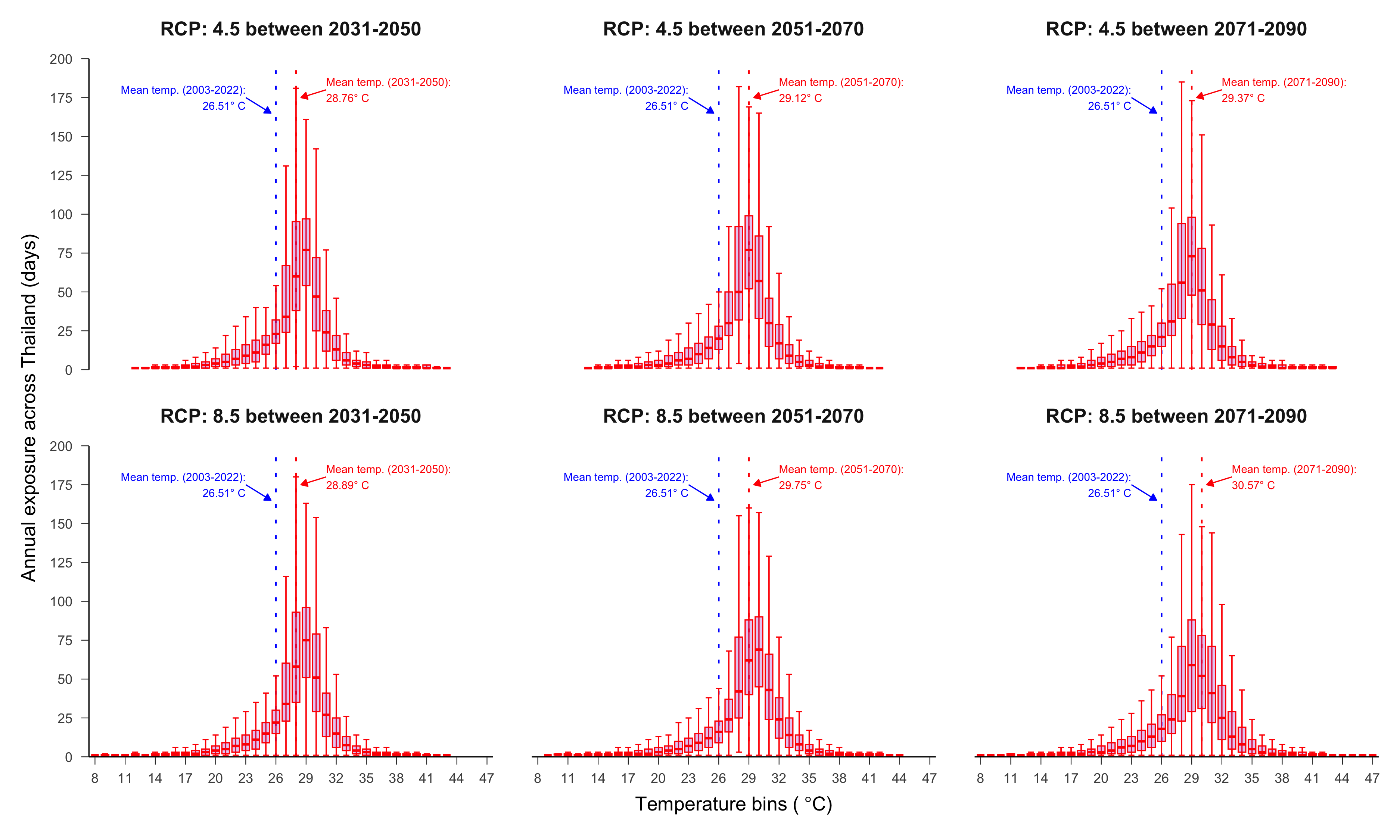}
\caption{\textbf{Future climate projections under different emission scenarios.} The boxplots display the projected distribution of the number of days that the population in 77 provinces will be exposed to each 1$^\circ$C temperature bin in three future periods: 2031-2050 (left column), 2051-2070 (middle column), and 2071-2090 (right column). The box represents the 25\%-75\% range, and the middle line within each box indicates the median across all provinces and years. The whiskers indicate the minimum and maximum exposure to each interval. The blue lines indicate population-weighted average temperatures observed between 2003-2022, while red lines indicate projected population-weighted average temperatures for the respective periods. Upper row shows future climate projections under RCP4.5, and those under RCP8.5 are shown in the lower row.}
\label{fig:summary_future_climate}
\end{figure}

Climate projections under both RCP4.5 and RCP8.5 emission scenarios were downloaded from \href{https://www.nccs.nasa.gov/services/data-collections/land-based-products/nex-gddp}{The NASA Center for Climate Simulation}. Each emission scenario is based on different assumptions about greenhouse gas emissions and mitigation policies, among other factors \citep{meinshausen2011rcp}. Following \cite{burke2015global}, I used the "business as usual" RCP8.5 emission scenario which assumes an intensive growth in fossil fuel emission, and is generally taken as the worst-case climate change scenarios. \cite{hausfather2020emissions} however argued that RCP8.5 should instead be labelled as unlikely worst cases, rather than as business
This suggests that climate-impact studies should include scenarios that reflect more plausible outcomes, such as SSP2-4.5, SSP4-6.0, and SSP3-7.0. Therefore, I added an RCP4.5 emission scenario, which is described by the Intergovernmental Panel on Climate Change (IPCC) as an intermediate scenario with modest mitigation policies. These downscaled projections provide maximum and minimum temperatures, and mean precipitation at daily frequency of 0.25$^{\circ}$ x 0.25$^{\circ}$ (~25 km x 25 km at the equator). Only seven (out of 21 models) covering a wide range of projections are used to project future impacts on economic output due to constraints in computational resources. Specifically, ACCESS1-0, BNU-ESM, CCSM4, INMCM4, MIROC5, MPI-ESM-MR, and NorESM1-M. Both temperature and precipitation estimates were processed using the same procedure as the historical weather data, as described in the main text. To construct the population-weighted average of the climate projection, I fixed the population distribution to be the same as that in 2022 and assumed that the land use pattern remained fixed at that of 2019. Figure \ref{fig:summary_future_climate} displays projections of the distribution of average annual exposure across 77 provinces in Thailand in three future periods: 2031-2050 (left column), 2051-2070 (middle column), and 2071-2090 (right column). The shift toward higher temperature bins in the temperature distribution was observed under both emission scenarios. By 2090, average temperature in provinces in Thailand are projected to rise by 2.86$^{\circ}$ under RCP 4.5 emission scenario, and by 4.06$^{\circ}C$ under RCP8.5. Although the temperatures above 42$^{\circ}C$, which have never been observed historically, become more frequent, the consequent increases in the annual exposure, relative to other temperature bins, do likely not substantively affect the projection of economic impacts.

\begin{figure}[t!]
\centering
\captionsetup{width=1\textwidth}
\includegraphics[scale=.2225]{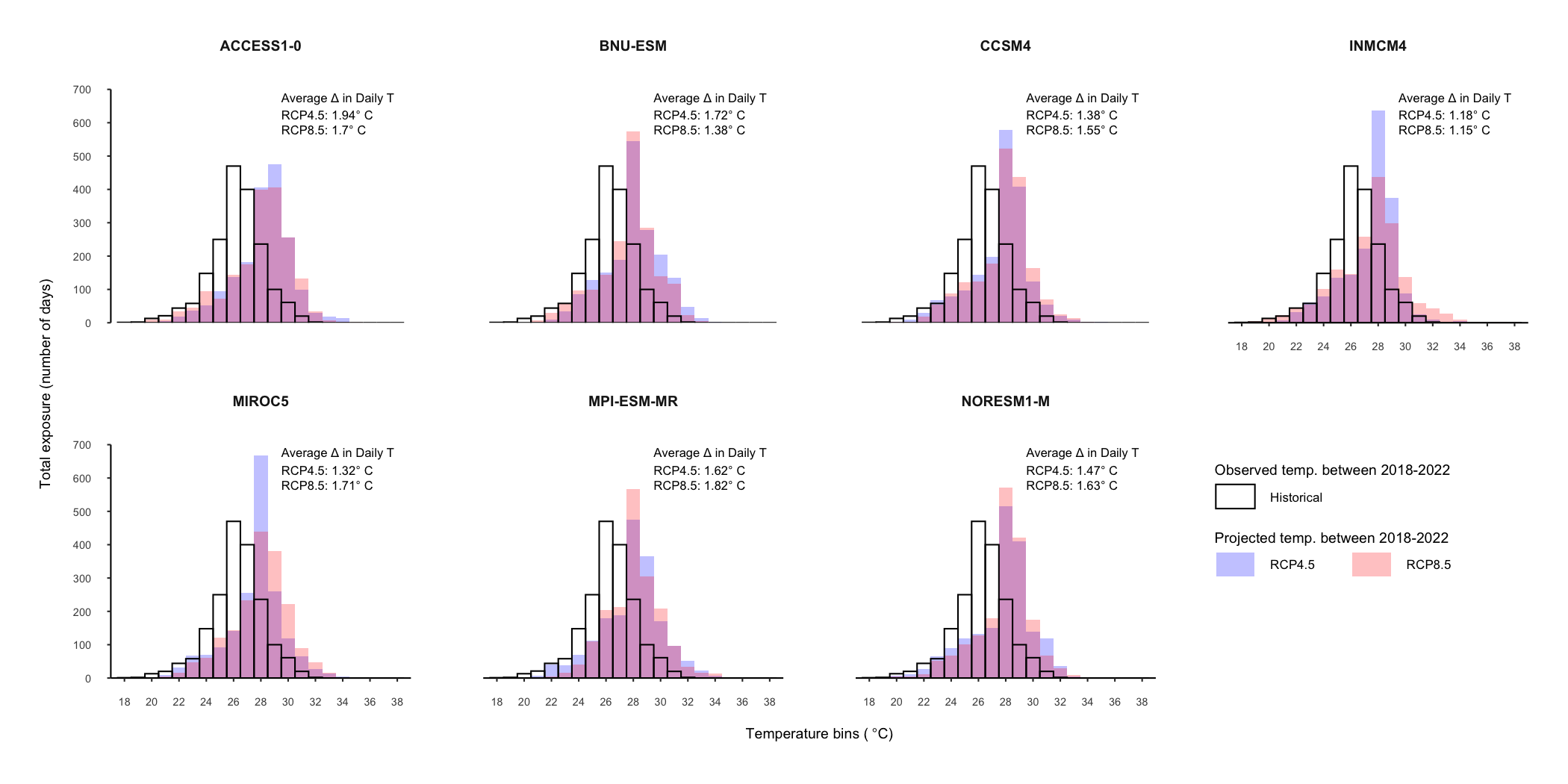}
\caption{\textbf{Projected versus observed distribution of daily mean temperature between 2018-2022}. The histograms display the full distributions of daily mean temperature bins across Thailand between 2018-2022. The black solid outlines are the historical temperatures observed during 2018-2022, whereas blue shaded areas are the corresponding projections in seven climate models under RCP4.5, and RCP8.5 are red-shaded. The height of each bar corresponds to total number of days that the average person experiences in each temperature bin.}
\label{fig:bias_in_projected_climate}
\end{figure}

Figure \ref{fig:bias_in_projected_climate} presents the full distributions of daily mean temperature bins across Thailand between 2018-2022. The black solid outlines are the historical temperatures observed during 2018-2022, whereas the blue shaded areas are the corresponding projections in seven climate models under RCP4.5, and RCP8.5 are red-shaded. The height of each bar corresponds to the total number of days that the average person experiences in each temperature bin. An examination of the figure highlights that the projected temperature distributions under both RCPs are potentially upward-biased compared to the observed ones in all seven climate models. The average differences in daily temperature are between 1.2-1.9$^\circ$C, depending on the climate model and emission scenario. Therefore, it is very challenging to address these potential biases before proceeding to an exercise of climate-impact projections. Section \ref{sec:projection} in the main text discusses the strategy used to estimate the bias-corrected impact projections.

\subsection{Methods} \label{si:methods}
\begin{figure}[t!]
\centering
\captionsetup{width=1\textwidth}
\includegraphics[scale=.078]{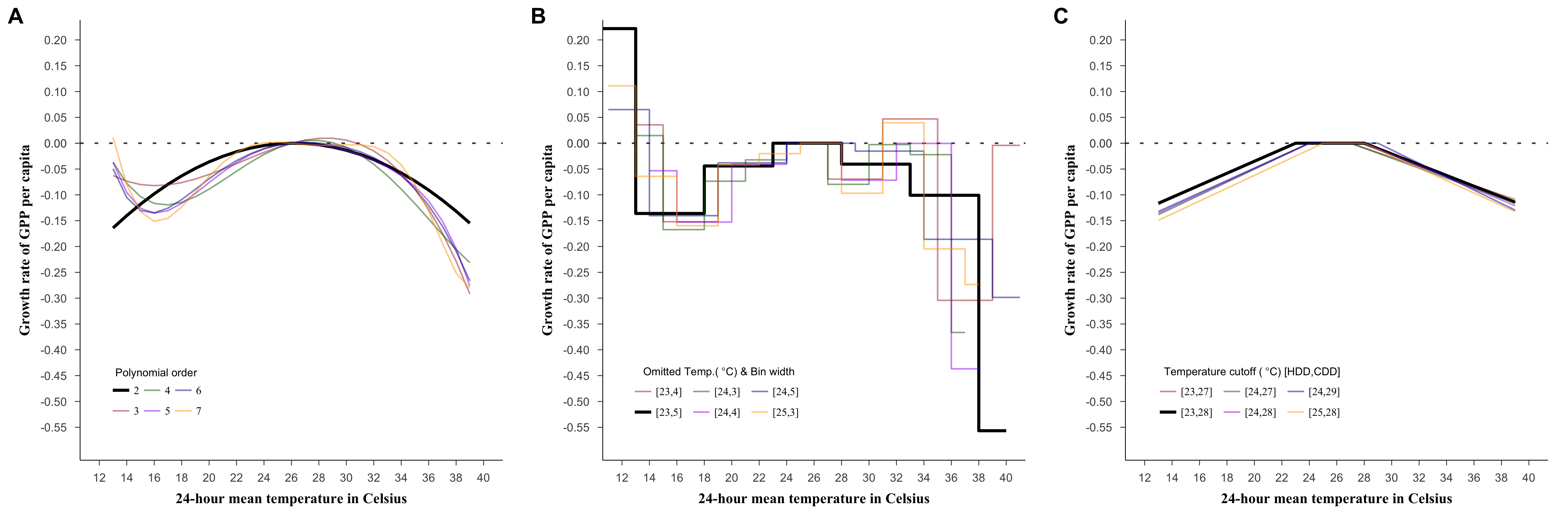}
\caption{\textbf{Specification selection for alternative functional form $h(T)$}. The graphs show the growth-temperature response functions estimated using various specifications of all three functional forms, as detailed in Section \ref{sec:econ_temp_relationship}. \textbf{Panel A} shows the responses, using 24-hour average $M$th order polynomial in hourly average temperature, relative to a day with an average temperature of 26$^\circ$C. Candidates are up to the seventh-order polynomial. \textbf{Panel B} shows the responses, using a vector of 24-hour average temperatures with 3-5$^\circ$C temperature bins, relative to their corresponding omitted bins. The legend $[23,5)$, for example, indicates a 5$^\circ$C temperature interval and the omitted bin of $[23,28)^\circ$C. Note that all candidates of the omitted bins and temperature intervals were intentionally designed to cover Thailand's median temperature of (approximately) 26$^\circ$C. \textbf{Panel C} shows the responses, using 24-hour sum of degree hours, relative to different temperature cutoffs. For comparison and interpretation, the lower and upper bin edges of each omitted bin considered in the binned functional form are directly used as the temperature thresholds for heating and cooling degree days, respectively. The legend $[23,28]$, for example, indicates the 24-hour sum of heating degree hours below 23$^{\circ}$C and 24-hour sum of cooling degree hours above 28$^{\circ}$C, summed across the year.}
\label{fig:model_specs}
\end{figure}

Figure \ref{fig:model_specs} displays the resulting response functions of economic growth to temperature, using the same set of fixed effects and controls as described in general functional form Equation \ref{eqn:general_spec}. Panel A shows the effect of temperature on growth, as estimated using various polynomial orders (up to the 7th order). All polynomial orders exhibit similar patterns in the data, suggesting an inverted-U shape of the response function. I select a quadratic polynomial in hourly average temperature (i.e., $h(T) = \beta_1 T_{py} + \beta_2 T_{py}^2$) as the main specification because it is the most parsimonious specification. Panel B shows the results of the stepwise function Equation \ref{eqn:binned_function}, using a variety of combinations of omitted bins and temperature intervals. The results show a similar inverted-U pattern in the data as uncovered by the polynomial functional form. The binned regression results in a discontinuous response function, which makes it difficult to clearly select the optimal specification, especially in the higher temperature bins. Therefore, I apply the concept of cross validation which is commonly used in machine learning (ML) models \citep{hastie2009elements} to search for the optimal combination of omitted bin and temperature interval.

I applied cross-validation in two separate spaces: (i) time and (ii) location. Firstly, I split the sample into two different sample periods: 1982-2014 and 2015-2022. The model was then fitted using the sample period of 1982-2014. The resulting coefficients were used to predict the \textit{holdout} sample period of 2015-2022. The out-of-time (OOT) root-mean-squared error (RMSE) was calculated to measure the model performance. Next, I adapted the group $k$-fold method by splitting the sample period of 1982-2014 into $k$-fold where $k$ equals to 77 provinces in Thailand. Each province served as the holdout sample once, with the rest being used to fit the model. The out-of-sample (OOS) RMSE was then calculated for model evaluation. Table \ref{tab:cross_validation} reports the RMSE values of both the OOT and OOS examinations for each candidate combination of the lower bin edge and temperature interval. The results show that a combination of the lower temperature bin of 23$^{\circ}$C with 5$^{\circ}$C interval (i.e. the omitted bin of $[23,28)^\circ$C) outperform other candidate models in terms of predictive performance (that is, the model with lowest predictive error) in both OOT and OOS examinations. Based on these results, I selected $[23,28)^\circ$C as an omitted bin for the binned regressions in subsequent analyses.

\begin{table}[t!]
\centering
\begin{threeparttable}
\captionsetup{width=1\textwidth}
\caption{\textbf{Selection of optimal omitted bins.} This table report root-mean-squared error (RMSE) of OOT and OOS examinations for each candidate combination of lower bin edge and temperature interval.}
\label{tab:cross_validation}
%\resizebox{0.5\textwidth}{!}{%
%\begin{tabular}{@{}cccc@{}}
\begin{tabularx}{1\textwidth}{llXXX}
\toprule
Lower bin edge & Temperature interval & Omitted bin & RMSE - OOT & RMSE - OOS \\ \midrule
23$^{\circ}$C & 4$^{\circ}$C & [23,27)$^{\circ}$C & 8.38 & 9.18 \\
23$^{\circ}$C & 5$^{\circ}$C & [23,28)$^{\circ}$C & 8.32*** & 9.15*** \\
24$^{\circ}$C & 3$^{\circ}$C & [24,27)$^{\circ}$C & 8.84 & 9.90 \\
24$^{\circ}$C & 4$^{\circ}$C & [24,28)$^{\circ}$C & 8.64 & 9.79 \\
24$^{\circ}$C & 5$^{\circ}$C & [24,29)$^{\circ}$C & 8.78 & 10.00 \\
25$^{\circ}$C & 2$^{\circ}$C & [24,27)$^{\circ}$C & 9.79 & 10.23 \\
25$^{\circ}$C & 3$^{\circ}$C & [25,28)$^{\circ}$C & 8.87 & 9.83 \\
25$^{\circ}$C & 4$^{\circ}$C & [25,29)$^{\circ}$C & 9.34 & 10.12 \\
26$^{\circ}$C & 1$^{\circ}$C & [26,27)$^{\circ}$C & 9.62 & 10.35 \\
26$^{\circ}$C & 2$^{\circ}$C & [26,28)$^{\circ}$C & 8.95 & 9.99 \\
26$^{\circ}$C & 3$^{\circ}$C & [26,29)$^{\circ}$C & 9.25 & 9.98 \\ \bottomrule
\end{tabularx}%
%}
\begin{tablenotes}[flushleft]
\item \textit{Notes:} All estimates using baseline specification which include province FE and year FE and account for the effects of changes in precipitation. *** indicates the optimal candidate combination of lower temperature bin and temperature interval.
\end{tablenotes}
\end{threeparttable}
\end{table}

Panel C shows the growth-temperature response functions, which were estimated using the heating/cooling degree days functional form with different temperature cutoffs. All results clearly exhibit an inverted-U shape pattern in the data and are broadly consistent in magnitude, particularly for the cooling degree (i.e., higher temperatures). As suggested by \cite{carleton2022valuing}, binned regression is an important benchmark because it is closest to being non-parametric. Therefore, the lower and upper edges of the omitted temperature bins were directly used as the thresholds of the heating and cooling degree days, respectively, to allow the results to be comparable with those of the binned regression.

\section{Additional econometric results and robustness checks}
This section provides additional results of the main econometric estimation discussed in the main text and discusses additional robustness checks for the main results.

\subsection{Common growth-temperature response function across provinces}
Table \ref{tab:robustness_results_common} reports the robustness checks of the estimated temperature effects on the growth rate of GPP per capita across provinces. To be comparable with the main estimation results, the regression estimates of those models in Figure \ref{fig:baseline_model} are repeated in the first column of Table \ref{tab:robustness_results_common}. I first dropped precipitation (column 2) to investigate whether the changes in precipitation affect the temperature estimates. Columns 3-6 demonstrate models that use alternative set of controls, relative to the baseline model. In column 3 I added region-by-year FE, in column 4 I adapted specification of \cite{burke2015global} by adding quadratic province-specific time trends, in column 5 I adapted specification of \cite{dell2012temperature} by replacing year FE with region-by-year FE and poor-year FE, in column 6 I replaced year FE with quadratic time trends. Column 7 estimated the baseline model using a balanced sample where all provinces were present in the sample for the entire period. Column 8 estimated the baseline model plus one lag of the per capita growth rate to account for potential time-varying omitted variables \citep{burke2015global}. Still, three functional form assumptions were estimated for each robustness check. 

\begin{table}[t!]
\centering
\captionsetup{width=1\textwidth}
\caption{\textbf{Regression estimates for robustness checks, as estimated assuming identical response across provinces.} \textbf{Panel A} shows regression estimates, as estimated using 24-hour average second-order polynomial in hourly average temperature, summed across the year. \textbf{Panel B} shows regression estimates, as estimated using a vector of 24-hour average temperatures with 5$^\circ$C temperature bins, summed across the year, relative to the omitted $[23,28)^\circ$C bins. \textbf{Panel C} shows regression estimates, as estimated using 24-hour sum of heating degree hours below 23$^{\circ}$C and 24-hour sum of cooling degree hours above 28$^{\circ}$C, summed across the year.}
\label{tab:robustness_results_common}
\resizebox{\textwidth}{!}{
\begin{threeparttable}
\begin{tabular}{@{}lllllllll@{}}
\toprule\toprule
 & \multicolumn{1}{c}{\begin{tabular}[c]{@{}l@{}}Baseline\\ (1)\end{tabular}} & \multicolumn{1}{c}{\begin{tabular}[c]{@{}l@{}}NoPrecip.\\ (2)\end{tabular}} & \multicolumn{1}{c}{\begin{tabular}[c]{@{}l@{}}Region Yr.\\ (3)\end{tabular}} & \multicolumn{1}{c}{\begin{tabular}[c]{@{}l@{}}BHM (2015)\\ (4)\end{tabular}} & \multicolumn{1}{c}{\begin{tabular}[c]{@{}l@{}}DJO (2012)\\ (5)\end{tabular}} & \multicolumn{1}{c}{\begin{tabular}[c]{@{}l@{}}TimeTrends\\ (6)\end{tabular}} & \multicolumn{1}{c}{\begin{tabular}[c]{@{}l@{}}Balanced\\ (7)\end{tabular}} & \multicolumn{1}{c}{\begin{tabular}[c]{@{}l@{}}Lagged DV\\ (8)\end{tabular}} \\ \midrule
\multicolumn{9}{l}{} \\
\multicolumn{9}{l}{\hspace*{-2mm}\textbf{Panel A: Second-order polynomial}} \\
\multicolumn{9}{l}{} \\
Temp. & 0.0494*** & 0.0524*** & 0.0612*** & 0.0363** & 0.0590** & 0.0602*** & 0.0531*** & 0.0474*** \\
 & (0.0147) & (0.0152) & (0.0222) & (0.0168) & (0.0227) & (0.0137) & (0.0151) & (0.0144) \\
Temp. Sq. & -0.0009*** & -0.0010*** & -0.0012*** & -0.0007** & -0.0011*** & -0.0011*** & -0.0010*** & -0.0009*** \\
 & (0.0003) & (0.0003) & (0.0004) & (0.0003) & (0.0004) & (0.0002) & (0.0003) & (0.0003) \\
 &  &  &  &  &  &  &  &  \\
Observations & \multicolumn{1}{l}{3,086} & \multicolumn{1}{l}{3,086} & \multicolumn{1}{l}{3,086} & \multicolumn{1}{l}{3,086} & \multicolumn{1}{l}{3,086} & \multicolumn{1}{l}{3,086} & \multicolumn{1}{l}{2,952} & \multicolumn{1}{l}{2,933} \\
$R^2$ & \multicolumn{1}{l}{0.330} & \multicolumn{1}{l}{0.328} & \multicolumn{1}{l}{0.465} & \multicolumn{1}{l}{0.361} & \multicolumn{1}{l}{0.475} & \multicolumn{1}{l}{0.093} & \multicolumn{1}{l}{0.327} & \multicolumn{1}{l}{0.333} \\
Within $R^2$ & \multicolumn{1}{l}{0.008} & \multicolumn{1}{l}{0.005} & \multicolumn{1}{l}{0.004} & \multicolumn{1}{l}{0.054} & \multicolumn{1}{l}{0.004} & \multicolumn{1}{l}{0.077} & \multicolumn{1}{l}{0.009} & \multicolumn{1}{l}{0.009} \\ \midrule
\multicolumn{9}{l}{} \\
\multicolumn{9}{l}{\hspace*{-2mm}\textbf{Panel B: Temperature bins}} \\
\multicolumn{9}{l}{} \\
Days \textless 13$^\circ$C & 0.2219 & 0.2560* & -0.0408 & 0.3631** & -0.0404 & -0.5860** & 0.2283 & 0.2400* \\
 & (0.1383) & (0.1380) & (0.2189) & (0.1482) & (0.2043) & (0.2702) & (0.1430) & (0.1389) \\
Days 13-18$^\circ$C & -0.1361** & -0.1558*** & -0.1030 & -0.1022* & -0.1051 & -0.0634 & -0.1491** & -0.1433** \\
 & (0.0576) & (0.0569) & (0.0983) & (0.0611) & (0.0997) & (0.0601) & (0.0595) & (0.0599) \\
Days 18-23$^\circ$C & -0.0441* & -0.0486* & -0.0849*** & -0.0377 & -0.0819*** & -0.0720*** & -0.0502* & -0.0450* \\
 & (0.0262) & (0.0248) & (0.0229) & (0.0346) & (0.0262) & (0.0265) & (0.0263) & (0.0254) \\
Days 28-33$^\circ$C & -0.0406 & -0.0335 & -0.1046*** & 0.0008 & -0.1071*** & 0.0209 & -0.0444 & -0.0420 \\
 & (0.0325) & (0.0339) & (0.0334) & (0.0363) & (0.0329) & (0.0342) & (0.0323) & (0.0345) \\
Days 33-38$^\circ$C & -0.1008*** & -0.1010** & -0.0795* & -0.0708* & -0.0775* & -0.0999*** & -0.1063*** & -0.0902** \\
 & (0.0380) & (0.0396) & (0.0433) & (0.0399) & (0.0424) & (0.0257) & (0.0385) & (0.0394) \\
Days \textgreater 38$^\circ$C & -0.5567 & -0.5322 & -0.2215 & -0.3723 & -0.2628 & 0.0949 & -0.5608 & -0.6208* \\
 & (0.3344) & (0.3488) & (0.4024) & (0.3576) & (0.4410) & (0.3795) & (0.3390) & (0.3132) \\
 &  &  &  &  &  &  &  &  \\
Observations & \multicolumn{1}{l}{3,086} & \multicolumn{1}{l}{3,086} & \multicolumn{1}{l}{3,086} & \multicolumn{1}{l}{3,086} & \multicolumn{1}{l}{3,086} & \multicolumn{1}{l}{3,086} & \multicolumn{1}{l}{2,952} & \multicolumn{1}{l}{2,933} \\
$R^2$ & \multicolumn{1}{l}{0.332} & \multicolumn{1}{l}{0.329} & \multicolumn{1}{l}{0.467} & \multicolumn{1}{l}{0.363} & \multicolumn{1}{l}{0.477} & \multicolumn{1}{l}{0.094} & \multicolumn{1}{l}{0.329} & \multicolumn{1}{l}{0.334} \\
Within $R^2$ & \multicolumn{1}{l}{0.011} & \multicolumn{1}{l}{0.006} & \multicolumn{1}{l}{0.008} & \multicolumn{1}{l}{0.058} & \multicolumn{1}{l}{0.008} & \multicolumn{1}{l}{0.077} & \multicolumn{1}{l}{0.011} & \multicolumn{1}{l}{0.012} \\ \midrule
\multicolumn{9}{l}{} \\
\multicolumn{9}{l}{\hspace*{-2mm}\textbf{Panel C: Heating/Cooling degree days}} \\
\multicolumn{9}{l}{} \\
HDD (<23$^\circ$C) & -0.0116** & -0.0126** & -0.0146 & -0.0065 & -0.0146 & -0.0179*** & -0.0127** & -0.0123** \\
 & (0.0054) & (0.0055) & (0.0104) & (0.0061) & (0.0107) & (0.0052) & (0.0057) & (0.0055) \\
CDD (>28$^\circ$C) & -0.0104** & -0.0100** & -0.0144** & -0.0055 & -0.0141** & -0.0020 & -0.0110*** & -0.0086** \\
 & (0.0040) & (0.0041) & (0.0066) & (0.0044) & (0.0065) & (0.0024) & (0.0041) & (0.0043) \\
 &  &  &  &  &  &  &  &  \\
Observations & \multicolumn{1}{l}{3,086} & \multicolumn{1}{l}{3,086} & \multicolumn{1}{l}{3,086} & \multicolumn{1}{l}{3,086} & \multicolumn{1}{l}{3,086} & \multicolumn{1}{l}{3,086} & \multicolumn{1}{l}{2,952} & \multicolumn{1}{l}{2,933} \\
$R^2$ & \multicolumn{1}{l}{0.328} & \multicolumn{1}{l}{0.327} & \multicolumn{1}{l}{0.464} & \multicolumn{1}{l}{0.360} & \multicolumn{1}{l}{0.474} & \multicolumn{1}{l}{0.084} & \multicolumn{1}{l}{0.326} & \multicolumn{1}{l}{0.331} \\
Within $R^2$ & \multicolumn{1}{l}{0.005} & \multicolumn{1}{l}{0.004} & \multicolumn{1}{l}{0.003} & \multicolumn{1}{l}{0.052} & \multicolumn{1}{l}{0.003} & \multicolumn{1}{l}{0.068} & \multicolumn{1}{l}{0.006} & \multicolumn{1}{l}{0.006} \\ \bottomrule\bottomrule
\end{tabular}
\begin{tablenotes}[flushleft]
\item \textit{Notes:} Unless otherwise specified, all models included province FE and year FE and account for the effects of changes in precipitation (see text). Columns: (1) baseline specification, (2) as in column 1 but excluding precipitation, (3) as in column 1 and adding region-by-year fixed effects (FE), (4) as in column 1 and adapts specification of \cite{burke2015global} by adding quadratic province-specific time trends, (5) adapts specification of \cite{dell2012temperature} by dropping year FE and adding region-by-year FE and poor-year FE, (6) as in column 1 but dropping year FE and adding quadratic time trends, (7) as in column 1 but using a balanced sample where all provinces are present in the sample for the entire period, (8) as in column 1 but adding 1 lag of per capita growth rate. The standard errors in parentheses were clustered by province.\\
***, **, * indicate significance level of 1\%, 5\%, and 10\%, respectively.
\end{tablenotes}
\end{threeparttable}
}%
\end{table}

When not controlling for the changes in precipitation, the point estimates in column 2 shows that the temperature effects on growth are similar in magnitude to those in the baseline results. These results suggest that whether or not accounting for precipitation does not substantively affect the temperature estimates. Notably, the effects of precipitation are typically statistically insignificant and broadly consistent across a range of alternative model specifications for both the polynomial (Panel A) and the temperature bins (Panel B) functional forms. Nonetheless, the effects of precipitation in quadratic province-specific time trends (column 4), balanced sample (column 7), and lagged dependent variable (column 8) specifications are statistically significant and almost identical to the baseline results for the degree days functional form (Panel C).\footnote{The effects of annual precipitation on growth are not shown to conserve space. Estimates are available upon request.} 

The results in columns 3-6 of the polynomial functional form (Panel A) are broadly consistent in magnitude and statistical significance as I change the set of controls. In contrast, the temperature estimates and statistical significance in the binned functional form (columns 3-6 of Panel B) appear to be sensitive to alternative sets of controls, suggesting that they should be interpreted with caution. These results are not unexpected since the binned model is the method that is demanding of the data and columns 3-6 all have more control variables, compared to baseline specification. The results in columns 3-6 of the degree days functional form (Panel C) are somewhat mixed. The HDD estimates in columns 3, 5, and 6 are comparable in magnitude to baseline specification. However, for CDD estimates, only results in column 3 and 5 are similar in magnitude to baseline specification. The results in column 4, which is demanding of the data as the model incorporates province-specific time trends as well as province FE and year FE, appear to be statistically insignificant and smaller in magnitude than the baseline results in both HDD and CDD estimates. The results in columns 8 of all three functional forms are close in both magnitude and standard errors to those in the baseline results, suggesting that the results are not affected by not controlling for time-varying omitted variable. Column 7 shows how the temperature estimates change under alternative samples. Both point estimates and standard errors of all three functional forms are close to those in the baseline specification.

\subsection{Robustness of estimates in the growth-temperature response function to alternative formulation} \label{si:alternative_formulation}

\begin{table}[t!]
\small
\centering
\begin{threeparttable}
\captionsetup{width=1\textwidth}
\caption{\textbf{Alternative formulation of the average growth-temperature response function}. \textbf{Panel A} shows regression estimates of Equation \ref{eqn:alternative_formulation}. \textbf{Panel B} reports marginal effects of temperature on growth evaluated at various average temperatures. Given that the effect evaluated at country average income, the marginal effect on growth at some average temperature $\bar{T}^*$ is then $\beta_1 + \beta_2 \cdot \bar{T}^*$.}
\label{tab:alternative_formulation}
%\resizebox{\textwidth}{!}{%
\begin{tabularx}{1\textwidth}{lXXX}
%\begin{tabular}{@{}llll@{}}
\toprule\toprule
 & (1) & (2) & (3) \\ \midrule
%  &  &  &  \\
\textbf{Panel A: Coefficients} &  &  &  \\
% &  &  &  \\
$T_{py}$ & 0.0564*** & 0.0307 & 0.0350 \\
         & (0.0193) & (0.0211) & (0.0226) \\
$T_{py}\cdot \Bar{T}_p$ & -0.0022*** & -0.0013 & -0.0014* \\
                        & (0.0007) & (0.0008) & (0.0008) \\
$T_{py}\cdot \Bar{Y}_{p}/10^5$ &  & -0.0038** &  \\
                            &  & (0.0015) &  \\
$T_{py}\cdot ln(\Bar{Y_p})$ &  &  & -0.0045* \\
                        &  &  & (0.0023) \\ \midrule
% &  &  &  \\
\textbf{Panel B: Marginal effects} &  &  &  \\
% &  &  &  \\
15$^{\circ}$C & 0.0236 & 0.0117 & 0.0136 \\
              & (0.0186) & (0.0203) & (0.0217) \\
20$^{\circ}$C & 0.0126 & 0.0054 & 0.0065 \\
              & (0.0178) & (0.0195) & (0.0209) \\
30$^{\circ}$C & -0.0093 & -0.0073 & -0.0077 \\
              & (0.0171) & (0.0187) & (0.0201) \\
35$^{\circ}$C & -0.0202 & -0.0136 & -0.0148 \\
              & (0.0164) & (0.018) & (0.0193) \\
40$^{\circ}$C & -0.0312** & -0.0200 & -0.0219 \\
              & (0.0156) & (0.0172) & (0.0184) \\
 &  &  &  \\
Observations & 3,086 & 3,086 & 3,086 \\
$R^2$ & 0.328 & 0.332 & 0.331 \\
Within $R^2$ & 0.005 & 0.010 & 0.009 \\ \bottomrule\bottomrule
\end{tabularx}%
%}
\begin{tablenotes}[flushleft]
\item \textit{Notes:} Columns: (1) include temperature, temperature interacted with province average temperature, precipitation, precipitation interacted with average precipitation, province FE, and year FE, (2) as in column (1) and adds temperature and precipitation which both interacted with province average GPP per capita over the sample period ($\bar{Y}_p$) (divided by $10^5$ to coefficients are visible), (3) as in column (1) and adds temperature and precipitation which both interacted with province average GPP per capita over the sample period in logs ($ln(\bar{Y}_p)$). Both $\bar{Y}_p$ and $ln(\bar{Y}_p)$ variables are de-meaned so that the temperature terms can be interpreted at the country’s average income. The standard errors in parentheses were clustered by province.\\
***, **, * indicate significance level of 1\%, 5\%, and 10\%, respectively.
\end{tablenotes}
\end{threeparttable}
\end{table}

To understand the nonlinear response observed in Figure \ref{fig:baseline_model}, I investigated an alternative formulation of the growth-temperature response function $h(T)$. Following \cite{burke2015global}, I substituted temperature interacted with average temperature and temperature interacted with average GPP per capita for the quadratic temperature term in Equation \ref{eqn:polynomial_function}. Precipitation terms similarly interacted with the province’s average temperature and GPP per capita. Specifically, I estimate
\begin{equation} \label{eqn:alternative_formulation}
    g_{py} = \beta_1 T_{py} + \beta_2 (T_{py} \cdot \bar{T}_{p}) + \beta_3 (T_{py} \cdot \bar{Y}_{p}) + \rho_1 R_{py} + \rho_2 (R_{py} \cdot \bar{R}_{p}) + \rho_3 (R_{py} \cdot \bar{Y}_{p}) + \alpha_p + \alpha_y + \epsilon_{py}
\end{equation}
Panel A of Table \ref{tab:alternative_formulation} reports the estimation results of Equation \ref{eqn:alternative_formulation}. In the absence of the interaction with province average income, column 1 shows strong evidence of  a nonlinear and concave temperature response similar to the baseline results, with $\beta_1 > 0$ and $\beta_2 < 0$ and each is statistically significant. The results in columns 2 and 3 show the effects of income differences. As suggested by \cite{burke2015global}, if the nonlinear response is driven by differences in income rather than differences in temperature, then the inclusion of the temperature-income interaction should result in $\beta_2 = 0$, $\beta_1 < 0$, and $\beta_3 > 0$. The results in columns 2-3 remain consistent in sign (that is, $\beta_1 > 0$ and $\beta_2 < 0$), but appear to be statistically insignificant and smaller in magnitude than the results of column 1. In contrast to \cite{burke2015global}, the estimated magnitudes of the temperature-income interactions (i.e., $\beta_3$) are large and statistically significant. Panel B of Table \ref{tab:alternative_formulation} reports the estimated marginal effects of temperature on growth at various provincial average temperatures, as estimated from regressions with (columns 2-3) and without (column 1) the inclusion of the temperature-income interaction. Although the results of all three columns analogously indicate a nonlinear temperature response (that is, positive marginal effects at low temperatures and negative marginal effects at higher temperatures), the marginal effects when the temperature-income interaction is included (columns 2-3) are smaller in magnitude, suggesting that growth-temperature responses are driven by differences in average temperature and affected by differences in income.

\begin{table}
\centering
\captionsetup{width=1\textwidth}
\caption{\textbf{Regression estimates for robustness checks, as estimated allowing high- and low-income provinces to respond differently to temperature changes.} \textbf{Panel A} shows regression estimates, as estimated using 24-hour average second-order polynomial in hourly average temperature, summed across the year. \textbf{Panel B} shows regression estimates, as estimated using a vector of 24-hour average temperatures with 5$^\circ$C temperature bins, summed across the year, relative to the omitted $[23,28)^\circ$C bins. \textbf{Panel C} shows regression estimated, as estimated using 24-hour sum of heating degree hours below 23$^{\circ}$C and 24-hour sum of cooling degree hours above 28$^{\circ}$C, summed across the year.}
\label{tab:robustness_heterogenous_response}
\resizebox{\textwidth}{!}{%
\begin{threeparttable}
\begin{tabular}{@{}lllllllll@{}}
\toprule \toprule
\multicolumn{1}{c}{} & \multicolumn{1}{c}{\begin{tabular}[c]{@{}l@{}}Baseline\\ (1)\end{tabular}} & \multicolumn{1}{c}{\begin{tabular}[c]{@{}l@{}}NoPrecip.\\ (2)\end{tabular}} & \multicolumn{1}{c}{\begin{tabular}[c]{@{}l@{}}Region Yr.\\ (3)\end{tabular}} & \multicolumn{1}{c}{\begin{tabular}[c]{@{}l@{}}BHM (2015)\\ (4)\end{tabular}} & \multicolumn{1}{c}{\begin{tabular}[c]{@{}l@{}}DJO (2012)\\ (5)\end{tabular}} & \multicolumn{1}{c}{\begin{tabular}[c]{@{}l@{}}TimeTrends\\ (6)\end{tabular}} & \multicolumn{1}{c}{\begin{tabular}[c]{@{}l@{}}Balanced\\ (7)\end{tabular}} & \multicolumn{1}{c}{\begin{tabular}[c]{@{}l@{}}Lagged DV\\ (8)\end{tabular}} \\ \midrule
\multicolumn{9}{l}{} \\
\multicolumn{9}{l}{\hspace*{-2mm}\textbf{Panel A: Second-order polynomial}} \\
\multicolumn{9}{l}{} \\
Temp. x Low-income & 0.0503*** & 0.0556*** & 0.0618*** & 0.0383** & 0.0444* & 0.0527*** & 0.0538*** & 0.0461*** \\
 & (0.0171) & (0.0170) & (0.0232) & (0.0186) & (0.0225) & (0.0138) & (0.0176) & (0.0166) \\
Temp. x High-income & 0.0435 & 0.0443 & 0.0449 & 0.0565 & 0.0817* & 0.0798** & 0.0425 & 0.0408 \\
 & (0.0313) & (0.0323) & (0.0408) & (0.0381) & (0.0489) & (0.0312) & (0.0316) & (0.0313) \\
Temp.Sq. x Low-income & -0.0009*** & -0.0010*** & -0.0012*** & -0.0007** & -0.0008** & -0.0009*** & -0.0010*** & -0.0009*** \\
 & (0.0003) & (0.0003) & (0.0004) & (0.0003) & (0.0004) & (0.0003) & (0.0003) & (0.0003) \\
Temp.Sq. x High-income & -0.0008 & -0.0008 & -0.0009 & -0.0010 & -0.0016* & -0.0014** & -0.0008 & -0.0008 \\
 & (0.0006) & (0.0006) & (0.0007) & (0.0007) & (0.0008) & (0.0006) & (0.0006) & (0.0006) \\
 &  &  &  &  &  &  &  &  \\
Observations & \multicolumn{1}{l}{3,086} & \multicolumn{1}{l}{3,086} & \multicolumn{1}{l}{3,086} & \multicolumn{1}{l}{3,086} & \multicolumn{1}{l}{3,086} & \multicolumn{1}{l}{3,086} & \multicolumn{1}{l}{2,952} & \multicolumn{1}{l}{2,933} \\
$R^2$ & \multicolumn{1}{l}{0.331} & \multicolumn{1}{l}{0.328} & \multicolumn{1}{l}{0.465} & \multicolumn{1}{l}{0.362} & \multicolumn{1}{l}{0.475} & \multicolumn{1}{l}{0.097} & \multicolumn{1}{l}{0.328} & \multicolumn{1}{l}{0.333} \\
Within $R^2$ & \multicolumn{1}{l}{0.009} & \multicolumn{1}{l}{0.005} & \multicolumn{1}{l}{0.005} & \multicolumn{1}{l}{0.055} & \multicolumn{1}{l}{0.005} & \multicolumn{1}{l}{0.081} & \multicolumn{1}{l}{0.010} & \multicolumn{1}{l}{0.011} \\ \midrule
\multicolumn{9}{l}{} \\
\multicolumn{9}{l}{\hspace*{-2mm}\textbf{Panel B: Temperature bins}} \\
\multicolumn{9}{l}{} \\
Days \textless 13$^{\circ}$C x Low-income & 0.3343** & 0.3643*** & 0.0051 & 0.4505*** & 0.0037 & -0.5748* & 0.3626** & 0.3684*** \\
 & (0.1363) & (0.1379) & (0.1852) & (0.1511) & (0.1858) & (0.2990) & (0.1407) & (0.1344) \\
Days \textless 13$^{\circ}$C x High-income & 0.0772 & 0.1950 & -0.1480 & 0.0498 & -0.3102 & -0.6927 & 0.0548 & 0.0255 \\
 & (0.3014) & (0.2860) & (0.5039) & (0.2720) & (0.5307) & (0.6907) & (0.3070) & (0.3098) \\
Days 13-18$^{\circ}$C x Low-income & -0.1425*** & -0.1555*** & -0.0840 & -0.1478** & -0.0772 & -0.0515 & -0.1591*** & -0.1478*** \\
 & (0.0529) & (0.0553) & (0.0941) & (0.0609) & (0.0936) & (0.0429) & (0.0533) & (0.0547) \\
Days 13-18$^{\circ}$C x High-income & -0.0227 & -0.0480 & 0.0616 & -0.0398 & 0.0450 & -0.1625 & -0.0115 & 0.0174 \\
 & (0.2097) & (0.2127) & (0.2584) & (0.2280) & (0.2746) & (0.2809) & (0.2086) & (0.2191) \\
Days 33-38$^{\circ}$C x Low-income & -0.1535*** & -0.1622*** & -0.1083* & -0.1123*** & -0.0632 & -0.0661* & -0.1590*** & -0.1314*** \\
 & (0.0404) & (0.0408) & (0.0562) & (0.0416) & (0.0570) & (0.0380) & (0.0415) & (0.0416) \\
Days 33-38$^{\circ}$C x High-income & -0.0511 & -0.0574 & -0.0723 & -0.0270 & -0.1262* & -0.0935** & -0.0557 & -0.0517 \\
 & (0.0486) & (0.0489) & (0.0517) & (0.0514) & (0.0643) & (0.0396) & (0.0489) & (0.0502) \\
Days \textgreater 38$^{\circ}$C x Low-income & -0.0882 & -0.0184 & 0.0990 & 0.0099 & 0.1941 & 0.3918 & -0.0890 & -0.1596 \\
 & (0.3451) & (0.3397) & (0.4848) & (0.4475) & (0.5121) & (0.4760) & (0.3510) & (0.3193) \\
Days \textgreater 38$^{\circ}$C x High-income & -1.2485*** & -1.2156*** & -0.7429** & -0.9016** & -1.0063** & -0.2697 & -1.2671*** & -1.3298*** \\
 & (0.3786) & (0.3895) & (0.3691) & (0.3542) & (0.4694) & (0.3620) & (0.3762) & (0.3810) \\
 &  &  &  &  &  &  &  &  \\
Observations & \multicolumn{1}{l}{3,086} & \multicolumn{1}{l}{3,086} & \multicolumn{1}{l}{3,086} & \multicolumn{1}{l}{3,086} & \multicolumn{1}{l}{3,086} & \multicolumn{1}{l}{3,086} & \multicolumn{1}{l}{2,952} & \multicolumn{1}{l}{2,933} \\
$R^2$ & \multicolumn{1}{l}{0.336} & \multicolumn{1}{l}{0.330} & \multicolumn{1}{l}{0.469} & \multicolumn{1}{l}{0.367} & \multicolumn{1}{l}{0.480} & \multicolumn{1}{l}{0.101} & \multicolumn{1}{l}{0.333} & \multicolumn{1}{l}{0.339} \\
Within $R^2$ & \multicolumn{1}{l}{0.016} & \multicolumn{1}{l}{0.008} & \multicolumn{1}{l}{0.012} & \multicolumn{1}{l}{0.062} & \multicolumn{1}{l}{0.013} & \multicolumn{1}{l}{0.085} & \multicolumn{1}{l}{0.017} & \multicolumn{1}{l}{0.018} \\ \midrule
\multicolumn{9}{l}{} \\
\multicolumn{9}{l}{\hspace*{-2mm}\textbf{Panel C: Heating/Cooling degree days}} \\
\multicolumn{9}{l}{} \\
HDD (\textless 23$^{\circ}$C) x Low-income & -0.0122** & -0.0133** & -0.0155 & -0.0070 & -0.0131 & -0.0161*** & -0.0133** & -0.0126** \\
 & (0.0054) & (0.0054) & (0.0098) & (0.0062) & (0.0095) & (0.0041) & (0.0056) & (0.0054) \\
HDD (\textless 23$^{\circ}$C) x High-income & -0.0057 & -0.0070 & -0.0038 & -0.0122 & -0.0147 & -0.0300 & -0.0050 & -0.0038 \\
 & (0.0181) & (0.0183) & (0.0225) & (0.0234) & (0.0259) & (0.0215) & (0.0181) & (0.0183) \\
CDD (\textgreater 28$^{\circ}$C) x Low-income & -0.0096** & -0.0104** & -0.0122* & -0.0057 & -0.0081 & 0.0005 & -0.0102** & -0.0073 \\
 & (0.0044) & (0.0044) & (0.0065) & (0.0047) & (0.0059) & (0.0028) & (0.0044) & (0.0045) \\
CDD (\textgreater 28$^{\circ}$C) x High-income & -0.0104** & -0.0091* & -0.0177** & -0.0054 & -0.0224** & -0.0031 & -0.0109** & -0.0091* \\
 & (0.0051) & (0.0050) & (0.0080) & (0.0056) & (0.0109) & (0.0043) & (0.0051) & (0.0054) \\
 &  &  &  &  &  &  &  &  \\
Observations & \multicolumn{1}{l}{3,086} & \multicolumn{1}{l}{3,086} & \multicolumn{1}{l}{3,086} & \multicolumn{1}{l}{3,086} & \multicolumn{1}{l}{3,086} & \multicolumn{1}{l}{3,086} & \multicolumn{1}{l}{2,952} & \multicolumn{1}{l}{2,933} \\
$R^2$ & \multicolumn{1}{l}{0.329} & \multicolumn{1}{l}{0.327} & \multicolumn{1}{l}{0.465} & \multicolumn{1}{l}{0.360} & \multicolumn{1}{l}{0.475} & \multicolumn{1}{l}{0.089} & \multicolumn{1}{l}{0.326} & \multicolumn{1}{l}{0.332} \\
Within $R^2$ & \multicolumn{1}{l}{0.007} & \multicolumn{1}{l}{0.004} & \multicolumn{1}{l}{0.004} & \multicolumn{1}{l}{0.052} & \multicolumn{1}{l}{0.004} & \multicolumn{1}{l}{0.072} & \multicolumn{1}{l}{0.007} & \multicolumn{1}{l}{0.008} \\ \bottomrule \bottomrule
\end{tabular}%
\begin{tablenotes}[flushleft]
\item \textit{Notes:} Unless otherwise specified, all models include province FE and year FE and account for the effects of changes in precipitation (see text). Columns: (1) baseline specification, (2) as in column 1 but excluding precipitation, (3) as in column 1 but adding region-by-year FE, (4) as in column 1 and adapts specification of \cite{burke2015global} by adding quadratic province-specific time trends, (5) as in column 1 and adapts specification of \cite{dell2012temperature} by dropping year FE and adding region-by-year FE and poor-year FE, (6) as in column 1 but dropping year FE and adding quadratic time trends, (7) as in column 1 but using a balanced sample where all provinces are present in the sample for the entire period, and (8) as in column 1 but adding 1 lag of per capita growth. A low-income dummy is defined as a province with below-median average inflation-adjusted GPP per capita across the sample period. Precipitation × Low-income and Precipitation × High-income coefficients are suppressed to save space. The standard errors in parentheses are clustered by province.\\
***, **, * indicate significance level of 1\%, 5\%, and 10\%, respectively.
\end{tablenotes}
\end{threeparttable}
}
\end{table}

\subsection{Heterogeneity in the growth-temperature response function}
Table \ref{tab:robustness_heterogenous_response} reports the results of estimating the baseline specification (column 1) and a variety of robustness checks (columns 2-8) for differential growth-temperature response functions. All alternative specification checks are analogous to those in Table \ref{tab:robustness_results_common} (see Section \ref{sec:pooled_model} for details). Regardless of statistical significance, the estimated parameters in the baseline specification (column 1) of all three alternative functional forms exhibit a nonlinear and concave structure of the growth-temperature response function $h(T)$ in both low- and high-income provinces. No interaction terms in the baseline specification of the polynomial (Panel A) and the degree days (Panel C) functional forms are statistically significant, but the interaction terms of the binned functional form (Panel B) are only statistically significant in the two highest temperature bins ($[33,38)^{\circ}$C and $>38^{\circ}$C bins). These results suggest that we cannot reject the hypothesis that high- and low-income provinces respond identically to changes in temperature, at least for the polynomial and degree days functional forms.

Analogous to the pooled growth-temperature models, the results in column 2 suggest that accounting for precipitation does not substantively affect the temperature estimates. The results in columns 3-6 of the polynomial functional form (Panel A) are broadly comparable in magnitude to those in the baseline specification. Including province, region-by-year, and low-income province-year fixed effects only (column 5) shows substantial effects in high-income provinces, with a substantial change from the baseline specification. In addition, when I include province fixed effects and use common country time trends instead of the year fixed effects, the estimates for high-income provinces also substantially change. As for the binned functional form (columns 3-6 of Panel B), the temperature estimates as well as the statistical significance for both lower income and higher income provinces appear to be sensitive to alternative sets of controls, as expected. The results in columns 3-6 of the degree days functional form (Panel C) are somewhat mixed. The HDD estimates in columns 3, 5, and 6 are comparable in magnitude to the baseline specification for low-income provinces and only in columns 3 and 6 for high-income provinces. For CDD estimates, only the results in columns 3 and 5 are similar in magnitude to the baseline specification for low-income provinces; estimates are substantially different in magnitude for high-income provinces. The CDD estimates in columns 4 and 6 for both low- and high-income provinces appear to be statistically insignificant and substantially different in magnitude from the baseline results. When all provinces are present in the entire period (column 7), the results in all three functional forms are similar in both magnitude and statistical significance to those in the baseline specification. The results in all three functional forms are broadly similar to those in the baseline specification, whether or not all provinces are present in the entire period (column 7) and whether the lagged dependent variable is included to account for potential time-varying omitted variables (column 8).

\subsection{Heterogeneity in growth-temperature response functions on the components of GPP}
\begin{figure}[b!]
\centering
\captionsetup{width=1\textwidth}
\includegraphics[scale=.145]{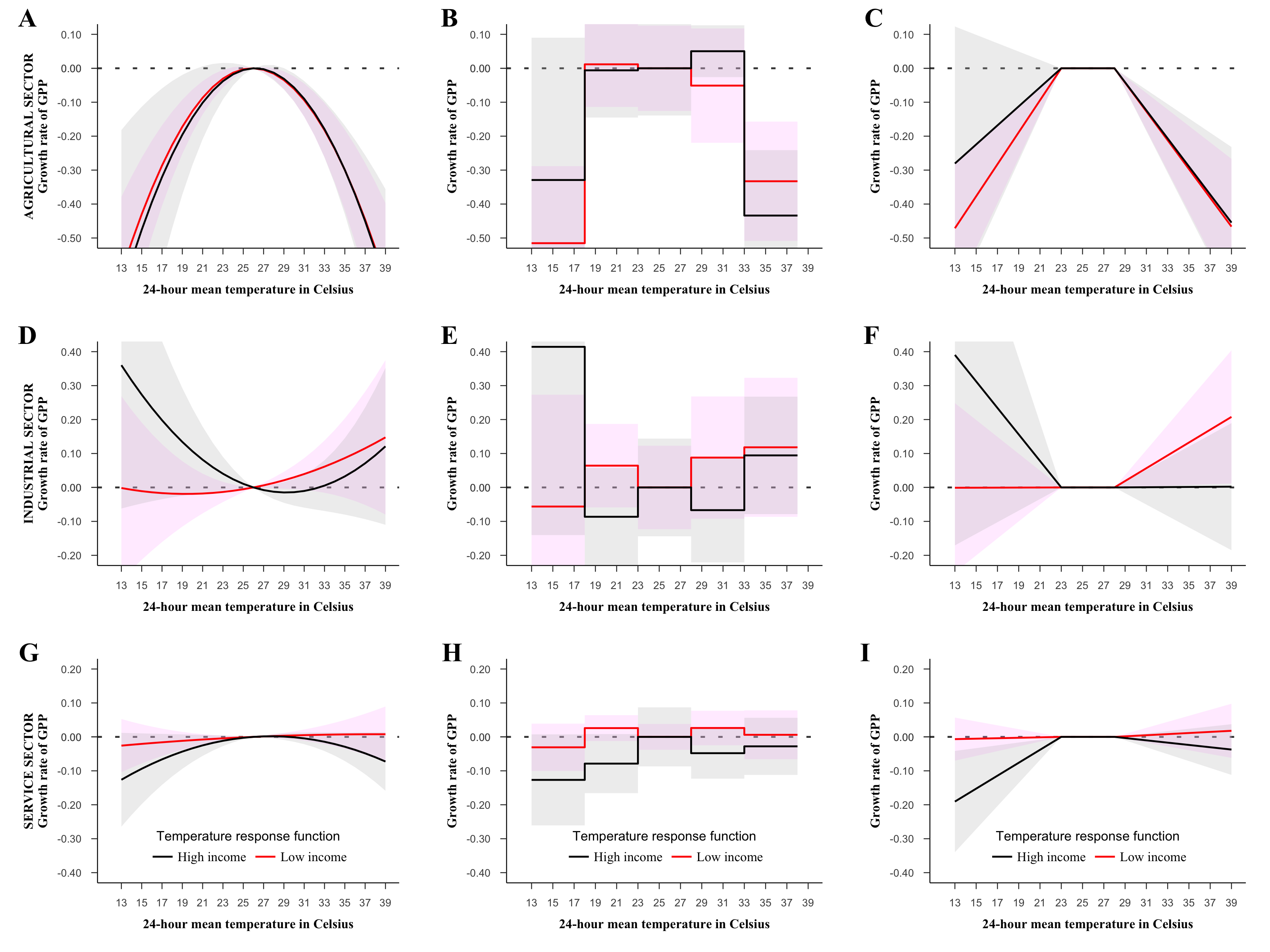}
\caption{\textbf{Heterogeneous temperature responses on the components of GPP}. The graphs show heterogeneity in the growth-temperature response functions of the GPP components, as estimated using three different functional forms. The upper row (Panels A, B, and C) presents the estimates of growth in agricultural output, the middle row (Panels D, E, and F) presents the estimates of growth in industrial output, and the lower row (Panels G, H, and I) presents the estimates of growth in service output. The black line shows the temperature effect in high-income provinces, and the gray areas represent the 95\% confidence interval. The red line shows the temperature effects of low-income provinces, and the pink areas represent the 95\% confidence interval. \textit{Panel A} shows the response, using 24-hour average second-order polynomial in hourly average temperature, relative to a day with an average temperature of 26$^\circ$C. \textit{Panel B} shows the response, using a vector of 24-hour average temperatures with 5$^\circ C$ temperature bins, relative to the omitted $[23,28)^\circ C$ temperature bin. \textit{Panel C} shows the response, using 24-hour sums of heating degree hours below 23$^{\circ}C$ and 24-hour sums of cooling degree hours above 28$^{\circ}C$, summed across the year. All response functions are only shown for 24-hour average temperatures that are usually experienced across Thailand over the sample period as illustrated in Figure \ref{fig:summary_weather}. All response functions were estimated with province-specific and year-specific fixed effects. See Section \ref{sec:econ_temp_relationship} in main text for details of each functional form.}
\label{fig:response_xPoor_sector}
\end{figure}

Figure \ref{fig:response_xPoor_sector} displays the heterogeneity in the growth-temperature response functions in agricultural output (Panels A, B, and C), industrial output (Panels D, E, and F), and service output (Panels G, H, and I), as estimated using the polynomial (left column), temperature bins (middle column), and degree days (right column) functional forms. The temperature effects were estimated using a regression that allows high- and low-income provinces to respond differently to changes in temperature.

Results in Panels A, B, and C consistently show the inverted-U shape in the pattern of temperature effects on the growth in agricultural output in both high- and low-income provinces. The temperature effects in all three functional forms are statistically significant at the hot end temperature distribution in both high- and low-income provinces, suggesting that we can reject the hypothesis that the growth-temperature responses in agricultural real value added are zero at all points in the temperature distribution for both high- and low-income provinces. Panels D, E, and F shows the temperature effects on the growth in industrial output. The pattern of temperature effects on the growth in industrial real value added in both high- and low-income provinces is somewhat convex in all three functional forms, although the estimates are not statistically significant, suggesting that we cannot reject the hypothesis of no temperature effects on the growth in industrial output in either high- or low-income provinces. The results for the service sector are shown in Panels G, H, and I. The temperature effects on the growth in service output are also inverted-U shape in high-income provinces and flatter in low-income provinces. The results are broadly consistent across the three functional forms. These effects are not statistically significant at the conventional confidence level, suggesting that we cannot reject the hypothesis of no temperature effects on the growth in service output in either high- or low-income provinces.

\section{Projections of future damages: Additional results and sensitivity}

\subsection{Sensitivity of projected impacts on GRP}
\begin{figure}[H]
\centering
\captionsetup{width=1\textwidth}
\includegraphics[scale=.0925]{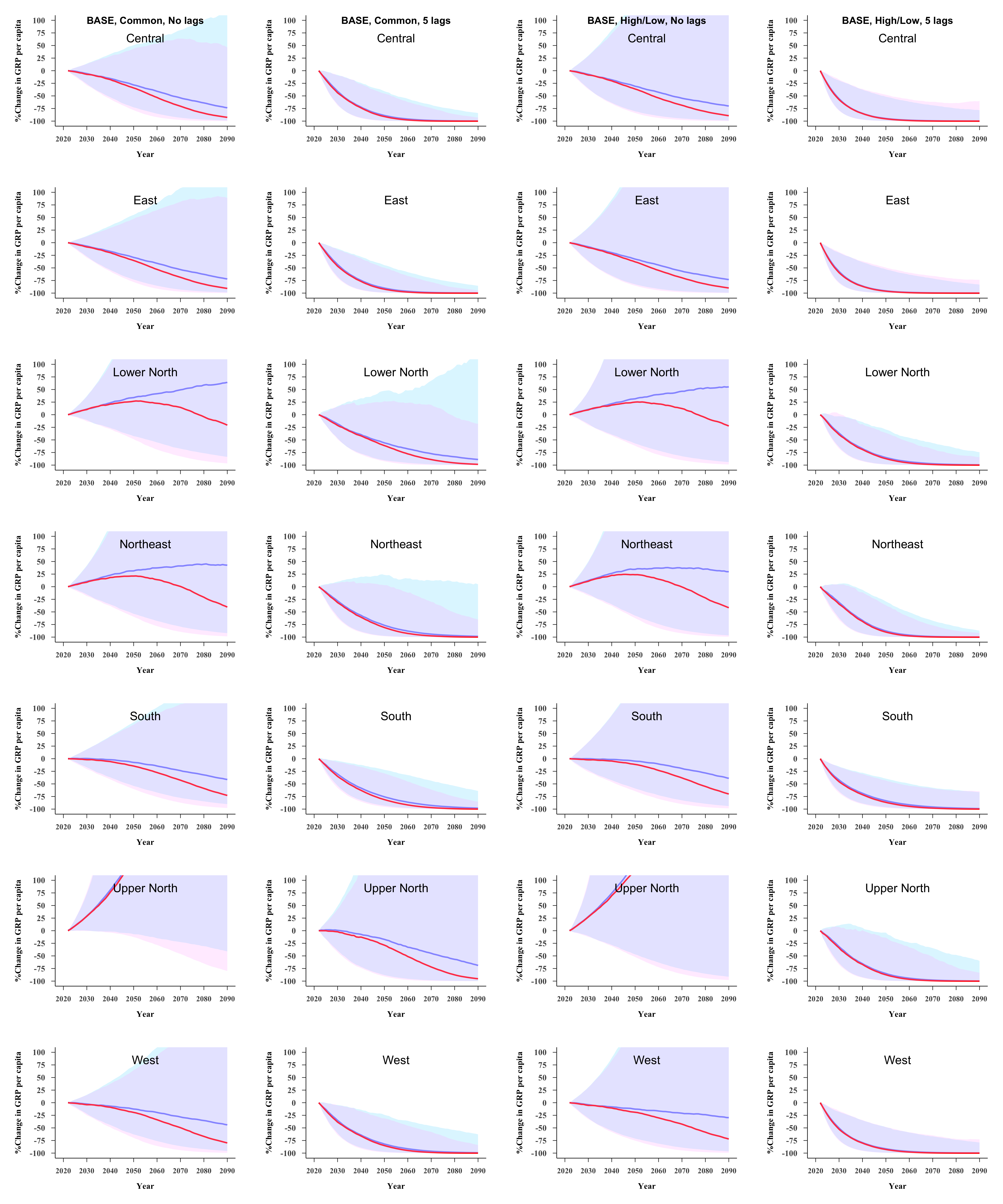}
\caption{\textbf{Projected impact of climate change "\textit{without bias-correction}" on regional output per capita of 7 regions between 2023-2090, relative to their GRP per capita absent climate change}. The graphs show the projected impact of climate change with baseline output growth assumption under RCP8.5 (pink shaded area) and RCP4.5 (light blue shaded area) emission scenarios for four historical growth-temperature response functions. The red lines are median projection under RCP8.5, while blue lines are under RCP4.5. The pink shaded area is 95\% confidence interval under RCP8.5, while light blue shaded area is under RCP4.5. Columns: (1) a common response function across high- and low-income provinces with no lags, (2) a common response function across high- and low-income with 5 lags, (3) the differentiated response functions between high- and low-income provinces with no lags, (4) the differentiated response functions between high- and low-income provinces with 5 lags.}
\label{fig:projected_impact_GRP_allRCPs_allSpecs_allGrowths}
\end{figure}

To examine whether the region-level projections are sensitive to different combinations of emission scenarios, specifications, and output growth assumptions, I projected the impacts of climate change on regional output per capita under both RCP4.5 and RCP8.5 emission scenarios between 2023-2090 using regression estimates from second-order polynomial Equation \ref{eqn:polynomial_function} with a baseline output growth assumption for each of the four historical growth-temperature response functions. Figure \ref{fig:projected_impact_GRP_allRCPs_allSpecs_allGrowths} displays the projected impact of climate change under both RCP8.5 (pink shaded area) and RCP4.5 (light blue shaded area) emission scenarios. The red lines represent the median projection under RCP8.5, and the blue lines represent that under RCP4.5. The pink shaded area is the 95\% confidence interval under RCP8.5, while the light blue shaded area is under RCP4.5.

Median estimates under both RCP4.5 and RCP8.5 are uniformly negative in models with delayed impacts, whether high- and low-income provinces are assumed to respond identically (column 2 in Figure \ref{fig:projected_impact_GRP_allRCPs_allSpecs_allGrowths}) or differently (column 4 in Figure \ref{fig:projected_impact_GRP_allRCPs_allSpecs_allGrowths}). Nonetheless, projections in the "Common, 5 lags" models under RCP4.5 (column 2 in Figure \ref{fig:projected_impact_GRP_allRCPs_allSpecs_allGrowths}) are more uncertain than those under RCP8.5 for all three output growth assumptions in the lower north, northeast, and upper north regions. This is because emission stabilization scenario RCP 4.5, which results in a lower temperature rise than that of RCP8.5, makes net economic impacts less negative for cold provinces, especially those in the Upper North region. The same explanation can also be applied to the more positive median impacts observed in the lower north and northeast regions in models under the RCP4.5 emission scenario that do not account for lagged effects (columns 1 and 3 in Figure \ref{fig:projected_impact_GRP_allRCPs_allSpecs_allGrowths}).

Projections are broadly similar in structure in both emission scenarios. Projected impacts in models with delayed impacts are less uncertain whether high- and low-income provinces respond identically (column 2) or differently (column 4) to temperature changes. The estimates in Tables \ref{tab:growth_level_effects_pooled} and \ref{tab:growth_level_effects_xPoor} provide the reason for this: as the effects of temperature are assumed to be cumulative and affect growth in a given year, cold provinces could initially benefit on net, and hotter provinces remain worse off with additional warming. However, as a given province's average temperature becomes warmer, the impacts of future warming climate worsens in each future which makes net economic impacts become negative for provinces that are initially cold. While in models that do not account for lagged effects, the projections become more uncertain whether high- and low-income provinces are assumed to respond identically (column 1) or differently (column 3). This is because the estimated growth-temperature response functions are substantially flatter than the response function with lagged effects. Cold provinces increasingly benefits from increased average temperature, while hotter provinces remain worse off under warming climate.

\subsection{Sensitivity of projected impacts on GDP}
\begin{figure}[t!]
\centering
\captionsetup{width=1\textwidth}
\includegraphics[scale=.145]{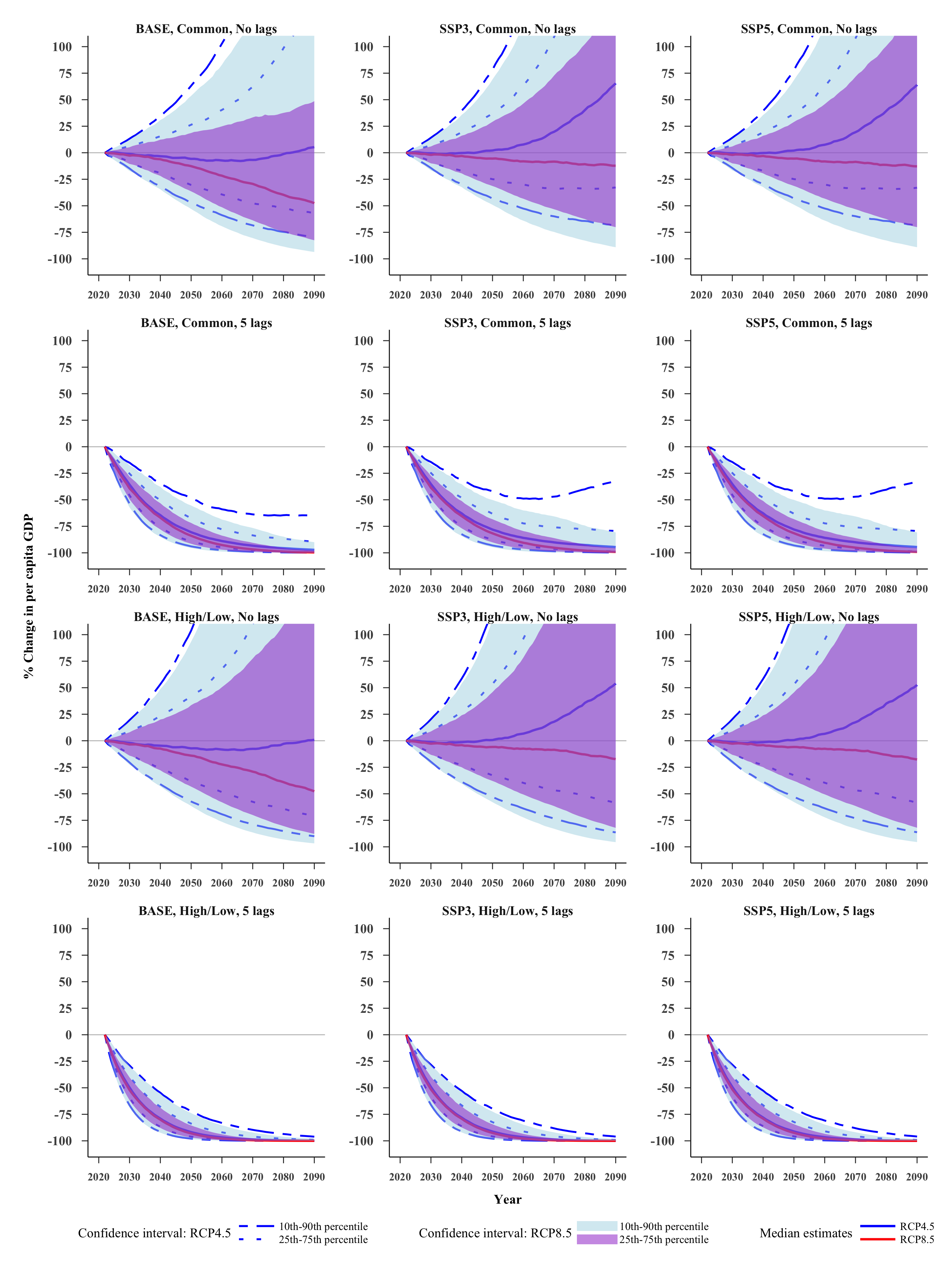}
\caption{\textbf{Projected impacts of climate change "\textit{without bias-correction}" on Thailand's GDP per capita, relative to GDP per capita absent climate change}. The graphs show the projected impacts of climate change under both RCP4.5 and RCP8.5 emission scenarios between 2023-2090 for each combination of four historical growth-temperature response functions and three output growth assumptions. Rows: (1) a common response function across high- and low-income provinces with no lags, (2) a common response function across high- and low-income with five lags, (3) the differentiated response functions between high- and low-income provinces with no lags, and (4) the differentiated response functions between high- and low-income provinces with five lags. Columns: (1) baseline scenario, (2) SSP3, and (3) SSP5.}
\label{fig:projected_impact_onGDP_bothRCPs}
\end{figure}

To explore the sensitivity of the projections in Figure \ref{fig:projected_impact_onGDP_rcp85_density} to different emission scenarios, I additionally projected the impacts of climate change on Thailand's GPD per capita under an intermediate scenario, RCP4.5. Figure \ref{fig:projected_impact_onGDP_bothRCPs} displays projections under both RCP4.5 and RCP8.5 emission scenarios between 2023-2090. The projections used estimates from the second-order polynomial Equation \ref{eqn:polynomial_function} for each combination of the four historical growth-temperature response functions and three output growth assumptions. 

Median estimates under both RCP4.5 and RCP8.5 are uniformly negative in models that allow lagged effects to persist over years, whether high- and low-income provinces are assumed to respond identically (row 2 in Figure \ref{fig:projected_impact_onGDP_bothRCPs}) or differently (row 4 in Figure \ref{fig:projected_impact_onGDP_bothRCPs}). Nonetheless, projections in the "Common, 5 lags" models under RCP4.5 are more uncertain than those under RCP8.5 for all three output growth assumptions (row 2 in Figure \ref{fig:projected_impact_onGDP_bothRCPs}). This is because emission stabilization scenario RCP 4.5, which results in lower temperature rise than that of RCP8.5, makes net economic impacts less negative for cold provinces, especially those in the Upper North region. The same explanation can also be applied to the more positive median impacts observed in models under the RCP4.5 emission scenario that do not account for lagged effects (rows 1 and 3 in Figure \ref{fig:projected_impact_onGDP_bothRCPs}). 

Similar to regional projections, output growth assumptions likely have little impact on the uncertainty of the projected change in GDP per capita. The projections under RCP4.5 exhibit patterns similar to those under RCP8.5. As observed in the regional projections, projections are less uncertain and fall off steeply in earlier future years in models that allow the effects of temperature to persist on output growth (rows 2 and 4 in Figure \ref{fig:projected_impact_onGDP_bothRCPs}). As described in the main text, cold provinces could initially benefit from net, and hotter provinces remain worse off with additional warming in models that allow the effects of temperature to persist over years. However, as a given province's average temperature becomes warmer, the impacts of future warming climate worsen in each future, which consequently makes net economic impacts negative for provinces that are initially cold. The projections are more uncertain whether high- and low-income provinces respond identically (row 1 in Figure \ref{fig:projected_impact_onGDP_bothRCPs}) or differently (row 3 in Figure \ref{fig:projected_impact_onGDP_bothRCPs}) in models that do not account for lagged effects. This is because the estimated growth-temperature response functions are substantially flatter than the response function with lagged effects. Cold provinces increasingly benefits from increased average temperature, while hotter provinces remain worse off under warming climate.

\end{document}